\documentclass[]{emulateapj}
  \usepackage{natbib}
\usepackage{amsfonts}
\usepackage{amssymb}
\usepackage{amsmath,amsthm}
\usepackage{enumerate}
\usepackage{amsmath, amssymb, mathrsfs}
\usepackage{graphics}
\bibpunct{(}{)}{;}{a}{}{,}
\def\teff{T$_{\rm eff}$}

\def\logg{$\log$\,{\it g}}
\def\vt{v$_{\rm t}$}

\def\kms{km s$^{-1}$}
\def\feh{$\rm[Fe/H]$}
\def\1{SMSS~J003055.81$-$482011.3}
\def\2{SMSS~J005953.98$-$594329.9}
\def\3{SMSS~J010651.91$-$524410.5}
\def\4{SMSS~J010839.58$-$285701.5}
\def\5{SMSS~J022423.27$-$573705.1}
\def\6{SMSS~J100251.13$-$000152.1}
\def\7{SMSS~J215805.81$-$651327.2}
\def\8{SMSS~J224843.95$-$543610.1}

\slugcomment{Accepted to The Astrophysical Journal 25 March 2015}
\shorttitle{Abundance Analysis of Metal-Poor Star Candidates from the SkyMapper Survey}
\shortauthors{Jacobson et al.}

\begin{document}
\title{High-Resolution Spectroscopic Study of Extremely Metal-Poor Star Candidates from the SkyMapper Survey\altaffilmark{*}}

\author{
Heather R.\ Jacobson\altaffilmark{1},
Stefan Keller\altaffilmark{2},
Anna Frebel\altaffilmark{1},
Andrew R. Casey\altaffilmark{3,2},
Martin Asplund\altaffilmark{2},
Michael S.\ Bessell\altaffilmark{2},
Gary S.\ Da Costa\altaffilmark{2},
Karin Lind\altaffilmark{4},
Anna F. Marino\altaffilmark{2},
John E.\ Norris\altaffilmark{2},
Jos\'{e} M.\ Pe\~{n}a\altaffilmark{1},
Brian P.\ Schmidt\altaffilmark{2},
Patrick\ Tisserand\altaffilmark{2},
Jennifer M.\ Walsh\altaffilmark{1,5},
David Yong\altaffilmark{2},
Qinsi Yu\altaffilmark{1}
}

\altaffiltext{*}{This paper includes data gathered with the 6.5 meter
  Magellan Telescopes located at Las Campanas Observatory, Chile.}
\altaffiltext{1}{Kavli Institute for Astrophysics and Space Research
  and Department of Physics, Massachusetts Institute of Technology, 77
  Massachusetts Avenue, Cambridge, MA 02139, USA}

\altaffiltext{2}{Research School of Astronomy \& Astrophysics, The
  Australian National University, Cotter Road, Weston, ACT 2611, Australia}

\altaffiltext{3}{Institute of Astronomy, University of Cambridge,
  Madingley Road, Cambridge CB3 0HA, UK}

\altaffiltext{4}{Division of Astronomy and Space Physics, Department
  of Physics and Astronomy, Uppsala University, Box 516, SE-75120,
  Uppsala, Sweden}

\altaffiltext{5}{Harvard University, Cambridge, MA 02138, USA}

\begin{abstract}
 The SkyMapper Southern Sky Survey is carrying out a search for
  the most metal-poor stars in the Galaxy.  It identifies
  candidates by way of its unique filter set that allows for
  estimation of stellar atmospheric parameters.  The set includes a
  narrow filter centered on the Ca II K 3933\AA\ line, enabling a
  robust estimate of stellar metallicity.  
Promising candidates are
then  confirmed
with spectroscopy.  We present the
analysis of Magellan-MIKE high-resolution spectroscopy of 122
metal-poor stars found by SkyMapper
 in the first two years of commissioning observations.  41 stars 
have $\rm [Fe/H] \leq -3.0$. Nine have $\rm [Fe/H] \leq
-3.5$, with three at $\rm [Fe/H] \sim -4$.   A 1D LTE
abundance analysis of the elements Li, C, Na, Mg, Al,
Si, Ca, Sc, Ti, Cr, Mn, Co, Ni, Zn, Sr, Ba and Eu shows these stars
have [X/Fe] ratios typical of other halo stars.  One star with
low [X/Fe] values appears to be ``Fe-enhanced,'' while another star
has an extremely large [Sr/Ba] ratio: $>$2. Only one other star is
known to have a comparable value.  Seven stars are
 ``CEMP-no'' stars 
($\rm [C/Fe] > 0.7$, 
{\bf $\rm [Ba/Fe] < 0$}).  
21 stars  exhibit mild r-process element enhancements
($0.3 \le \rm[Eu/Fe] < 1.0$), while four stars have 
$\rm[Eu/Fe] \ge 1.0$.
These results  demonstrate the ability to identify 
extremely metal-poor stars from SkyMapper
photometry, pointing to increased sample sizes and a better
characterization of the metal-poor tail of the halo metallicity
distribution function in the future.
\end{abstract}

\section{Introduction}\label{sec:intro}

The past few decades have seen many searches for the most chemically
primitive, metal-poor stars in the Galaxy.  
Stars with $\rm [Fe/H]\footnote{In the standard
  notation [A/B] = log$_{10}$(N$_{\rm A}$/N$_{\rm B}$) $-$
  log$_{10}$(N$_{\rm A}$/N$_{\rm B}$)$_{\odot}$, where $N_{\rm
    A}/N_{\rm B}$ is the
  ratio of elements A and B by number, relative to that in the Sun
  ($\odot$).}\lesssim-3.0$
are very rare and much coveted because of the information they provide
about conditions in the early universe.  These stars are likely some
of the first low-mass stars to form in the universe after the first
chemical enrichment episodes occurred with the supernova deaths of
metal-free Population III (Pop III) stars.  
Early theoretical work on the characterstics of Pop III stars
  indicated that they were short-lived, very massive ($\gtrsim$100
  M$_{\odot}$) objects \citep{abel_sci,brommARAA}.  More recent work
  has shown that the mass range of Pop III stars may have spanned
  $\sim$3 orders of magnitude, leading to the possibility that some
  low-mass ($\sim$1 M$_{\odot}$) stars may have survived to the present day 
  \citep{hirano2014,stacy14,susa2014}.  Independent of whether a relic Pop III
  star is ever found,
the chemical compositions of the
most metal-poor stars  in the local universe 
provide a record of this first stellar
generation.

The metallicity distribution function (MDF) of the most
  metal-poor stars in our Galaxy presents a history of the formation
  process of the Milky Way.  It is a key constraint of any chemical
  evolution model that attempts to describe this process (e.g.,
  \citealt{hartwick1976}).  Early surveys for the most metal-poor
  stars in the halo (see below) indicated that the number of stars
  smoothly declined with metallicity (a factor of 10 in decline for
  every 1 dex in \feh) down to at least \feh$\sim -$3.5.
  Lower than this, some samples indicated a sharp cut-off at \feh\ =
  $-$3.6, with very few stars more metal-poor than this value
  \citep{schoerck, li_mdf}.  However, this cut-off is not seen in
  other samples \citep{yong13_III}.  
  Searches for metal-poor stars
  in part have been driven to populate the most extreme metal-poor end
  of the MDF.  We refer the reader to Frebel \& Norris (2015, ARA\&A, in
  press) and references therein for an overview of the complexities
  involved in its interpretation.

Historically, surveys searched for extremely metal-poor (EMP) stars
with $\rm [Fe/H] \leq -3$ in the Galaxy halo.  These surveys 
 exploited the stars' tendency to have large proper motions (e.g.,
\citealt{Ryanetal:1991, carneyetal:1996}) or the wide-field
capabilities of Schmidt telescopes.  Objective prism observations of
millions of stars, carried out by such landmark surveys as the HK
Survey \citep{BPSII} and the Hamburg-ESO Survey \citep{hes4} on
Schmidt telescopes led to the medium-resolution spectroscopic
follow-up of thousands of EMP star candidates \citep{norrisUBVpho,frebel_bmps, 
schoerck, li_mdf, placco_gband}.  Of these,  of order 
several hundred have been followed up with high-resolution spectroscopy
and detailed element abundance analyses 
(e.g., \citealt{McWilliametal, norris96data, Ryan96, aoki_mg,
  Francois03, cohen04, cayrel2004, lai2008, hollek11, norris13_I,
  placco2013_magII, cohen2013, roederer_313stars}).

More recently, medium-resolution spectroscopy of $\sim$10$^{5}$
stars obtained by the 
Sloan Extension for Galactic Understanding and Exploration 
(SEGUE-I; \citealt{segueI})
 and SEGUE-II 
extensions of Sloan Digital
Sky Survey \citep{york_sdss} have led to the identification of
hundreds more EMP stars.
Dozens of these have been observed with high resolution
spectroscopy (e.g.,
\citealt{aoki2008,Bonifacio2012,aoki2013,caffau2013,toposI}).
A search for extremely metal-poor stars is also underway with
the Large sky Area Multi-object fiber Spectroscopic Telescope 
(LAMOST; \citealt{zhao_lamost,cui_lamost}), 
and high-resolution spectroscopic follow-up of
the first candidates has recently been reported \citep{LAMOST_emp}.

EMP star candidate selection in objective prism surveys is based on
the strength of the Ca II K line at 3933 \AA\ in stellar spectra.
This calcium line serves as a useful proxy for overall stellar
metallicity.  It is also possible to identify EMP candidates in pure
photometric searches, but the determination of metallicity
from broadband colors is difficult due their decreased sensitivity to
metallicity at low [Fe/H] (however, see \citealt{sc14}).
The SkyMapper Southern Sky Survey \citep{keller}, being carried out with the
SkyMapper 1.3m telescope at Siding Spring Observatory in Australia, is
a new survey that takes a rather hybrid approach.  It
combines the efficiency of an all-sky photometric survey with the
power of metallicity measurements through narrow-band photometry of
the Ca II K line, similar to what has been done in objective prism surveys.

SkyMapper's filter system
 is comprised of a 
 {\it ugriz} set with the addition of a narrow Str\"{o}mgren-like filter centered on the Ca II
 K line \citep{skymapper_filter}.  The combination of colors including this narrow filter 
 provides constraints
on stellar effective temperature, surface gravity and metallicity.  A
``metallicity color index'' therefore allows for the identification of
metal-poor candidates from the photometry of the $\sim$5
billion stars potentially observable by the survey.   For more details about the
survey techniques and candidate selection, see Keller et al.\ (2015,
in preparation).

The most promising SkyMapper metal-poor candidates are selected for
follow-up spectroscopic observation at both medium- and
high-resolution.
3198 candidates were selected from 5,452,735 stars in 195 SkyMapper
fields. 1127 were followed up with medium-resolution
spectroscopy, and 259 with high-resolution spectroscopy.  Indeed, one such candidate
was verified as being the most Fe-poor star to-date via medium and high-resolution
spectroscopic follow-up.  The discovery and abundance pattern of SMSS~
J031300.36$-$670839.3, with $\rm [Fe/H] < -7.1$, has already been reported \citep{keller_thestar}. 
The addition of this star raises the number of stars known to have
$\rm [Fe/H] \leq -4.5$ to six \citep{HE0107_Nature, HE1327_Nature,
  he0557, caffau2011, keller_thestar, hansen2014}, and the SkyMapper EMP
star candidate selection technique shows the promise of finding more
of these stars.  A corresponding survey for the most metal-poor and
oldest stars in the Galactic bulge using SkyMapper photometry and
AAOmega/AAT multi-object spectroscopy is also underway (P.I.\
M.\ Apslund; see \citealt{howes2014} for first results).

In this work, we present results of the high resolution
spectroscopic follow-up of other metal-poor stars candidates identified by
SkyMapper from 2011 to 2013 November.  As described in S.\ Keller
et al.\ (2015, in preparation), the SkyMapper photometry used to select
these candidates was obtained during the commissioning phase of the survey.
Section~\ref{sec:obs} describes the target selection, observations and data,
and Section~\ref{sec:analysis} describes our method of analysis.
Results are presented in Section~\ref{sec:results}, while a discussion and
summary are given in Sections~\ref{sec:disc} and \ref{sec:summary}.

\section{Target Selection, Observations and Data Reduction}\label{sec:obs}
Most of the candidate metal-poor stars were first observed with the
Wide Field Spectrograph on the ANU 2.3m telescope, providing 
medium-resolution (R$\sim$3000) optical spectra.
Stellar parameters and metallicities were estimated based on a
  comparison of the spectra to a library of synthetic spectra
(S.\ Keller et al., in preparation).
However,
for some early Magellan observing runs (2012 February, 2012 May), 
candidates were selected based on metallicity estimates from their
SkyMapper photometry alone.  At that time, the color-metallicity calibration of the
SkyMapper filter set was still being developed and improved, so
a number of the candidates turned out to be metal-rich.  
A preliminary analysis of 160 stars from these early campaigns found 46 stars to have $\rm[Fe/H]
\ge -1$, 85 stars with $-2 \le \rm[Fe/H] < -1$, and 29 stars to have
$\rm[Fe/H]< -2$.
We have therefore made a metallicity cut, and here are presenting
the results only for 24 stars with $\rm [Fe/H]\lesssim-2.2$, as measured
from high-resolution spectra.  For the later observing runs
(2012 September forward), all candidates were selected from
medium-resolution spectroscopy, and we analyzed all stars observed in
these runs including the metal-rich ones.  (Ten stars have \feh\ $> -2.2$.)  
Nine candidates were selected based on preliminary photometry that did
not pass subsequent quality cuts.  As such, these objects do not have SMSS
photometry or coordinates and instead we have adopted 2MASS identifiers and
coordinates \citep{2MASS}.

\begin{figure*}[!ht]
\begin{center}
   \includegraphics[clip=true,width=18cm]{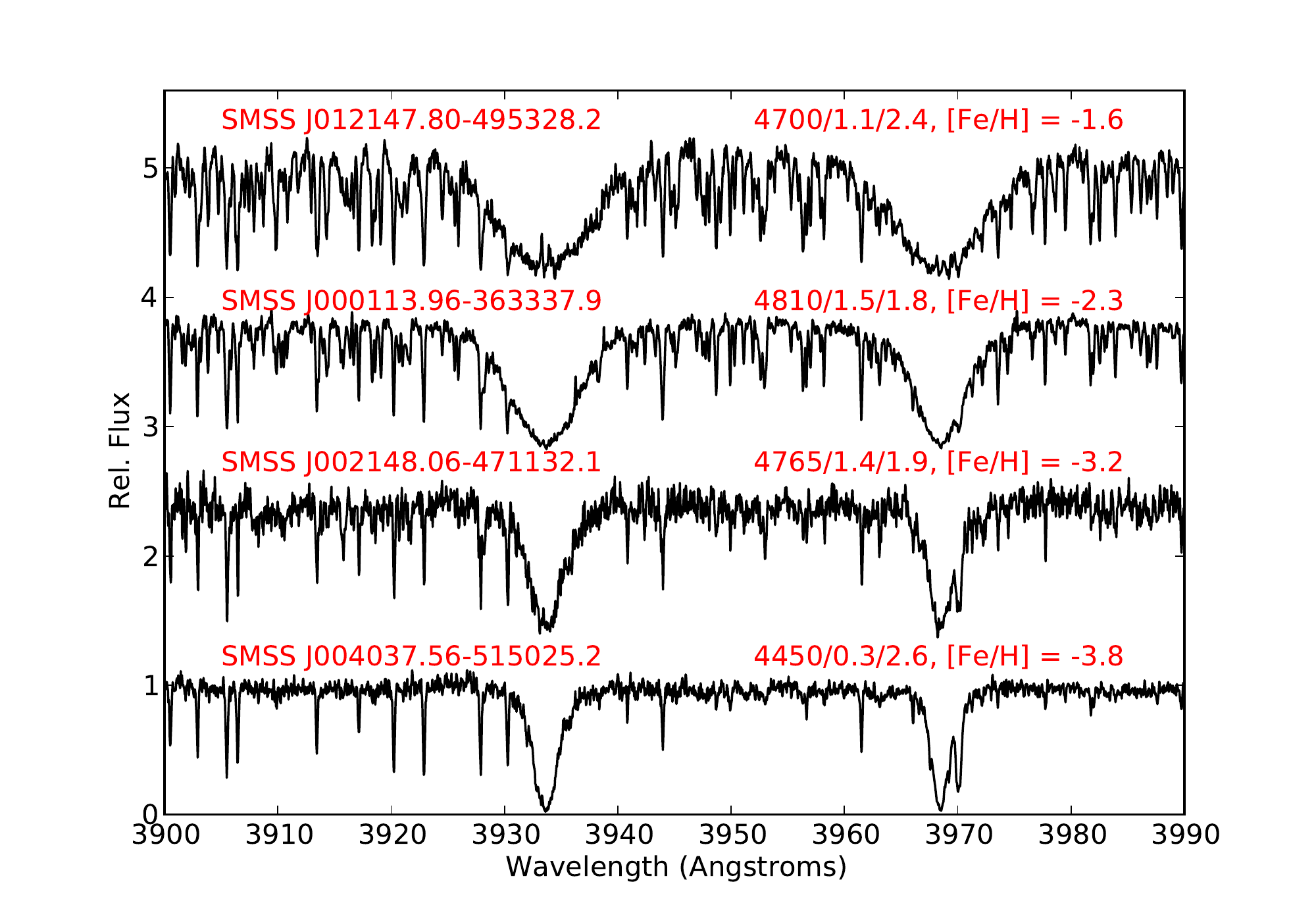} 
     \figcaption{ \label{spec}
       Portions of the MIKE spectra for four
       stars in our sample around the Ca II H and K lines.  LTE stellar
       parameters and metallicities determined in our analysis are
       also indicated (``\teff/\logg/\vt, \feh'').  Note the variations in line strength with
       decreasing [Fe/H] (top to bottom).}
 \end{center}
\end{figure*}

The spectra analyzed in this work were obtained with the {\it Magellan}
Inamori Kyocera Echelle (MIKE) spectrograph on the 6.5m {\it Magellan} Clay
Telescope at Las Campanas Observatory \citep{mike}. Observations
spanned multiple
campaigns from 2011 to 2013 November.  Depending on
sky conditions, spectra were obtained with either the $0\farcs7$ or
$1\farcs0$ slits, resulting in spectral resolutions ($\rm R \equiv
\lambda/\Delta\lambda$) of R$\sim$35,000 in the blue and R$\sim$28,000
in the red, and R$\sim$28,000 in the blue and R$\sim$22,000 in the
red, for the smaller and larger slit sizes, respectively.   
Exposure
times generally ranged from 300 to 1800 s, depending on the
brightness of the target, to obtain ``snapshot'' spectra with which
to confirm EMP star candidates.  
For some of the fainter, more promising
targets, multiple (2$-$4) exposures were obtained to increase the 
signal-to-noise ratios (S/N).  All spectra
span nearly the full optical range, 3350-9000\AA, but the
S/N was
generally too low blueward of $\sim$3800\AA\ for analysis. 
Details of the observations are given
in Table~\ref{Tab:obs}, including  star ID number, J2000
coordinates, {\it g} magnitude, and {\it g$-$i} color from the
SkyMapper observations (see S. Keller et al.\ 2015, in preparation for details of the
SkyMapper photometric system).  Also included are 
the UT date, the total
exposure time in seconds, the slit size used, the measured radial
velocity (see next Section), and the S/N ratios (per pixel) measured at
4500 and 6000\,\AA. The average~($\pm$1$\sigma$) S/N ratios of the sample
are 35($\pm$15) and 55($\pm$21) at 4500 and 6000\,\AA, respectively.

All spectra were reduced using the CarPy data reduction
pipeline\footnote{See http://code.obs.carnegiescience.edu/mike.}
described in \citet{kelson03}. 
Individual exposures were combined to increase S/N, 
individual orders were merged, and the blue and red spectra combined and
then continuum normalized to create one continuous spectrum
per star.

Example MIKE spectra for four stars are presented in
Figure~\ref{spec}, all obtained with the $0\farcs7$ slit.  Their stellar
parameters and metallicities, as determined in our analysis, are also
indicated.  All told, the total sample of analyzed stars is 122.
For candidates selected based on metallicity
estimates from  medium-resolution 
spectroscopy, agreement between [Fe/H]$_{\rm MRS}$ and 
[Fe/H]$_{\rm HRS}$ measured
in this work generally agree at the 
0.3 dex level (Keller et al.\ 2015, in preparation).

\section{Analysis}\label{sec:analysis}

Our abundance analysis software \citep{smh}
incorporates the Castelli \& Kurucz 1D LTE hydrostatic model atmosphere grid
\citep{castelli_kurucz} and the version of the LTE abundance analysis program MOOG
that includes treatment of Rayleigh scattering
(\citealt{moog, sobeck11}).  Our linelist was that compiled by
\citet{roederer10}, though we only considered lines within the
wavelength range 3500$-$6500\AA.  

Radial velocities for our stars were determined
via cross-correlation of the Ca triplet ($\lambda$8450-8700) and/or Mg
Ib ($\lambda$5150-5200) region of the spectra against that of 
a high S/N, rest frame MIKE spectrum of the metal-poor
standard star HD~140283.
Velocity errors due to the cross-correlation technique are generally
small (0.1$-$0.3 \kms).  Based on repeat observations of standard stars
such as HD~122563, G64$-$12 and HD~13979 during each observing run, we
estimate a zero-point offset to our radial velocity scale of 1$-$2 \kms,
with our values being smaller than ones in the literature for these
stars.

Four stars were observed in two separate observing runs, and two stars 
were observed in three different campaigns.  The radial
velocities determined from the different spectra of these stars also
show differences
of 1$-$2 \kms, with the exception of the star SMSS~J022410.38$-$534659.9,
 which
shows a variation of $\sim$14 \kms.  This star may be a single-lined
spectroscopic binary, however, its stellar parameters (see next Section)
place it on the edge of the instability strip in the
Hertzsprung-Russell diagram, so it may instead be a variable.
For those stars observed with the same slit size in multiple
observing runs, all the spectra were combined to increase S/N before
the abundance analysis. For those stars observed with different slit
sizes, the spectrum with the highest S/N was analyzed; in cases where
S/N levels were comparable, the $0\farcs7$ spectrum was
analyzed\footnote{Generally, these candidates were photometrically
  selected more than once and given
  two different identifiers, and were only identified as duplicates
  during the high-resolution spectroscopic analysis.}.

Heliocentric
radial velocities for each of the stars are given in
Table~\ref{Tab:obs}.  As can be seen, many have large heliocentric
 velocities as 
expected for halo stars. All spectra were shifted
to rest wavelength for the abundance analysis described in the next
section.

\subsection{Determination of Stellar Parameters}\label{sec:stelpars}

The stellar parameters for each star were determined solely from its 
MIKE spectrum 
using the standard spectroscopic techniques: effective temperature (\teff) 
by removal of any slope of Fe\,I abundance with excitation 
potential (E.P.), \logg\ 
by matching Fe\,I and Fe\,II abundances, microturbulence
(\vt) by removal of
any slope of Fe\,I abundance with reduced equivalent width (REW).
In this process, individual lines with abundances $\sim$2$\sigma$ 
away from
the mean were visually inspected, reassessed for measurement quality,
and if necessary, rejected (due to blending, uncertainty in continuum
placement, etc.).  Our general tolerances were as follows: slope of
log $\epsilon$(Fe\,I) versus E.P. $<$ 0.005 dex/eV; [Fe\,II/H] $-$
[Fe\,I/H] $<$ 0.05 dex, and slope of log $\epsilon$(Fe\,I) versus log(RW) $<$ 0.005.
The [M/H] of the model atmosphere was set to [Fe\,I/H]$+$0.25, as
described in \citet{teff_calib}.

Spectroscopic effective
temperatures are generally cooler than photometric temperatures
due to departures from local thermodynamic
equilibrium (LTE;
\citealt{johnson2002_23stars,cayrel2004,lai2008,hollek11,lind2012,cohen2013}).
The use of 1D models as opposed to time-dependent 3D or temporally 
and spatially averaged 3D ($<$3D$>$) models can also lead to this effect
\citep{asplund_araa,bergemann12}.  Too-cool
temperatures translate into smaller \logg\ and larger \vt\ values
than would be found using photometric temperatures.  
We have adopted the
effective temperature correction presented in \citet{teff_calib}.  It places
spectroscopically-determined temperatures on a scale similar to that found by
photometric temperature methods.  This correction is appropriate for the 
program stars, and the majority of them span the metallicity range for which the correction has
been 
tested ($-3.3<\rm [Fe/H]<-2.5$).\footnote{Recall the caveat in that paper
  that the calibration may not be valid for stars with 
$\rm [Fe/H]<-4.0$.  The lowest metallicity of any star in this
  sample is $-$4.  We note that although this correction was
  not tested for stars with $\rm [Fe/H]>-2.5$, such stars have been
  known to have similar discrepancies between their photometric and
  spectroscopic \teff\ values (see, e.g., \citealt{johnson2002_23stars}).}
The final adopted
(corrected) spectroscopic temperatures were checked by visual inspection of
H$\alpha$ and H$\beta$ line profiles, in comparison to stars
of previously determined 
effective temperature.  Surface gravity and \vt\ were then adjusted to
maintain ionization balance and remove any trend of Fe\,I abundance
with line strength, as necessary.

\begin{figure}[!ht]
\begin{center}
   \includegraphics[clip=true,width=8cm]{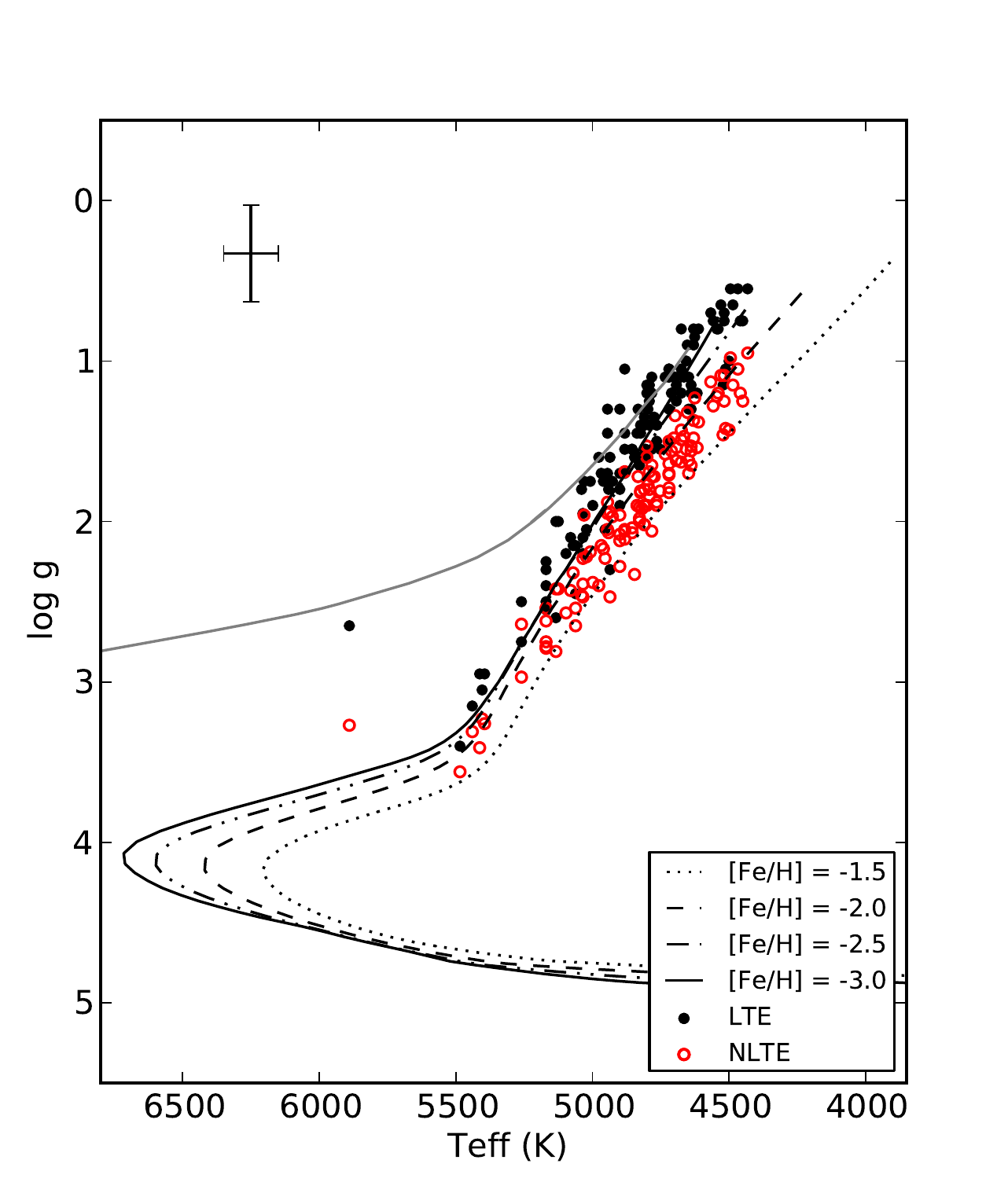}   
    \figcaption{ \label{f_isochrone}
      The Hertzsprung-Russell diagram for the current sample, plotted
      with isochrones from \citet{Y2_iso} and a horizontal branch
      isochrone from \citet{basti_HB}. LTE parameters are shown as
      filled circles, and NLTE parameters are open red circles.}
 \end{center}
\end{figure}

Stellar parameters for our program stars are presented in
Table~\ref{Tab:stellpar}.  Note that the metallicities are relative to
the solar abundance from \citet{asplund09}.  We have also calculated
1D non-LTE (NLTE) \logg\ and \feh\ values for the stars following the method
described in \citet{ruchti2013}, and using the NLTE grid of
\citet{lind2012}. These values are also given in Table~\ref{Tab:stellpar}.

Figure~\ref{f_isochrone} shows the positions of the stars in this
study in the Hertzsprung-Russell diagram. 12
Gyr $\alpha$-enhanced isochrones with [Fe/H] = $-$3.0, $-$2.5, $-$2.0,
and $-$1.5 from \citet{Y2_iso}, and a 12-Gyr, [Fe/H] = $-$2.2
BaSTI horizontal branch isochrone \citep{basti_HB} are also shown.   
Filled symbols indicate the LTE \logg\ values, and open circles
represent the NLTE \logg\ values.  As expected, \logg\ values
calculated in NLTE are $\sim$0.4-0.5 dex larger than the LTE gravities.

\begin{figure*}[ht!]
\begin{center}
   \includegraphics[clip=true,width=15cm]{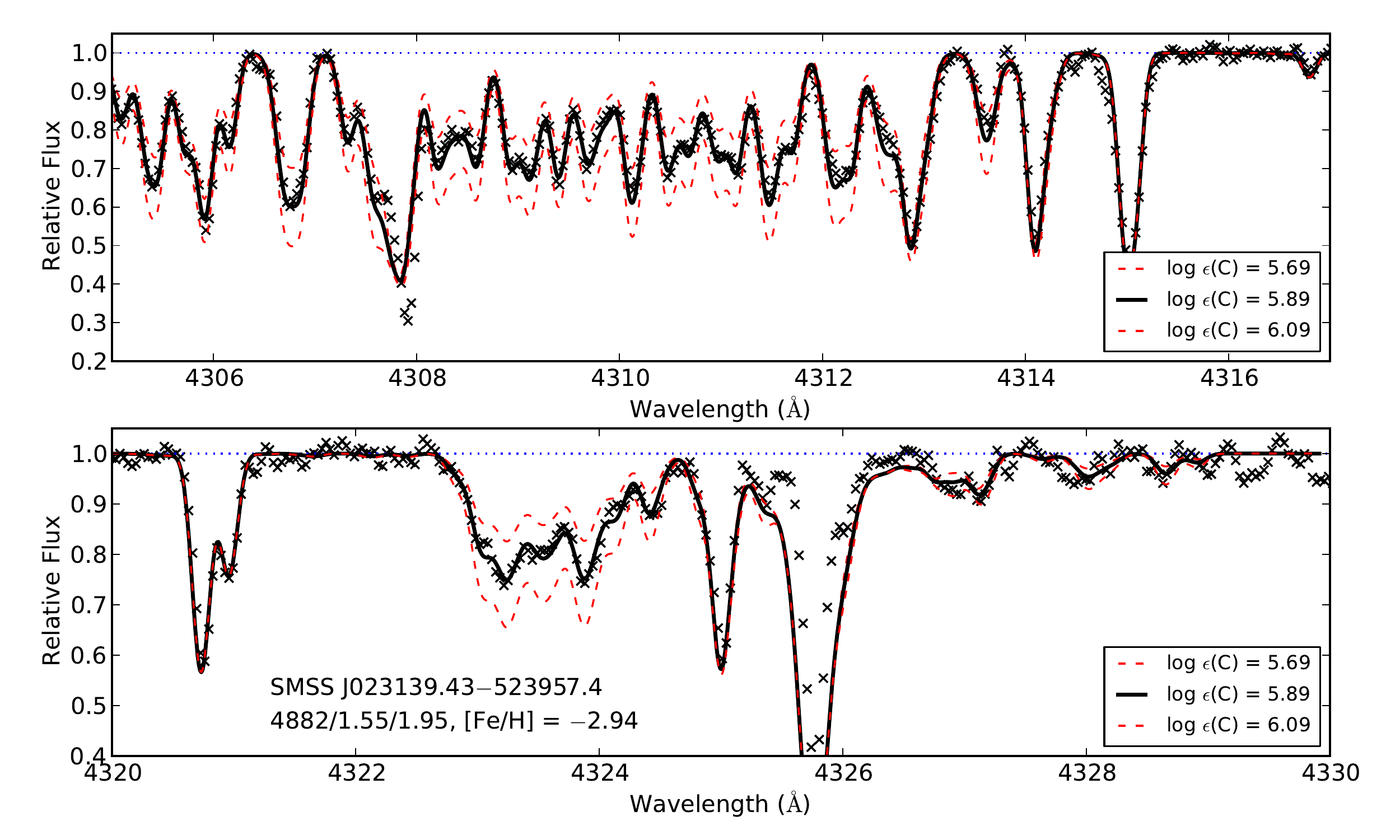} 
     \figcaption{ \label{f_csynth}
     Syntheses of the CH bands at 4305$-$4317\AA\ (top) and
     4320$-$4330\AA\ (bottom) in the star SMSS~J023139.43$-$523957.4.  
     Its stellar parameters are also listed as ``\teff/\logg/\vt''.
The
     synthetic spectra have C abundances varying in steps of 0.2 dex.
     The best fit, log $\epsilon$(C) = 5.89, or [C/Fe] = $+$0.40, is
     the solid black line.  Spectra with C abundances $\pm$ 0.2 dex
     around this value are shown as red dashed lines.}
 \end{center}
\end{figure*}

All the stars in this sample are giant stars, many of which lie
on the upper part of the giant branch.  While this means that the
abundances of elements modified by stellar evolution (i.e., Li, C, N,
O, s-process) will not reflect
the primordial values of some of the stars, we do not have to worry about
systematic differences in abundances between dwarfs and giants as found
in some studies (see, e.g., \citealt{yong13_II}).  
The likely reasons for the lack of dwarfs in this sample are discussed
in Keller et al.\ (2015, in prep.).

\subsection{Element Abundance Determination}\label{sec:abund_analysis}

Equivalent widths (EW) of lines in our line list were
measured via 
fits of Gaussian profiles to absorption features in the spectra of the
program stars.  All measures were visually inspected.  Any that were
identified as outliers in the analysis process were further
scrutinized and, where appropriate, adjusted or rejected.
The Fe\,I and Fe\,II EWs were used to determine the
stellar parameters (see Section~\ref{sec:stelpars}); in high quality
spectra of generally more metal-rich stars,
$\sim$200 Fe\,I and $\sim$20 Fe\,II lines were
used, while in the lower S/N
star spectra, as few as 22 Fe\,I and $\sim$2 Fe\,II lines
could be measured.  \citet{teff_calib} compared EW measures for lines
in our line list using the same technique in this current work to
literature measures for the standard star HD~122563.  The agreement
was excellent (differences less than 0.25 m\AA\ in the mean).

The stars subject to repeat observations also allow for a
quantitative estimate of the robustness of our EW measures.  For each
star, the EW's measured from each spectrum were directly compared.  
The mean difference, in the sense 
$\rm \Delta EW = (EW1 - EW2)\pm(\sigma/\sqrt N))$, 
ranged from $0.5\pm1.3$ m\AA\ to $1.7\pm0.5$ m\AA, with standard
deviations ranging from 8 m\AA\ to 29 m\AA. 

The number of lines available for EW measurement varied widely for the
other elements, with as many as 29 Ti\,I and 46 Ti\,II lines available,
but typically only one line of, e.g., Al\,I and Si\,I.
  Chemical abundances for
elements were determined using the measured EW values and the stellar
parameters found from the iron lines.  
Table~\ref{tab_ew_stub} gives the EW measures for all program
  stars, along with the measured log$\epsilon$ abundances.

The following absorption features  were
analyzed via spectrum synthesis: 4313\AA, 4323\AA\ (CH, G-band); 
4246\AA\ (Sc\,II); 4030\AA, 4033\AA, and 4034\AA\ (Mn\,I);
4077\AA, 4215\AA\ (Sr\,II); 4554\AA, 4934\AA\ (Ba\,II); and 4129\AA\ (Eu\,II).  
Additional Mn\,I and Sc\,II lines were synthesized for 54 stars in the
sample, along with the Al\,I 3944\AA\ feature.
In each synthesized region, the
abundances of elements other than the one of interest were fixed to
the value determined via EW.

Synthetic spectra were generated with
MOOG and then convolved with a Gaussian to match the resolution of the
MIKE spectra.  Where necessary, the continuum-placement of the data
 was adjusted and the spectrum radial-velocity shifted to
correct for subtle wavelength differences.
Abundances were then determined by minimizing the difference between
the observed and synthetic spectra by eye.  
The uncertainty of the spectral matching  was then
determined by decreasing the stepsize (in abundance space) between
three synthetic spectra until the best match could no longer be
uniquely 
identified.  Example syntheses are given in Figure~\ref{f_csynth}.  
 Abundance results from spectrum synthesis (excluding CH) are also 
given in Table~\ref{tab_ew_stub}.

Element abundances of the stars in this sample are presented in
Table~\ref{tab_abund}, relative to the solar abundances of
\citet{asplund09}.
Abundance upper limits obtained via spectrum synthesis are
  identifed as such in the Table.

\subsection{Error Analysis}\label{sec:errors}

The abundance uncertainties in our analysis are a combination of both
random uncertainties (e.g., in the EW measures,
etc.) and systematic uncertainties (due to continuum-placement, the adopted temperature
scale, model atmosphere grid, etc.).  Based on the spectroscopic
techniques used here to determine the stellar parameters, we estimate
their uncertainties to be $\sim$100 K, 0.3 dex and 0.2 \kms\ for
\teff, \logg\ and \vt, respectively.  The contribution of each of
these to the abundance uncertainty of each element was determined by
varying each parameter by its uncertainty and recalculating the
abundance.    
Table~\ref{tab_unc} lists the abundance uncertainties of
individual elements for both a warmer and cooler example star from our sample.
The random uncertainty ($\sigma$) listed in the third column is
the 
standard error in the mean of individual line abundances for each
element.  
In the cases where only one
line was measureable, this value is placed conservatively at 0.2
dex, as appropriate for the low S/N ratio of many of our ``snapshot''
spectra.  The last column shows the quadratic sum of the individual
uncertainties.

\section{Results}\label{sec:results}

In this section, we review the abundance results for this SkyMapper
sample for individual elements, divided roughly by group in the
periodic table.   Unless stated otherwise, the
  abundances presented here are LTE abundances.

\begin{figure}[!ht]
\begin{center}
   \includegraphics[clip=true,width=8cm]{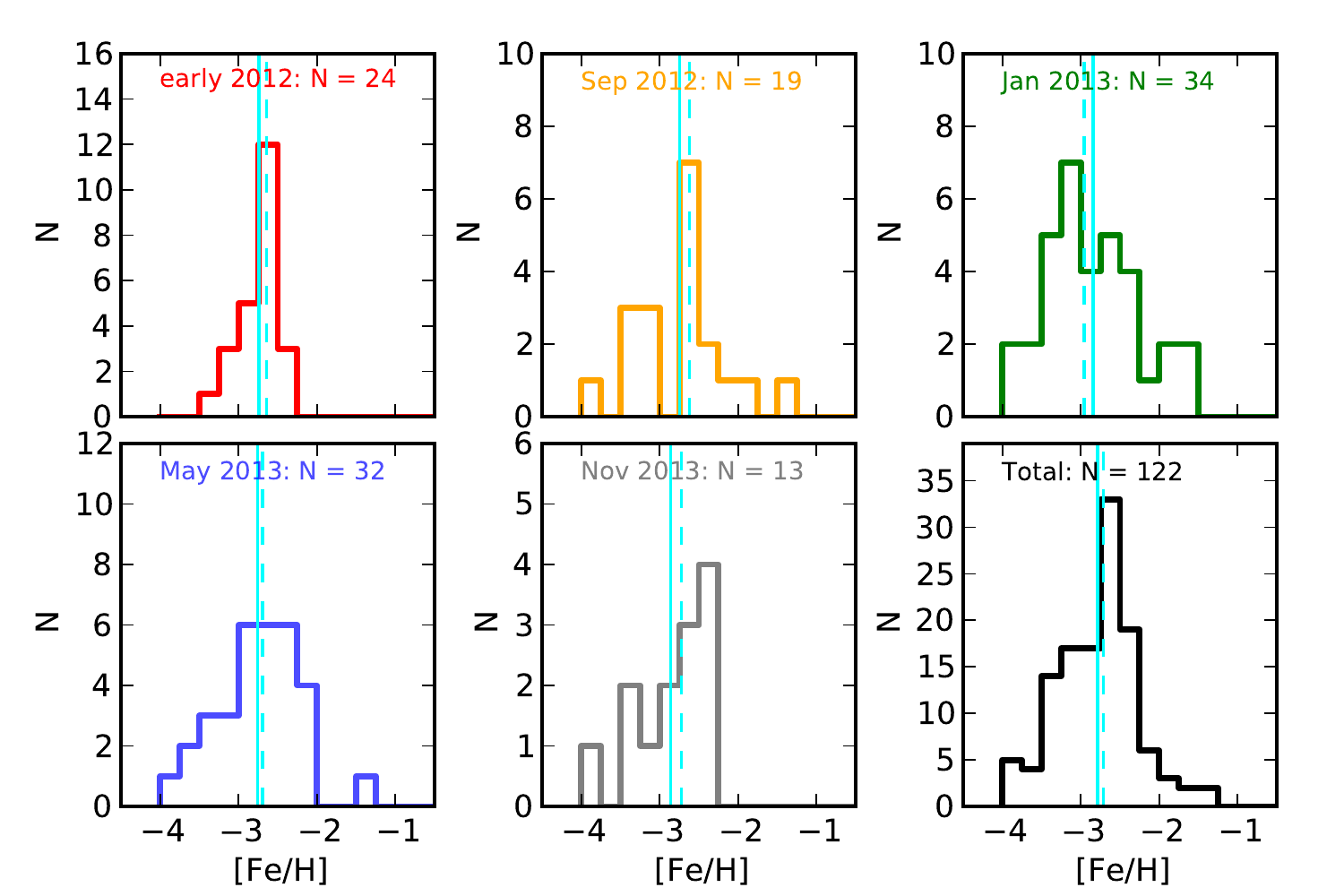} 
     \figcaption{ \label{f_mdf}
     The (LTE) metallicity distribution function of the Skymapper sample,
     separated by observation date, with the total sample in the
     bottom right hand panel.  The bin size is 0.25 dex everywhere.
     The mean and median $\rm[Fe/H]$ values are indicated by solid and
     dashed cyan lines, respectively.  Note that the ``early 2012''
     distribution is incomplete at $\rm[Fe/H] \gtrsim -2.5$.}
 \end{center}
\end{figure}

First though, Figure~\ref{f_mdf} shows the distribution of
metallicities of the SkyMapper metal-poor candidates in different
observing campaigns.  
The top left panel shows only 24 most metal-poor
stars of the first 160 followed up with high-resolution spectroscopy,
as previously discussed.  The remaining panels save the last include all of the
stars observed in each campaign.  Stars observed in multiple campaigns (see
  Table~\ref{Tab:obs}) are distributed according to their first
  observation date.
  The bottom right panel shows the distribution of
the entire sample.  In each panel, the mean and median \feh\ values
are indicated by cyan solid and dashed lines, respectively.

As can be seen, the mean and median \feh\ values do not vary much from
\feh\ $\sim -$2.8 in each panel, though the metal-poor tail is
especially evident in the total sample (bottom right panel).  
Despite the fact that the photometric
candidate selection technique was improved during the accumulation of
this sample, no obvious improvement is seen in the distributions of the individual panels of
Figure~\ref{f_mdf}.  This is largely due to the relative rarity of
stars with \feh\ $< -3.5$ in the Milky Way halo and the necessity of
observing
more metal-rich targets due to the lack of more interesting metal-poor
candidates in some runs (e.g., 2013 May).  
That said, 92 of the 122 stars (75\%) have
\feh\ $\le -2.5$; 51 have $-3 < \feh \le -2.5$ (42\%); 32 have $-3.5 < \feh
\le -3$ (26\%); and nine have \feh $\le -3.5$ (7\%).  Keep in mind
these numbers have not been corrected for any biases.  Indeed,
since these candidates were selected from commissioning data, the
distributions in Figure~\ref{f_mdf}
should not be interpreted as {\it the} metallicity distribution function of
stars identified in the SkyMapper Survey.  This will be the subject of
future work.

\subsection{Lithium}

\begin{figure}[!ht]
\begin{center}
   \includegraphics[clip=true,width=8cm]{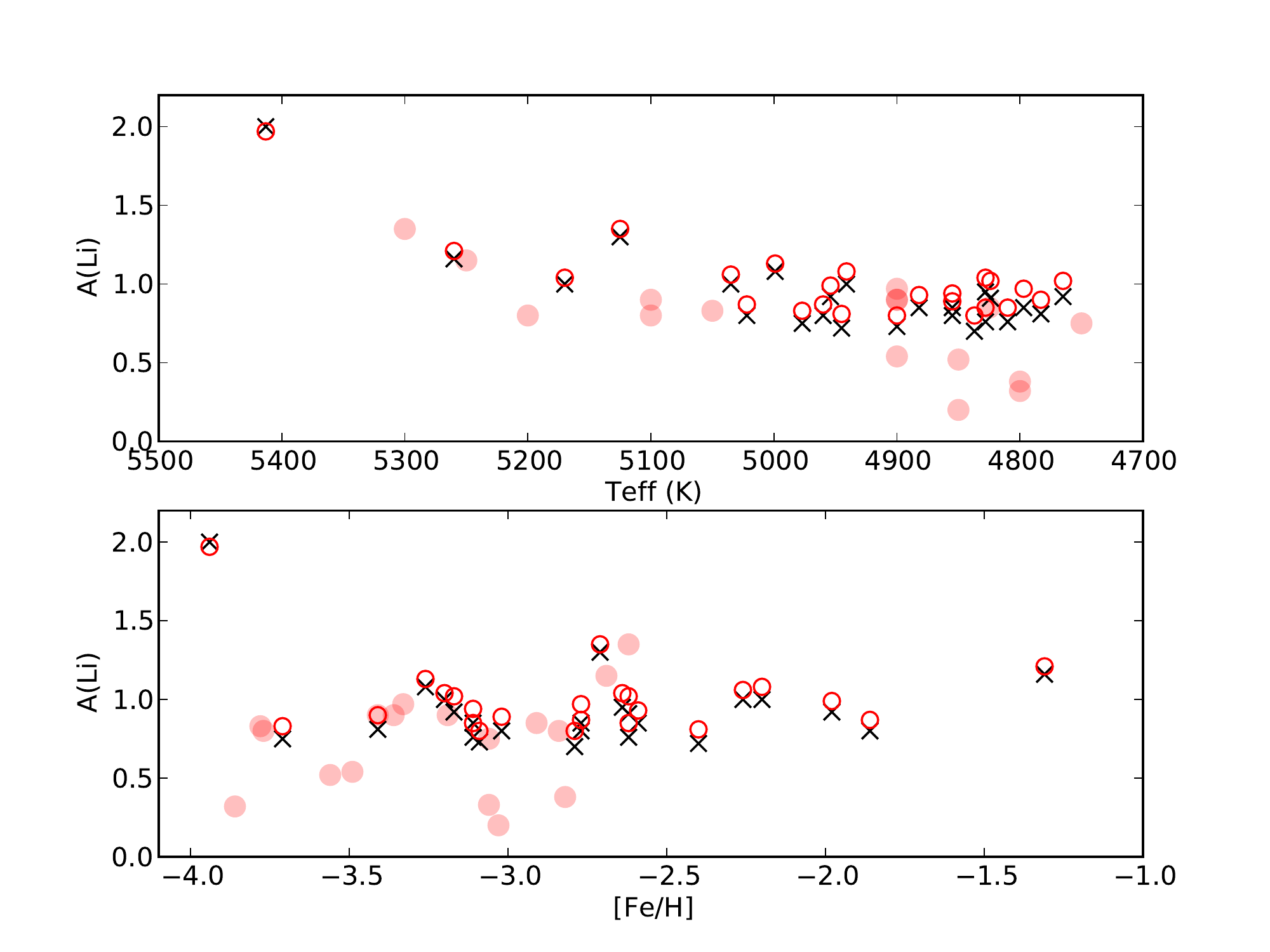} 
     \figcaption{ \label{f_li}
     The Li abundances of stars showing $\lambda$6707 absorption as a
     function of T$_{\rm eff}$ (top panel) and LTE $\rm [Fe/H]$ (bottom
     panel).  LTE and NLTE abundances are shown as crosses and open
     circles, respectively.  Giant stars with Li measures from the
     sample of \citet{spite2005} are shown as filled red circles.}
 \end{center}
\end{figure}

The Li\,I 6707\AA\ feature was detected in the spectra of 24 stars.
We determined LTE Li abundances via spectrum synthesis using the line
list of \citet{hobbs99} and assuming a pure $^{7}$Li component.
These are given in Table~\ref{tab_li}, along with 
measured EWs (in m\AA) and NLTE Li abundances
calculated using the grid of \citet{lind_li}.  
Figure~\ref{f_li} shows log $\epsilon$(Li) = A(Li) (LTE: crosses, NLTE: open circles) as a
function of \teff\ and \feh, along with those of giant stars from the sample of
\citet{spite2005}.  (We only show their stars with Li measures, no upper
limits.)  As can be seen, the distributions of Li abundances for the
two samples are similar.  The second most metal-poor star in our
sample, \2\ (\feh $\sim -$4), has the largest Li abundance, A(Li) =
2.0 (LTE), but at a depletion level appropriate for its \teff.  
Note that NLTE corrections to the Li abundances are small
for these stars: no more than 0.12 dex.

\subsection{Carbon}\label{sec:carbon}
During the ascent up the red giant branch, the surface C abundance of
a star 
decreases due to dredge-up of CN-processed material.
\citet{placco14} provide C abundance 
corrections as a function of surface gravity and metallicity
to take this effect into account. 
We have corrected the carbon abundances of our sample stars 
accordingly since we are interested in the stars' natal carbon 
abundances and whether their birth gas clouds were particularly 
enhanced in carbon. In Table~\ref{tab:cfe_corr} we provide uncorrected and 
corrected [C/Fe] values.  The corrections from \citet{placco14}
  were calculated adopting [N/Fe] = 0.0.

\begin{figure}[!ht]
\begin{center}
   \includegraphics[width=8cm]{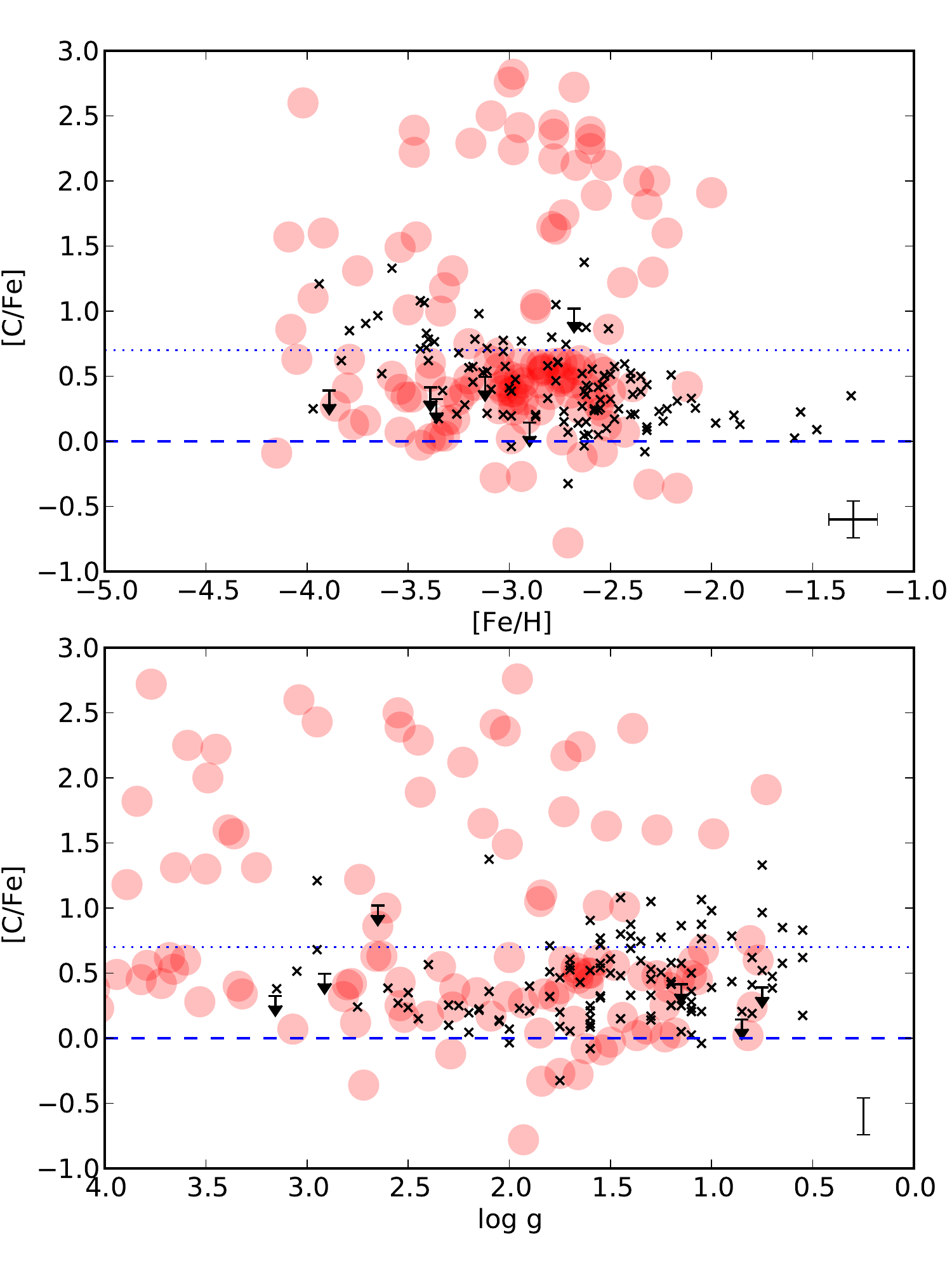} 
     \figcaption{ \label{f_carbon} Top panel:
     Corrected [C/Fe] abundances versus (LTE) [Fe/H] for our sample (crosses)
     compared to the sample of
     \citet{yong13_II} (red circles; also corrected). Upper limits are denoted by
     arrows.  
     Bottom panel: Corrected [C/Fe] abundances versus surface gravity.  
     The CEMP definition of
     \cite{aoki_cemp_2007} is indicated by a dotted line.}
 \end{center}
\end{figure}

Corrected carbon abundances are shown as crosses in
Figure~\ref{f_carbon}, along with upper limits as arrows.  
For comparison, in this and the following figures we
plot our results against those of the {\it giant} stars 
in \citet{yong13_II} (red circles, also corrected).\footnote{We consider here the giant stars from
  both their literature compilation and their own sample.
We note that the stellar parameter determination by \citet{yong13_II}
and the line list they used were different from those used here and so
systematic differences between results may exist.}
 The bottom panel of Figure~\ref{f_carbon} shows [C/Fe] for this
sample plotted against surface gravity. 
The carbon abundances exhibited by the SkyMapper sample 
are overall typical for stars found in the halo. 

\begin{figure*}
\begin{center}
   \includegraphics[clip=true,width=18cm]{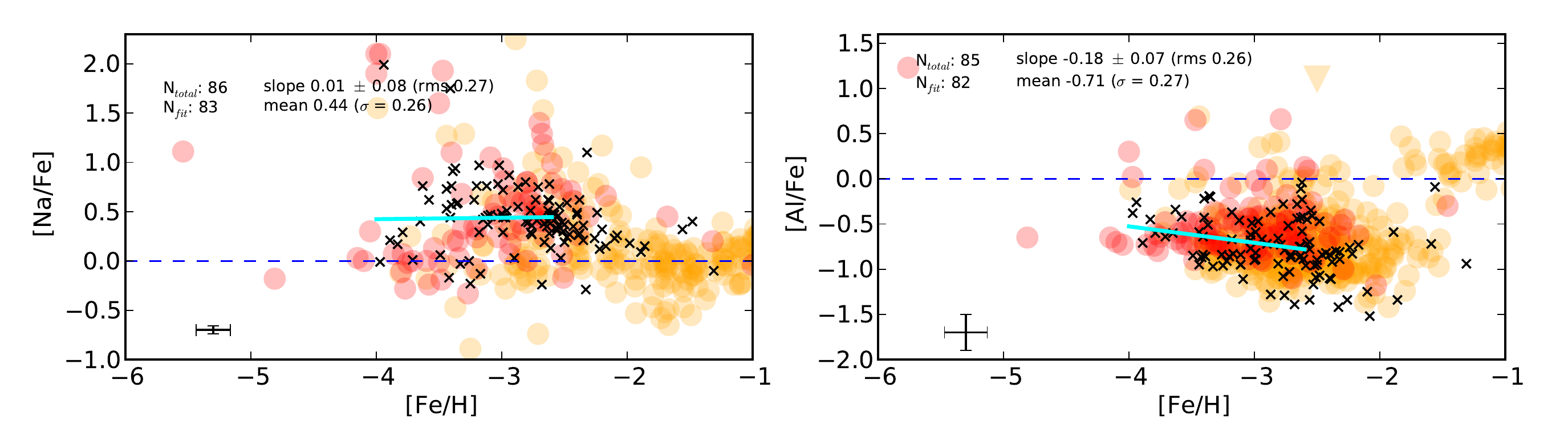} 
     \figcaption{ \label{f_light}
     LTE [Na/Fe] and [Al/Fe] ratios versus (LTE) \feh\ for the SkyMapper sample (crosses) 
     compared to the sample of \citet{yong13_II} (red circles) and
     the literature compilation of \citet{frebel10} (orange circles).  A
     least-squares fit to SkyMapper stars with (LTE) $\rm [Fe/H] \leq -2.5$,
     excluding 2$\sigma$ outliers, is indicated by the cyan line.
     Parameters of the least-squares fit are also given.}
 \end{center}
\end{figure*}

Also indicated in the Figure is the definition for carbon enhanced metal-poor
(CEMP) stars of $\rm [C/Fe]\ge0.7$ (dotted line), following \citet{aoki_cemp_2007}.
Considering the corrected carbon abundances, we determine the 
frequency of CEMP stars to be 20\% (24/120) for the total sample. 
The frequency is 21\% (24/113) for stars with $\rm [Fe/H] \le -2.0$, 26\% (24/91) 
for $\rm [Fe/H] \le -2.5$, 39\% (16/41) for $\rm [Fe/H] \le -3.0$, and 
56\% (5/9) for stars with $\rm [Fe/H] \le -3.5$. 
For comparison, using 505 metal-poor stars from the literature 
with $\rm [Fe/H] \le -2.0$ and corrected carbon abundances, 
\citet{placco14} determined these frequencies to be 20\%, 24\%, 43\%
and 60\%, 
respectively. Our values agree very well with theirs. 
We note that CEMP-s and CEMP-rs stars have been excluded from the 
Placco et al. sample; since our sample does not contain any of these 
stars either, the comparison between these samples is 
appropriate.  

Interestingly, our sample contains only seven stars with $\rm [C/Fe]>1.0$,
of which five have $\rm [C/Fe]\sim1.0$. Star SMSS~J173823.36$-$145701.0 has a corrected carbon 
abundance of $\rm [C/Fe] = 1.33$, which is the highest in the sample.
 (Its uncorrected $\rm [C/Fe]$ is 0.60; the 
gravity is low, \logg\ = 0.75, which leads to the large correction.) 
It has $\rm [Fe/H] =-3.58$. \2, 
at \feh\ = $-$3.94, has the second highest [C/Fe] ratio of $+$1.21 
(no carbon correction because of \logg\ = 2.95). These two, together 
with the other five CEMP stars, are thus prominent examples of CEMP-no 
group: they lack enhanced neutron-capture element 
abundances, and their other [X/Fe] ratios are comparable to those of
typical halo stars at similar metallicity
 \citep{fnARAA}. Since five of the seven
stars have \feh\ $< -$3.4, we confirm that CEMP-no stars 
preferentially appear at the lowest metallicities, i.e., below \feh\ $< -$3.0.

It is worthwhile asking why no stars with [C/Fe] $\gtrsim$2 appear in
our sample.  Such stars typically exhibit large enhancements of
s-process elements like Ba (CEMP-s stars, mentioned above).  Very
strong G-band absorption may change the colors of a star, moving it
out of the range used for candidate selection.  This is currently under
investigation (S.\ Keller et al.\ 2015, in preparation).

\subsection{Na and Al}

Figure~\ref{f_light} presents the LTE [X/Fe] ratios for Na and Al versus
LTE [Fe/H] for our sample. Also shown are the giant star sample from
\citet{yong13_II} and the Milky Way halo star literature sample from
\citet{frebel10} (orange circles).\footnote{In this and following figures that include
  both the \citet{yong13_II} and \citet{frebel10} samples, all stars
  in the former sample have been excluded from the latter.  The
  \citet{frebel10} sample is a mixture of dwarf and giant stars.}  
In this and following figures, we have performed a linear regression
analysis on the [X/Fe] versus [Fe/H] distributions of our sample in
order to compare our results to other studies in the literature.
Following \citet{yong13_II}, we restricted the sample to 
 $\rm [Fe/H] \leq -2.5$ and calculated the rms scatter of
points about that fit.  Stars with [X/Fe] ratios more than 2$\sigma$
away from the fit were excluded and the linear regression redone.  The
resulting line of best fit, excluding the 2$\sigma$ outliers, is shown
in each panel (cyan line), along with its slope, the slope error, and the
rms scatter about the slope.  Also shown are the mean [X/Fe] ratios
and standard deviations, the
total number of stars, and the number of stars used in the fit.

As can be seen, the SkyMapper stars exhibit
the $>$1 dex spread in [Na/Fe] found in other studies of metal-poor stars.
The [Na/Fe]$\sim$0 for stars with $\rm [Fe/H] > -2$ is also
consistent with other stars in this metallicity range.  There is no
significant change in [Na/Fe] as a function of [Fe/H], as indicated by
the flat slope for stars with $\rm [Fe/H] \leq -2.5$.  A star's
spectrum is often contaminated with Na D absorption from the
interstellar medium.  The 5889/5895\AA\ Na lines of stars with very small
($\lesssim10$ \kms) radial velocities were inspected for possible
contamination with ISM features and were discarded when necessary.

\begin{figure}[!h]
\begin{center}
   \includegraphics[clip=true,width=8cm]{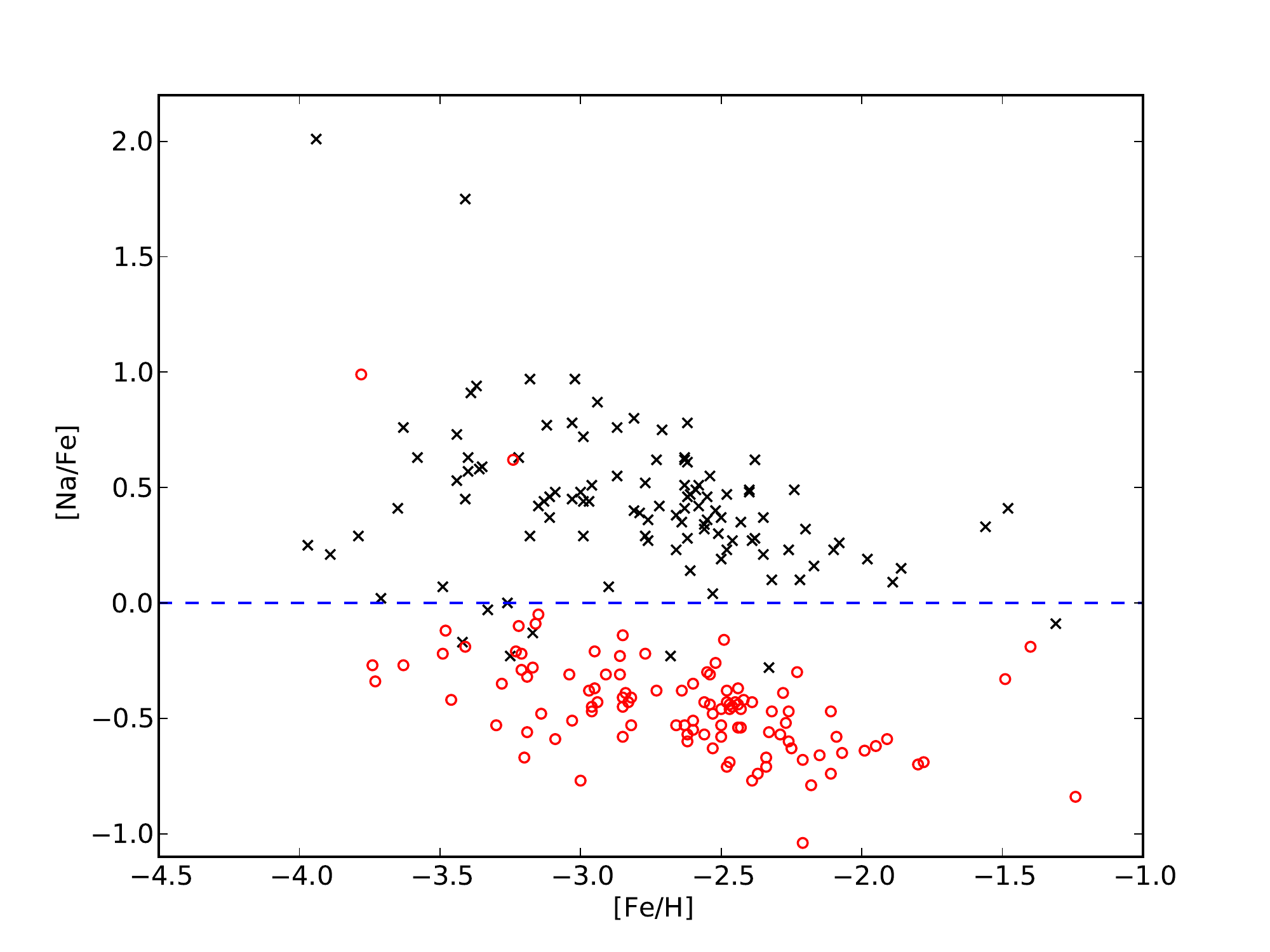} 
     \figcaption{ \label{f_na_lte_nlte}
     LTE [Na/Fe] (crosses) and NLTE [Na/Fe] (red open circles) plotted
     as a function of LTE [Fe/H] for the SkyMapper sample.  Note not all
     stars have NLTE abundances.  Differences
     between LTE and NLTE abundances are of order 0.7$-$1 dex.  See text
     for more information.}
 \end{center}
\end{figure}

We have calculated NLTE Na abundances as described in \citet{lind_na}.  
Five
stars fell outside the grid of \citet{lind_na}, and therefore do not
have NLTE abundances.  We have confirmed with detailed calculations
for a few stars that these NLTE abundances
are appropriate at the level of $\sim$0.05 dex, in spite of differences in
model atmospheres use  in this work \citep{castelli_kurucz}, 
and in \citealt{lind_na} (MARCS; \citealt{marcs}).  LTE abundances of this work were
found to differ by up to 0.3 dex from those calculated using the
method of \citet{lind_na} for
strong lines.  Figure~\ref{f_na_lte_nlte} plots LTE and NLTE [Na/Fe]
values (crosses and open circles, respectively), versus LTE [Fe/H].
(Note that the NLTE [Na/Fe] values were calculated using both NLTE Na
and Fe abundances.)  The large ($\sim$0.7-1 dex) differences reflect
the negative (NLTE-LTE) corrections for Na and the positive (NLTE-LTE)
corrections for Fe.  

The Al abundances for our stars are based only on measurement of the
3961 \AA\ Al\,I line for roughly half of the sample, while the
other half included measurement of the 3944 \AA\ feature.  No 
systematic abundance offsets were found between stars with one and two measured features.
Based on Figure~\ref{f_light}, the LTE
[Al/Fe] ratios of our sample are also comparable to those of
\citet{yong13_II}, though the
standard deviation in [Al/Fe] is roughly 1.5 times as large as in
their work.  This is not surprising given the low S/N of
some of our spectra, especially below 4000 \AA.
\citet{al_nlte} found NLTE corrections as large as $+$0.65 dex are
necessary for Al abundances of cool, metal-poor stars.  A correction
of this magnitude would bring
[Al/Fe] values in Figure~\ref{f_light} within $\sim$0.1 dex of solar.
Such ratios are more consistent with predictions of chemical evolution
models (e.g., \citealt{kobayashi2006}) than the LTE stellar
abundances, as has been noted before.

\subsection{$\alpha$-Elements}\label{sec:alpha}

 \begin{figure*}
 \begin{center}
    \includegraphics[clip=true,width=18cm]{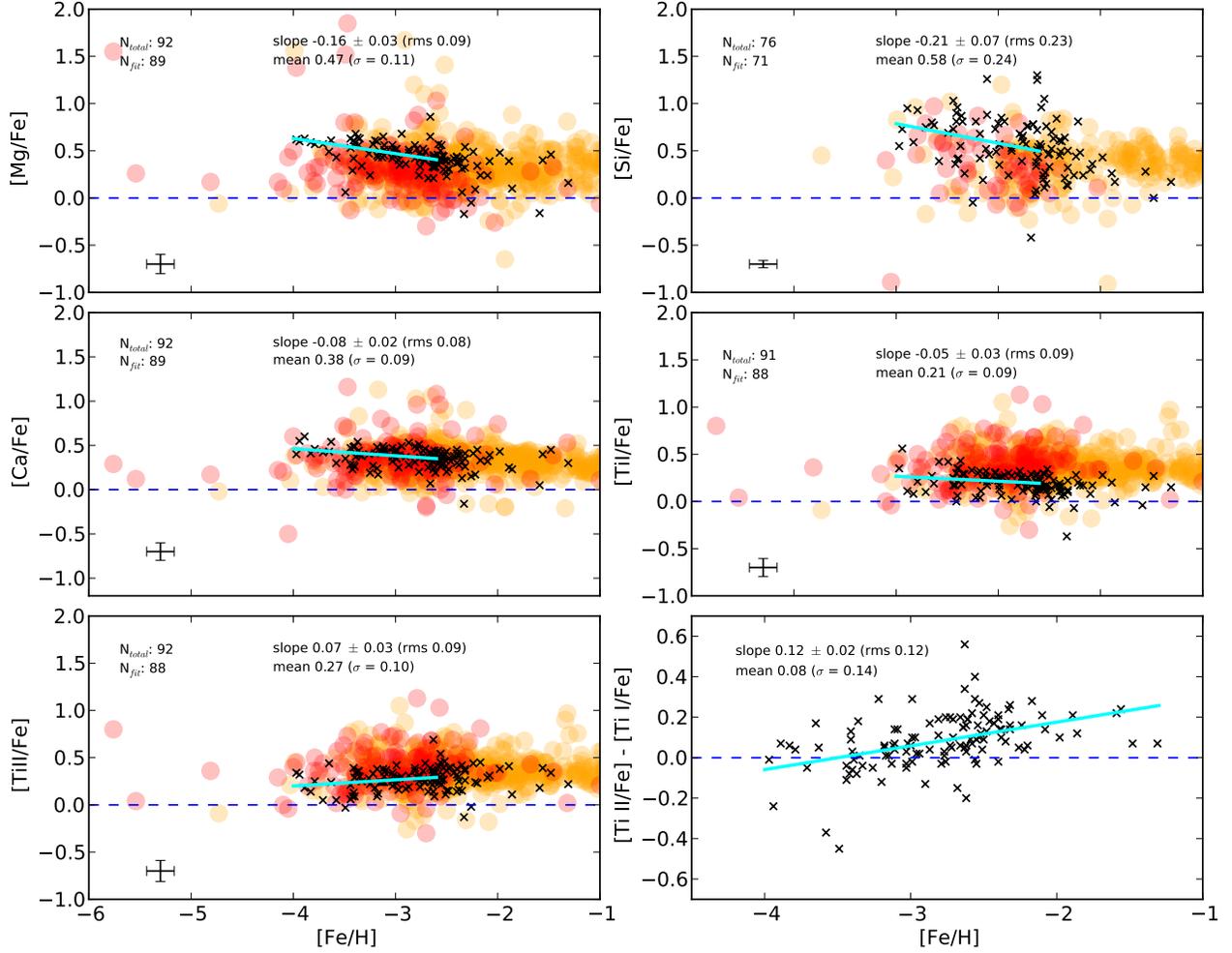} 
      \figcaption{ \label{f_alpha}
       Same as Figure~\ref{f_light}, but for the $\alpha$-elements.}
  \end{center}
 \end{figure*}

\begin{figure}[!h]
 \begin{center}
    \includegraphics[clip=true,width=8cm]{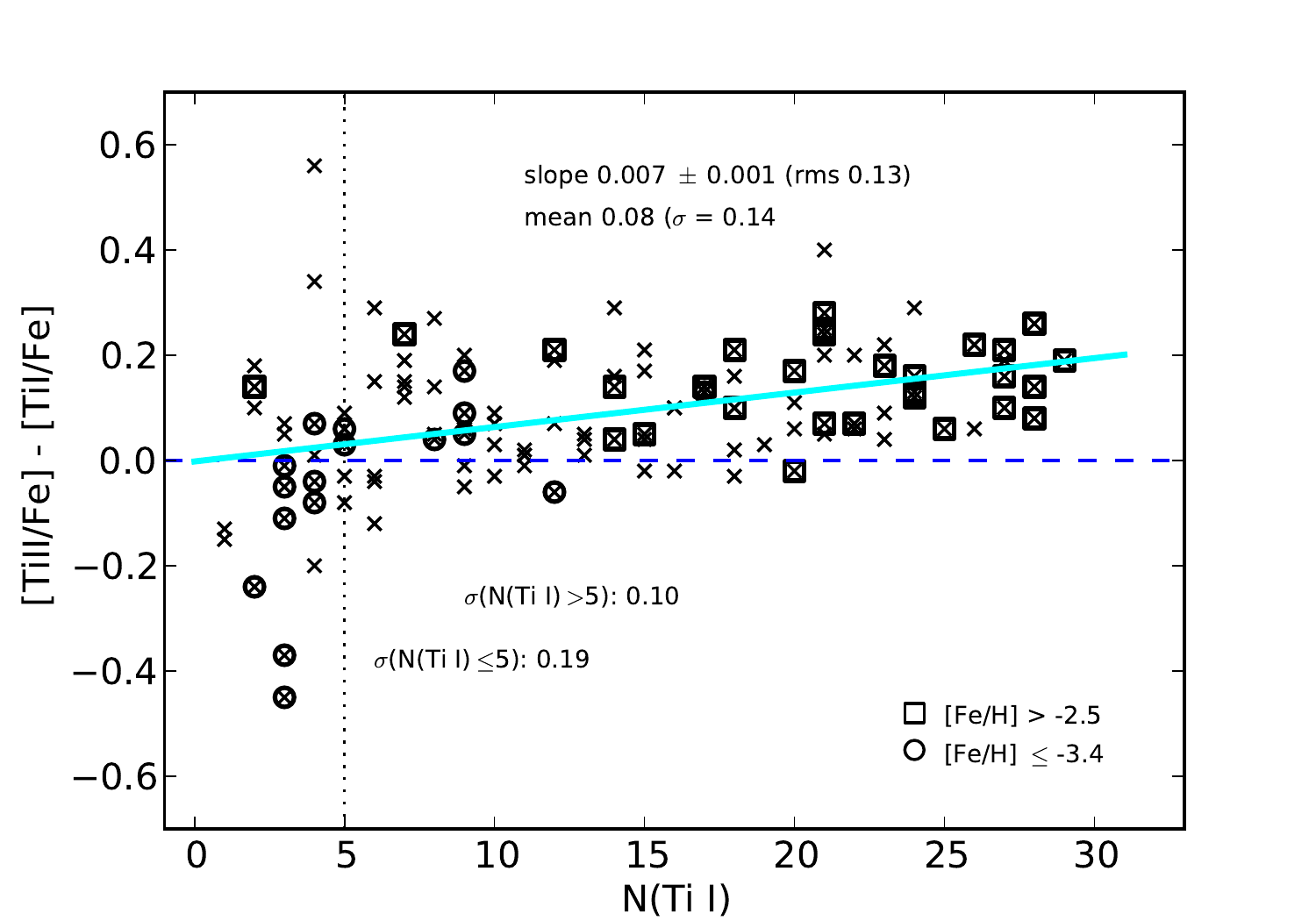} 
      \figcaption{ \label{f_tidiff_ntiI}
       [TiII/Fe] $-$ [TiI/Fe] differences versus the number of Ti\,I
       lines measured in each star.  There is a marked increase in
       scatter when N(Ti\,I) $\leq$ 5 (dotted line).  The most
       metal-poor (\feh $\leq -3.4$) and metal-rich (\feh $>-2.5$)
       stars are indicated by circles and squares, respectively.  See
       text for more information.} 
  \end{center}
 \end{figure}

The LTE [X/Fe] ratios for the $\alpha$-elements (Mg, Ca, Si, Ti) versus
LTE [Fe/H] are presented in Figure~\ref{f_alpha}.  
Ti\,I and Ti\,II abundances\footnote{Strictly, these are
  [Ti/Fe]$_{\rm Ti\,I}$ and [Ti/Fe]$_{\rm Ti\,II}$, but we denote them as
  [TiI/Fe] and [TiII/Fe] for convenience.}
 are plotted separately, with an additional panel that
 shows the difference between them as a function of metallicity.
As
can be seen, the agreement between them is good, with the mean
difference comparable to the dispersion about the means of both species.
This is similar to the agreement found by \citet{yong13_II} for their
giant star sample.  

Looking more closely, Ti\,I and Ti\,II abundances have different slopes in
  Figure~\ref{f_alpha}, while the difference between them at low \feh\ 
  is different from that at high \feh\ (bottom right).  The cause of these features is
  illustrated in Figure~\ref{f_tidiff_ntiI}, where the difference
  between Ti\,II and Ti\,I abundances is plotted against the number of
  Ti\,I lines measured per star.  As can be seen, the scatter in
  $\Delta$[X/Fe] increases by a factor of two when N(Ti\,I) $\leq$5,
  and that the most metal-poor stars preferentially have fewer
  measurable Ti\,I lines.  A star in our sample has an average of 30
  Ti\,II lines measured in its spectrum, compared to only 13 Ti\,I
  lines.  Consequently, the [Ti\,II/Fe] ratios in Figure~\ref{f_alpha}
  are more reliable.  For stars that have N(Ti\,I) $>$ 5, 
  ([Ti\,II/Fe] $-$ [Ti\,I/Fe]) = 0.11$\pm$0.10.  
 
We do not apply 
NLTE corrections to any $\alpha$-element abundances for our sample,
but summarize the magnitudes of corrections appropriate for our stars.  
NLTE corrections for Ti\,II abundances are
expected to be $\sim$0.05 dex or less, while
corrections for Ti\,I abundances are larger for metal-poor stars
($+$0.1$-$0.2 dex; \citealt{bergemann_ti}).  The difference
  between our LTE Ti\,II and Ti\,I abundances are consistent the
  magnitude of these corrections.
Uncertainties in atomic data for
individual lines likely also impact the scatter in abundances for both
species, though we note that improved atomic data are now available
\citep{lawler_tiI, wood_tiII}.
NLTE corrections for Mg are at the level of $\sim$0.1 dex
\citep{gehren2004_nlte, mashonkina_mg}, while for Ca\,I lines they can be as large as
$+$0.3 dex for stars like those in our sample
(\citealt{mashonkina_ca}, but see also \citealt{starkenburg10}).

\begin{figure*}
 \begin{center}
    \includegraphics[clip=true,width=18cm]{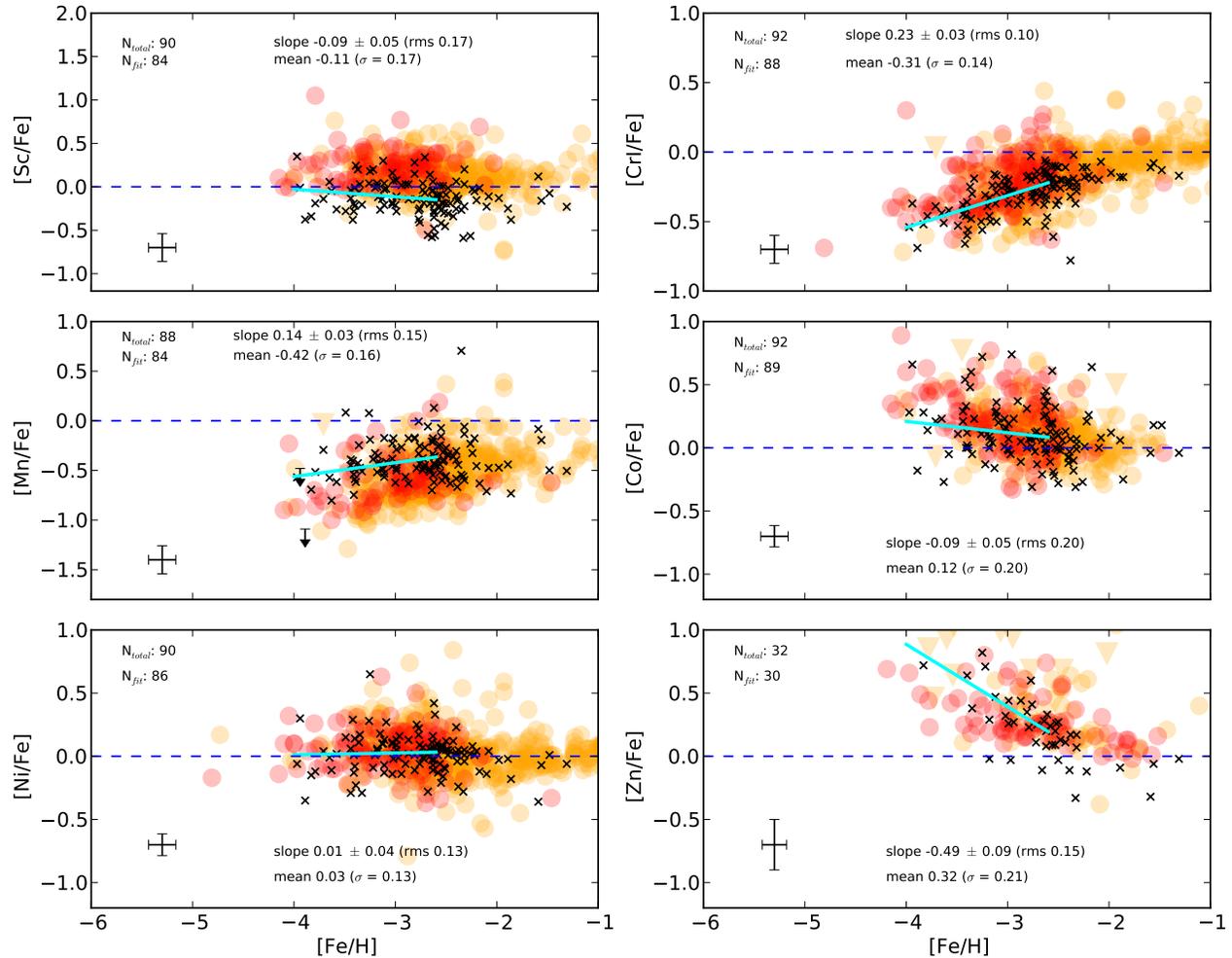} 
      \figcaption{ \label{f_fepeak}
       Same as Figure~\ref{f_light}, but for the Fe-peak elements.  
      For Zn (bottom right panel), the red symbols are 
      those of \citet{cayrel2004} and
      \citet{heresII}, while for the remaining elements they are from
      \citet{yong13_II}.  As in other figures, the orange symbols are
      from \citet{frebel10}.  Literature sample upper limits are shown as triangles.}
  \end{center}
 \end{figure*}

Overall,
the SkyMapper targets exhibit typical halo star abundance patterns,
with relatively small ($\sim$0.1 dex) dispersion in, e.g., Mg, Ca and Ti abundances, and
larger scatter in Si.  These dispersions are comparable to or smaller
than the
standard deviations of individual line abundances for most stars in
our sample.
The intrinsically small
dispersion in 
$\alpha$-element abundances over a wide range of metallicity is well-documented in the
literature (e.g., \citealt{cayrel2004}) and implies that their
nucleosynthetic yields have remained remarkably
constant throughout the earliest phases of chemical evolution in the universe.  
The larger
scatter in Si abundances is at least partly due to the difficulty in
obtaining a robust measure for this element; for many stars in our
sample it is based solely on one weak Si line (at 4102.9\AA, on the
wing of H$\delta$) that was not
measurable in all stars.  The blended 3905\AA\ Si\,I line was analyzed via spectrum
synthesis in a portion of our sample; no systematic offset between
abundances of stars based on one or both lines was observed.
In addition to the very small ($<$0.1 dex) dispersion in [X/Fe] versus
[Fe/H] in metal-poor stars, previous studies have found that
the $\alpha$-elements show flat trends
with slopes consistent with zero.  For all the $\alpha$-elements 
in Figure~\ref{f_alpha} save Mg the magnitudes of the 
slopes are equivalent to the rms scatter.

\subsection{Fe-peak Elements}\label{sec:fepeak}

Figure~\ref{f_fepeak} shows the trends with [Fe/H] for the Fe-peak
elements Sc, Cr, Mn, Co, Ni and Zn.  

\begin{figure*}
\begin{center}
   \includegraphics[clip=true,width=15cm]{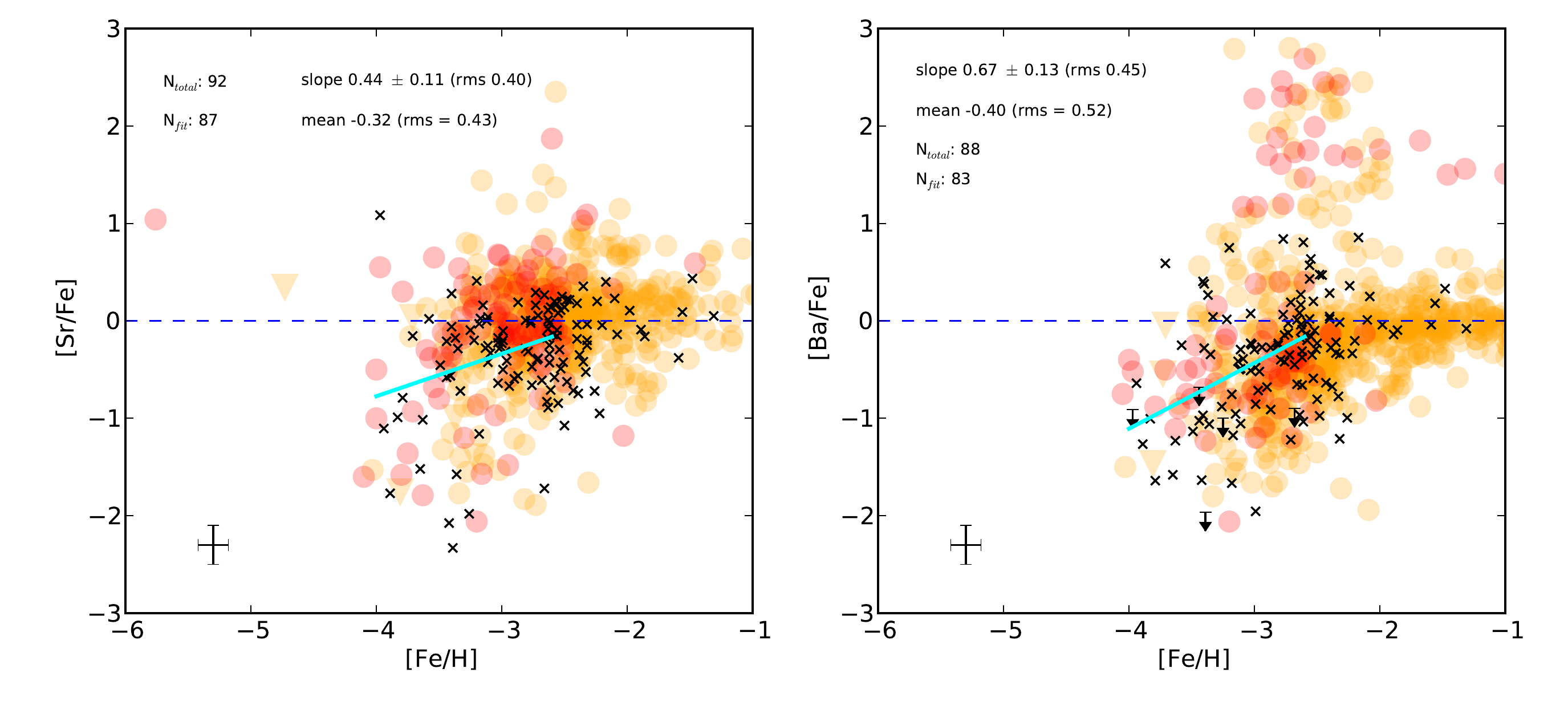} 
     \figcaption{ \label{f_ncap}
      Same as Figure~\ref{f_light}, but for Sr and Ba.  Upper limits
      are denoted by arrows and triangles for our sample and the
      literature sample, respectively.}
 \end{center}
\end{figure*}

As mentioned previously, Sc abundances are based
on spectrum synthesis of only the Sc\,II 4246\AA\ line for $\sim$40
stars in our sample, while for the remaining $\sim$50 as many as four
other lines were also analyzed.  A comparison of the Sc\,II abundance
determined from the 4246\AA\ line to the mean abundance of the other
lines found a 0.08 dex ($\sigma$=0.19) offset, in the sense that the
other line abundances were larger.  We have therefore added 0.08 dex
to the Sc\,II abundance for stars in which only the 4246\AA\ line was measured.
 
There is an unexplained systematic offset in
the zero-point of our Sc abundances compared to that of
\citet{yong13_II}: our mean [Sc/Fe] = $-$0.11 is $\sim$0.3 dex lower
than the value for their giant sample, although the values are comparable within the
standard deviations of the two samples (0.17 dex for ours, 0.14 dex
for theirs, see their Figure 22).  This offset is also visible
relative to the larger literature compilation and in our analysis of
the standard star HD~122563 compared to literature studies (see Section~\ref{sec:srbastar}).  \citet{yong13_II}
include hyperfine splitting in their Sc analysis, as we do here.
There is an 0.08 dex difference between their adoped log gf for Sc\,II
4246 \citep{kurucz&bell1995} and ours \citep{lawler_sc}, which is
accounted for by our 0.08 dex correction to that line's
abundance.  Differences in log gf values from the above sources for the other lines considered
here range from $+$0.03 to $-$0.20, and if anything, should make our
abundances slightly larger than those of \citet{yong13_II}.

\citet{cayrel2004} and \citet{lai2008}, among others, found that the Mn\,I
$\lambda$4030 resonance lines had lower abundances than other Mn\,I lines by
as much as 0.4 dex.  For the $\sim$40 stars in which we measured both
resonance and non-resonance Mn\,I lines, we found a difference of 
$\Delta$(non-res.$-$res.) = $+$0.44 (s.e.m. 0.03) dex.  We have
therefore applied a $+$0.44 dex correction to the abundances measured
from the Mn\,I 4030, 4033 and 4034\AA\ lines in all stars.
\citet{bergemann_mn} have demonstrated that the systematic offset
between resonance and non-resonance Mn\,I lines can be explained by
NLTE effects.  They found NLTE corrections for resonance lines as
large as $\sim$ $+$0.7 dex for warm, metal-poor stars, while corrections
for other Mn\,I lines as large as $+$0.4 dex are possible.  NLTE
[Mn/Fe] ratios for this SkyMapper sample would therefore be much
closer to the solar ratio.

As for the general abundance distributions shown in
Figure~\ref{f_fepeak}, the SkyMapper stars have
 the same trends of [X/Fe] versus [Fe/H] and the same scatter as
the literature samples.  The scatter with [Fe/H] is smallest for Cr and Ni,
while Mn and Co show (opposite to each other) trends of [X/Fe] with
[Fe/H].  \citet{cayrel2004} remarked upon the similar behavior of
[Cr/Fe] and [Mn/Fe] increasing with increasing [Fe/H] for their sample (both with quite small
scatter), and the same can be seen in the \citet{yong13_II} giant
sample.  
In our sample, [Cr/Fe] and [Mn/Fe] values show similar trends with
comparable scatter ($\sim$0.15 dex).  We also note that our mean
[X/Fe] values for Cr, Co and Ni agree very well with those of
\citet{yong13_II}, while our mean $\rm [Mn/Fe] =  -0.42$ is $\sim$0.15
dex larger than theirs.  
As \citet{yong13_II} did not include Zn, the literature sample we plot
in the bottom right panel of Figure~\ref{f_fepeak} is that of
\citet{cayrel2004} and \citet{heresII}.  The SkyMapper stars show a 
similar trend and scatter in [Zn/Fe] as in those samples, however, we
have found more stars exhibiting subsolar [Zn/Fe] ratios.

One star in Figure~\ref{f_fepeak} exhibits an [X/Fe] ratio very different from
the rest of the SkyMapper and literature samples.  SMSS~J093829.27$-$070520.9
appears to have $\rm [Mn/Fe] \sim +0.7$.  However, its spectrum has
S/N $\sim$10 at $\lambda$4000, and this abundance is based on
measurement of only
two Mn\,I resonance lines and has a standard deviation of 0.49 dex.
Consequently, its enhanced [Mn/Fe] ratio should be treated with skepticism.

\subsection{Neutron-Capture Elements}\label{sec:ncap}
The neutron-capture species considered in this analysis are Sr, Ba and
Eu.  The first two are predominantly formed via the s-process in
low-mass AGB stars, while Eu is almost entirely formed via the
r-process \citep{sneden_araa, jacobson13}.  The large variation of [X/Fe]
versus [Fe/H] ($>$1 dex) for neutron-capture elements, in strong constrast to the relative constancy of the $\alpha$-elements,
has also been well-established in the literature \citep{aoki05, heresII,
lai2008, roederer_ubiqrproc, yong13_II, cohen2013, spite_ncap2013,
roederer_ncap, roederer14_9stars}.  Our sample shows similar
behavior (Figure~\ref{f_ncap}).  Over the $\sim$2.5 dex range of
[Fe/H] spanned by our sample, there is evidence of the
dispersion in [X/Fe] increasing with decreasing [Fe/H] as found in the
literature (2$-$3 dex below $\rm [Fe/H] = -3$ compared to 1$-$2
dex at higher [Fe/H] for Sr and Ba in Figure~\ref{f_ncap}).
We have found no s-process  
stars in our sample, even though the mean [Fe/H] of our sample is that
of typical s-process metal-poor stars (e.g.,
\citealt{placco2013_magI}).  
This is consistent with the lack of stars with [C/Fe] $\gtrsim +$2, which
along with enhanced [s/Fe], is a signature of pollution from an AGB
companion (Section~\ref{sec:carbon}).

\begin{figure}
\begin{center}
   \includegraphics[clip=true,width=8cm]{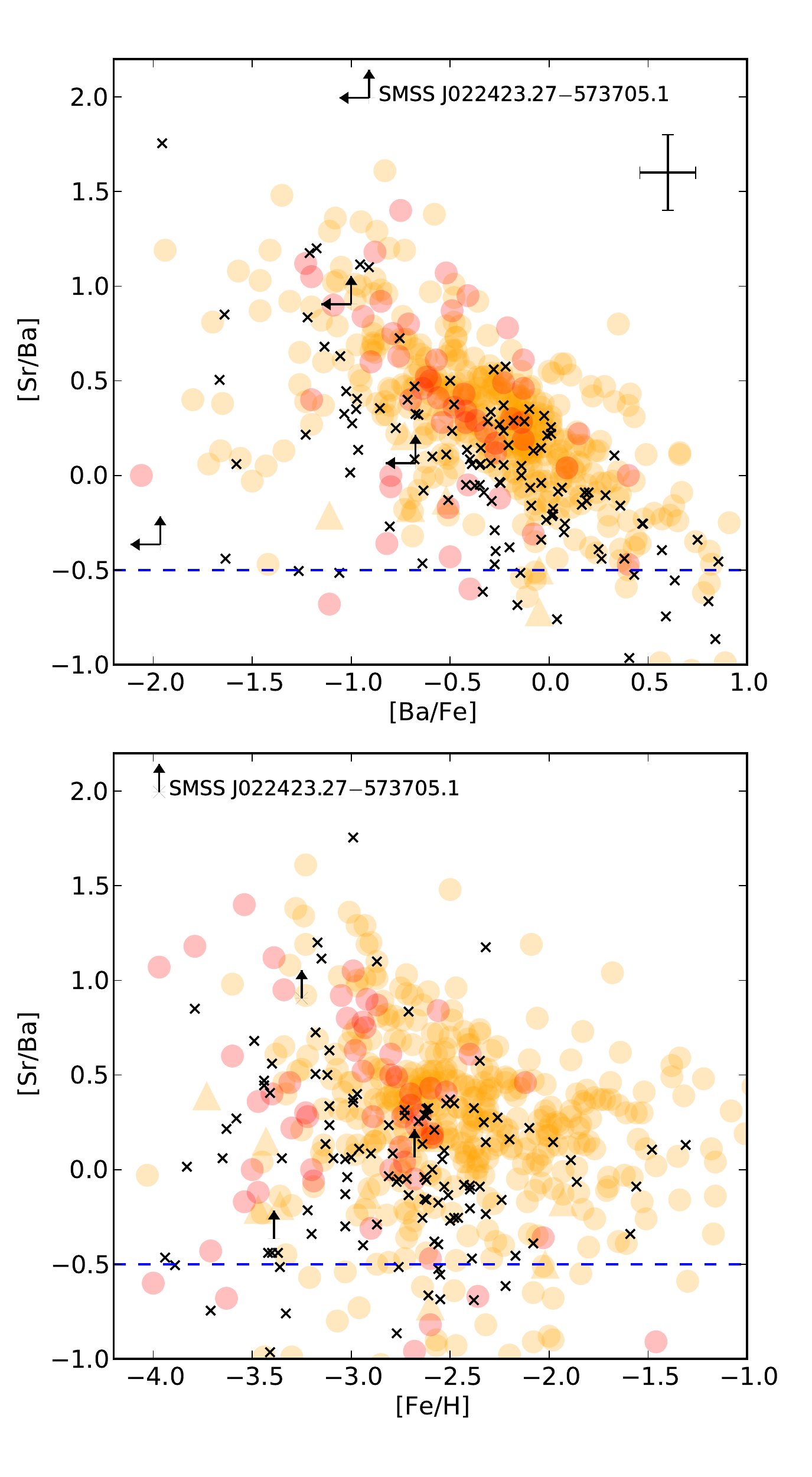} 
     \figcaption{ \label{f_srba}
      [Sr/Ba] versus [Ba/Fe] (top) and [Fe/H] (bottom) for our sample and literature stars.
      A conservative errorbar of 0.2 dex in [Ba/Fe] and 0.28 dex in
      [Sr/Ba] is indicated in the upper right.  
      The location of \5, which exhibits the largest [Sr/Ba]
      ratio of our sample (Section~\ref{sec:srbastar}) is labelled in both
      panels.
      The range of [Sr/Ba]
      increases with decreasing [Ba/Fe], but some Ba-poor stars with
      the solar r-process $\rm [Sr/Ba] = -0.5$ (dashed line) ratio are also
      present.  Though the number of stars with \feh\ $< -3.5$ is
      small, their presence is at odds with recent claims that there
      is a cut-off in [Sr/Ba] in this metallicity range
      \citet{aoki_srba}.}
 \end{center}
\end{figure}

The top panel of Figure~\ref{f_srba} shows the [Sr/Ba] ratios for our sample as a
function of their [Ba/Fe], which compares the relative abundances
of light and heavy neutron-capture elements.  
Except for the star \5\ which has [Sr/Ba]
$>$ 2 (see Section~\ref{sec:srbastar}),
our sample follows the same behavior as
those of, e.g., \citet{spite_ncap2013} and \citet{cohen2013}.  The Ba-poor objects
show the largest range of [Sr/Ba] ratios, while the most Ba-rich
objects show less scatter.  There are three Ba-poor stars
(with $\rm [Ba/Fe] < -1.0$) that exhibit the solar system r-process
[Sr/Ba] = $-$0.5.  The Eu 4129\AA\ line was not
measureable in any of their spectra; therefore, if they do follow
the solar system r-process pattern, their level of r-process element
enrichment would be extremely low.  The upper limits to their [Eu/Fe]
ratios are less than 0.4, at which level they would just be considered r-I
stars (see below).  

Based on a large literature sample, \citet{aoki_srba} recently claimed
that there is a dearth of stars with measurable [Sr/Ba] ratios below
\feh\ $< -3.5$.  \citet{placco2013_magII} suggested this was due to
the small/incomplete sample of stars in this metallicity regime; the
bottom panel of Figure~\ref{f_srba} lends support to this argument
(see also \citealt{LAMOST_emp}).
Roughly half of the stars below \feh $< -3.5$ exhibit large
($\gtrsim$1) [Sr/Ba] ratios.  

\begin{figure}
\begin{center}
   \includegraphics[clip=true,width=8cm]{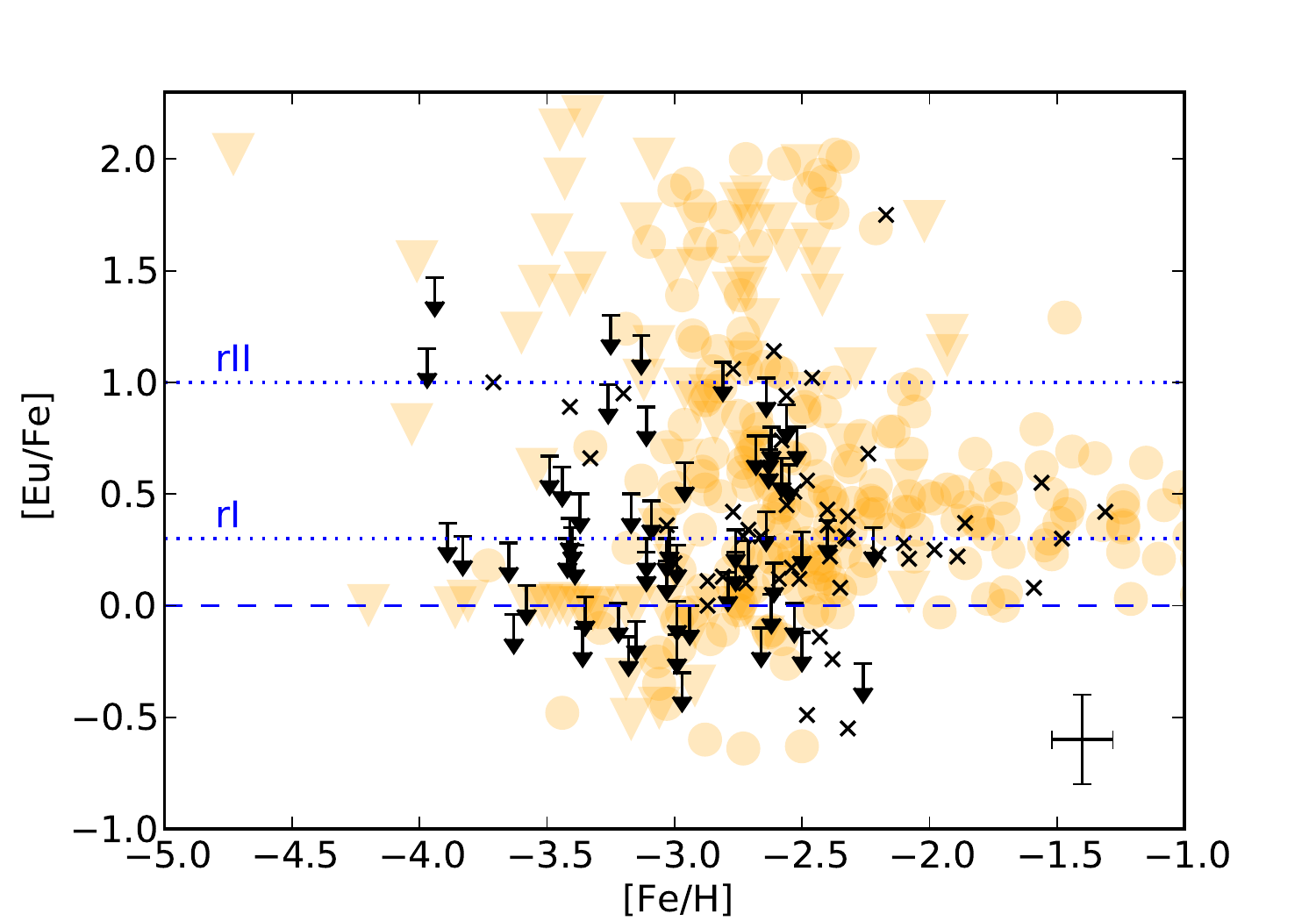} 
     \figcaption{ \label{f_eu}
      [Eu/Fe] versus [Fe/H] for our sample (crosses) compared to
      the \citet{frebel10} literature compilation (orange symbols).  Upper limits
      are indicated as arrows or triangles.  The [Eu/Fe] ranges for r-process
      enhanced (r-II and r-I) stars are indicated by dotted lines.}
 \end{center}
\end{figure}

We detected the Eu 4129\AA\ feature in a number of our MIKE spectra,
and obtained upper limits on the Eu abundances for all other stars for
which it was not detected.  These abundances are shown in
Figure~\ref{f_eu}.  As Eu abundances were not included in the study of
\citet{yong13_II}, we only include the \citet{frebel10} 
literature sample in Figure~\ref{f_eu}.  Again, we see a similar
distribution of [Eu/Fe] with [Fe/H] in our study to that in the
literature.  Note that most of our Eu abundances are upper limits
(denoted as arrows).

R-process enhanced stars are identified based on their Eu abundance:
strongly r-processed enhanced 
so-called r-II stars have $\rm [Eu/Fe] > 1.0$, while mildly r-process
enhanced r-I stars have $0.3 \leq \rm [Eu/Fe] \leq 1.0$ (and both classes
have $\rm [Ba/Eu] < 0$; \citealt{heresII}).  
These values are indicated with dotted lines
in Figure~\ref{f_eu}.  Of the stars in our sample for which we have
bona fide Eu measures, four have $\rm[Eu/Fe] \ge
1$, while another 22 qualify as r-I stars.  The metallicity range of
the r-II stars is $-2.77 \le \rm[Fe/H] \le -2.17$.  The star with the
largest enhancement ([Eu/Fe] = $+$1.75), SMSS~J175046.30$-$425506.9,
also happens to be the most metal-rich of the r-II stars.  Further
analysis of these r-process enhanced stars is ongoing.

We end the discussion with some remarks about NLTE effects on the
neutron-capture element abundances of metal-poor stars.
NLTE Sr\,II abundances are expected to differ from LTE values by no
more than 0.1 dex in the relevant stellar parameter regime
\citep{andrievsky_sr,bergemann_sr}.  NLTE corrections to Ba\,II
abundances (from, e.g., the $\lambda$4554 line) can range from roughly
$-$0.10 to $+$0.25 dex, and are dependent upon the Ba abundance
\citep{andrievsky_ba}.  However, as noted by, e.g., \citet{cohen2013} and
\citet{andrievsky_ba}, the magnitude of the scatter in metal-poor star Sr and Ba
abundances is far greater than can be attributed to NLTE effects, and
so they have little bearing on any interpretation of the data.
NLTE Eu abundances can be larger than the LTE
values by as much as $\sim$0.1 dex (\citealt{mashonkina2003}), though
to our knowledge Eu NLTE calculations have been done for dwarf stars only.

\subsection{Known Stars Recovered by SkyMapper}\label{sec:recover}
The coordinates of all the stars in Table~\ref{Tab:obs} were uploaded to the
Simbad\footnote{http://simbad.u-strasbg.fr/simbad/} database to check
for any that have been previously studied.  
We used a search radius of 30$\arcsec$ around the stellar coordinates.
Eight stars were found to have an entry in the database: four stars were
found in the RAdial Velocity
Experiment survey  (RAVE; data release 4) \citep{rave_dr4}, three were found in
various Hamburg-ESO survey studies, and the last is identified (as a star)
in the Millennium Galaxy Catalogue \citep{MGC03}.  Table~\ref{tab:reobs} lists these
stars along with their alternate identifications and reference
studies.

The two most metal-poor stars in our sample, \5\ and 
 \2\ (with \feh\ =
$-$3.97 and $-$3.94, respectively), are in fact rediscoveries.
\2\ was included in the sample of \citet{norris13_I} and \cite{yong13_II}, and
our stellar parameters and element abundances for this star are in excellent agreement
with their values.  \5\ was identified in the RAVE survey, but the
stellar parameters found by \citet{rave_dr4} are very different from
ours: \teff/\logg/\feh\ = 3600/4.5/$-$0.63 as opposed to
4846/1.60/$-$3.97.

Stellar parameters are determined from RAVE R$\sim$7500 spectra
($\lambda$8410-8795) using sophisticated algorithms that match the
data to a grid of synthetic spectra \citep{rave_dr4}.
\citet{rave_dr4} give a set of stellar parameters and data
characteristics (S/N, radial velocity measurement error, etc.) that
serves as quality checks to ensure the results of the RAVE pipeline
are robust and reliable.  \5\ fails to meet both the S/N 
($>$20 pixel$^{-1}$) and the \teff ($>$3800 K) requirements.  Of the
three other RAVE stars  in our sample (Table~\ref{tab:reobs}), two meet
all quality criteria while the RAVE pipeline did not converge for \4.
For the two stars that pass muster, our \teff\ values agree within 180 K of the
RAVE values and our \feh\ values agree within 0.15 dex.  
Differences between \logg\ values are
quite large, however: 0.6 and 2.4 dex for \1\ and \8, respectively.  
No systematic offset in any parameter is present.

\7\ and  \3\ were studied by \citet{cayrel2004} and \citet{heresII},
respectively.  For the former, our stellar parameters agree very well with those
of \citet{cayrel2004}, within 120 K, 0.15 dex, 0.25 \kms\ and 0.1 dex
in \teff, \logg, \vt\ and \feh, respectively.  The agreement with
\citet{heresII} is not as good in the case of \3: our \teff\ is 250 K
cooler, and our \logg\ and \feh\ values are 0.7 and 0.4 dex lower,
respectively.  

As for radial velocity measures, our value for \2\ agrees with that
found by \citet{norris13_I} within 1.4 \kms.  For \3\, our measure is
3.7 \kms\ larger than in \citet{heresII}, while  \citet{cayrel2004} does not
provide a radial velocity measurement for \7.  For those stars in
common with RAVE, our measures are $-$22 
(for \5)
to $+$23 (for \1) 
\kms\ different, in
the sense (This Study $-$ RAVE).  Our measure for 
\4\  
is 7.5
\kms\ smaller than RAVE's, while there is only 0.4 \kms\ difference
for\\ \8.   According to \citet{rave_dr4}, radial velocities measured
from RAVE spectra in the S/N range of these stars ($\sim$10-40) agree
within 5$-$8 \kms\ to literature values, though differences as large
as $\sim$20 \kms\ are possible (their Figure 34).  
Given the long base-line between our measures and theirs (the RAVE
observations were taken in 2004 and 2006), it is possible that at least
\1\ and \5\ are binary systems.

\subsection{Comparison to Literature Samples}

\begin{figure}
\begin{center}
   \includegraphics[clip=true,width=8cm]{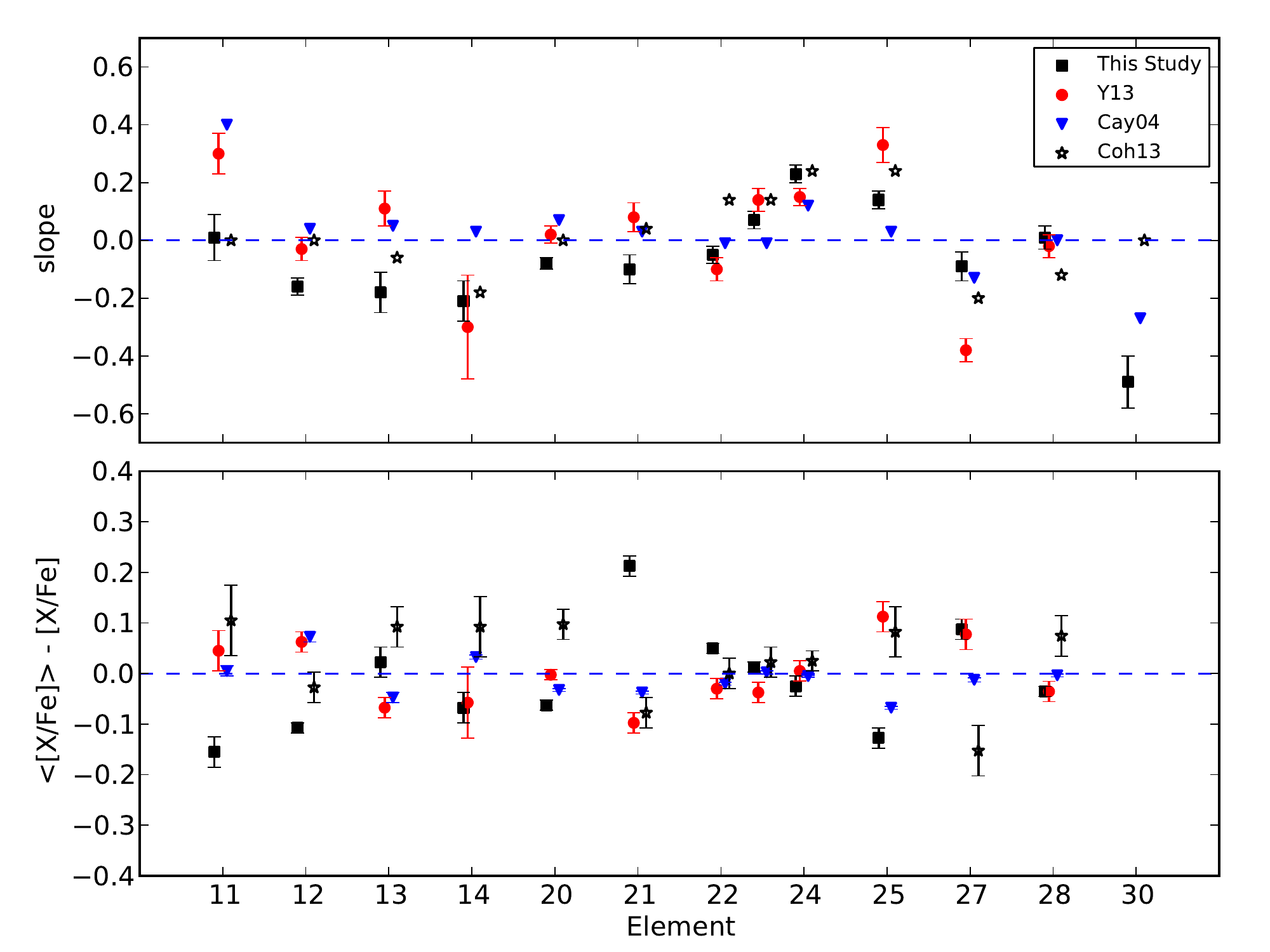} 
     \figcaption{ \label{f_slopes}
Top panel:
The slopes of lines of best fit for each element [X/Fe].  Here, our
linear regression analysis (black squares) is compared to that of
\citet{yong13_II} (red circles), \citet{cayrel2004} (blue triangles)
and \citet{cohen2013} (stars).  For all but the last sample, errorbars
on the points represent the slope uncertainties.  Note that the
uncertainties on the \citet{cayrel2004} slopes are smaller than the
symbols in the figure.  Bottom panel: the difference between
individual study mean [X/Fe] ratio for their stellar sample and the
mean [X/Fe] ratio of all four studies, for elements $\rm Z=11-30$.}
 \end{center}
\end{figure}

A quantitative comparison of our analysis to those of other studies
can be made by inspection of the linear regression analyses carried
out by different groups on different samples.  The results of the
regression analysis on this SkyMapper sample have been included in
Figures~\ref{f_light}--\ref{f_ncap}; for convenience, they are presented in
Table~\ref{tab_lsq} along with those of
\citet{yong13_II, cayrel2004, cohen2013}.  Figure~\ref{f_slopes}
presents the values from Table~\ref{tab_lsq} 
graphically.\footnote{All groups considered here confined their
  regression analysis to stars with $\rm [Fe/H] < -2.5$.  Note that
  the slopes given for \citet{cohen2013} in Table~\ref{tab_lsq}
  were calculated using [X/Fe] ratios at $\rm [Fe/H] = -3$ and $\rm
  [Fe/H] = -3.5$ for their CEMP-no stars (columns 3 and 4 in
  their Table 13).}
The errorbars on the points represent the uncertainty
of the slope, as given in this work and those of \citet{yong13_II} and
\citet{cayrel2004} (we note that the slope uncertainties in the latter
are smaller than the symbol in the figure).

Generally, the numerical values of the slopes in our analysis agree
with those of the literature studies within 2$\sigma$ for most of the
elements presented here.  Some elements show a large range of slopes:
namely Na, Mg, Al, Co and Zn.  Another way of comparing results from
different studies is to compare the mean [X/Fe] ratios found for stars
with $\rm [Fe/H] < -2.5$, and this is shown Table~\ref{tab_means}.
The bottom panel of Figure~\ref{f_slopes} shows the difference between
the [X/Fe] ratio found for a particular stellar sample and the mean
$<\rm [X/Fe]>$ ratio of the four studies in Table~\ref{tab_means}.  The
errorbars on the points are the standard errors of the mean.  Here one
can see evidence of the systematic offsets between our study and
others for some elements noted
earlier, namely for Sc.  For most of the elements, however, the
mean [X/Fe] ratios found by different studies agree within a factor of
two of their standard errors, though our Na and Mg values are higher
than those of the other samples considered here.

\section{SkyMapper Metal-Poor Stars of Interest}\label{sec:disc}

\subsection{A New ``Fe-enhanced'' Metal-poor Star}\label{sec:ferich}
One star, SMSS~J034249.53$-$284216.0 ($\rm [Fe/H] = -2.28$),
has subsolar [Mg/Fe], [Ca/Fe], [Sc/Fe], [Ti\,/Fe]
and [Ti\,II/Fe] ratios, the lowest of the entire sample.  In fact, it
is has $\rm [X/Fe]<0$ for all elements save Si and Eu.  Its
S/N ratio (mean $\sim$30) is less than the median value for the
sample, but by no means the lowest, and the abundances for
most elements are based on the measure of several lines, so these 
results are robust. 

There is a growing number of metal-poor stars in the literature that
show similar low [X/Fe] ratios
\citep{nissen_schuster, spite2000, ivans_alphapoor, 
   cayrel2004, honda04, cohen_huang2010, 
   bonifacio2011,  venn2012, caffau2013, yong13_II}.  
They have been called ``$\alpha$-poor''
or ``Fe-rich'' metal-poor stars.  The latter designation is likely
more appropriate for those stars that show deficiencies in numerous
other species in addition to the $\alpha$-elements.  Indeed the
element abundance patterns of such stars look similar to those of more
typical metal-poor stars, but shifted as a result of an additional Fe component.

\begin{figure}[!ht]
\begin{center}
   \includegraphics[clip=true,width=9cm]{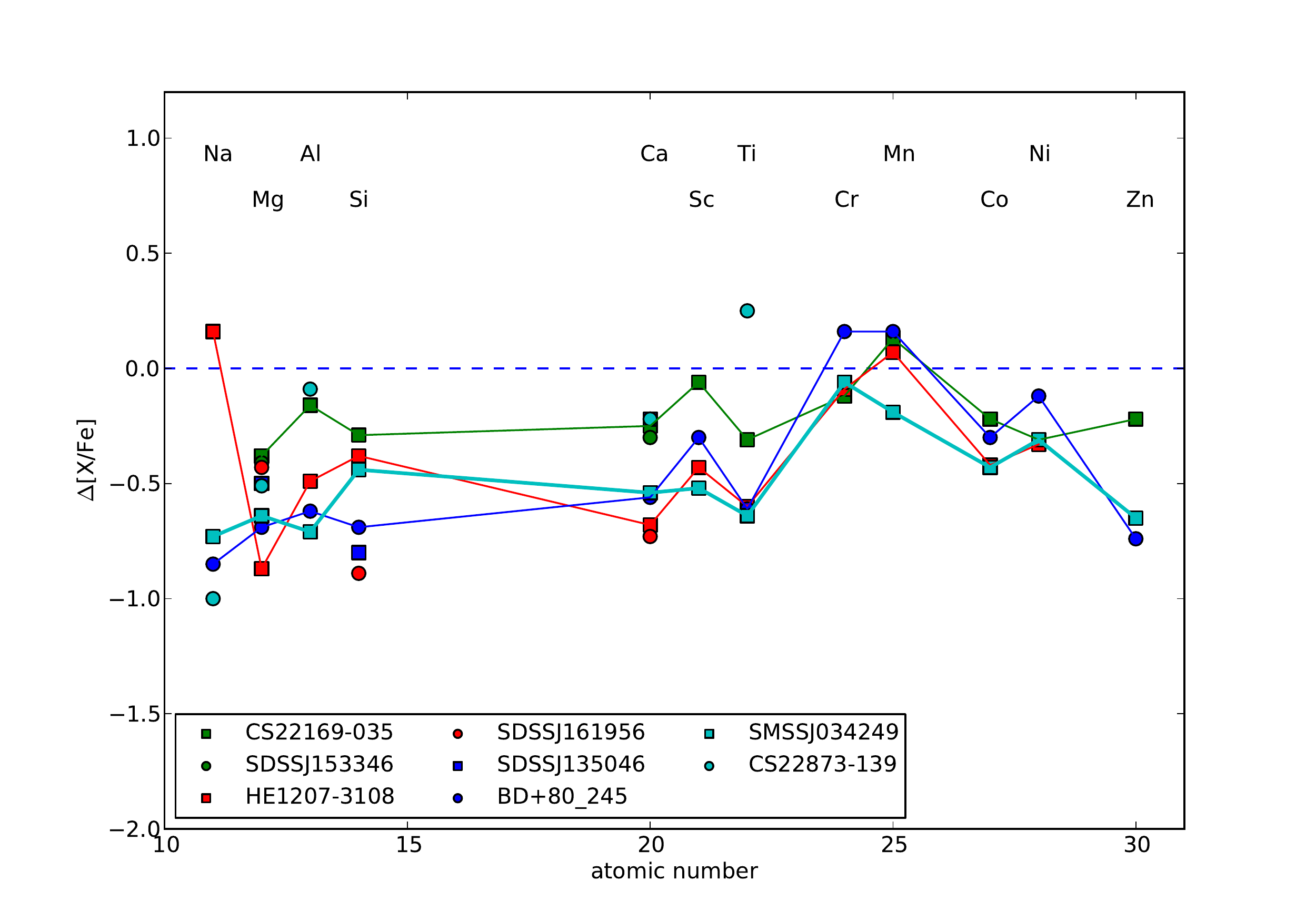} 
     \figcaption{ \label{f_ferich}
      The abundance difference, in the sense ([X/Fe] $-$ [X/Fe]$_{\rm Ref}$) for
      ``Fe-enhanced'' stars relative to that of the mean [X/Fe] ratios
      found in our study (Table~\ref{tab_means}).  The star in this study,
      SMSS~J034249.53$-$284216.0, is indicated by cyan squares and cyan
      bold line.  For simplicity, lines connecting individual element
      abundances are only drawn for stars where most of the species have
      been measured; some stars in this figure only have [$\alpha$/Fe]
      reported in the literature.  The patterns for all the stars are
      generally similar.   References for the
      literature sample include: 
      \citet{ivans_alphapoor,yong13_II,cayrel2004,caffau2013,spite2000,bonifacio2011}.}
 \end{center}
\end{figure}

Figure~\ref{f_ferich} plots the element
abundance pattern of SMSS~J034249.53$-$284216.0 (cyan squares, cyan bold line),
along with other stars exhibiting low [X/Fe] ratios in the
literature, relative to the mean abundances from the SkyMapper 
sample\footnote{These are taken to represent [X/Fe] ratios for typical
  halo stars with \feh\ $< -2.5$.}
(Table~\ref{tab_means}).  We restrict the literature stars in
Figure~\ref{f_ferich} to have \feh\ $< -2$, though we note that many
other ``Fe-enhanced'' stars exist in the literature at higher
metallicities (e.g., \citealt{nissen_schuster,
  ivans_alphapoor,cohen_huang2010, venn2012,ivans_alphapoor, bonifacio2011}).  
As has been noted
in the literature (e.g., \citealt{yong13_II}), there is some scatter in
the abundances of these stars.  The average [X/Fe] offset from the
mean SkyMapper sample abundances in Figure~\ref{f_ferich} is $-$0.40 dex, with a
1$\sigma$ scatter of 0.16 dex (for SMSS~J034249.53$-$284216.0, 
the offset is $-$0.52 dex).
While there is scatter in the abundance patterns, the stars in general
show sub-solar [X/Fe] ratios for all elements except for the Fe-peak
elements Cr and Mn.

A natural explanation for the Fe-enhancements exhibited by these
stars is that they formed from gas preferentially enriched with SNe Ia
products rather than just SNe II (e.g., \citealt{cayrel2004, caffau2013,
  yong13_II}).  Such environments exist in dwarf galaxies 
(indeed some known Fe-enhanced stars are in dwarf galaxies
\citep{venn2012, cohen_huang2010}), leading to the possibility that
the most metal-poor of the Fe-enhanced stars in the halo originated in dwarf galaxies.  That said, recent work by \citet{kobayashi2014_alphapoor} has shown
that the scatter and low element abundance ratios of stars in
\citet{caffau2013} and \citet{cohen2013} with $\rm [Fe/H] \leq -3$ are
well-matched by single core-collapse SN or hypernova yields, making
a dwarf galaxy origin unnecessary.  
This single enrichment scenario 
likely does not hold for the more metal-rich stars,
including SMSS~J034249.53$-$284216.0 with $\rm [Fe/H] = -2.3$. 
For these, the Fe-enhancements may be due to variations in
the progenitor masses and associated timescales of Type Ia supernovae.

For now, these few stars ($\sim$1\%$-$2\%\ of hundreds of halo stars so far
subject to high-resolution spectroscopic study)
indicate inhomogeneities in chemical evolution at the time
of their formation, in contrast to the apparent wide-spread
homogeneity in the bulk of metal-poor star formation (recall the small
scatter and lack of correlation in [$\alpha$/Fe] for the metal-poor
star sample of \citealt{cayrel2004}).
As more such stars are
found, it will be possible to investigate and better quantify the
degree of inhomogeneity in star formation and chemical evolution in
the early universe.

\subsection{A new $\rm [Fe/H] \sim -4$ star with high [Sr/Ba]}\label{sec:srbastar}
Although \5\ first appeared in the RAVE catalog (\citealt{rave_dr4}; see 
Section~\ref{sec:recover}), our work demonstrates for
the first time that it is an extremely metal-poor star, with 
 $\rm [Fe/H] = -3.97\pm0.14$.  With $\rm [C/Fe] = +0.07$ ($+$0.25
after applying the \citet{placco14} correction), it is not one of the CEMP
stars identified in Section~\ref{sec:carbon}, and its $\alpha$- and
Fe-peak element [X/Fe] ratios are normal (see Figures~\ref{f_alpha} and
\ref{f_fepeak}).  However, there are no barium lines detectable in its
fairly high S/N spectrum and an upper limit of $\rm [Ba/Fe] < -0.91$
was obtained.
In contrast, Sr lines are quite strong, giving a robust measure of
$\rm [Sr/Fe] = +1.08$. This [Sr/Fe] ratio is compatible with the most
Sr-rich stars of comparable metallicity as seen in
\citet{roederer_ncap} (his Figure 2).
The measured upper limit of [Eu/Fe] is $+$1.15 and
unfortunately not helpful in further constraining the origin of the
neutron-capture elements in this star.  Using Equation 6 of
\citet{hansen2014_rproc} to predict the Eu abundance from the Ba upper
limit, [Eu/Fe] $< +$0.07.

\begin{figure}[!ht]
\begin{center}
   \includegraphics[clip=true,width=9cm]{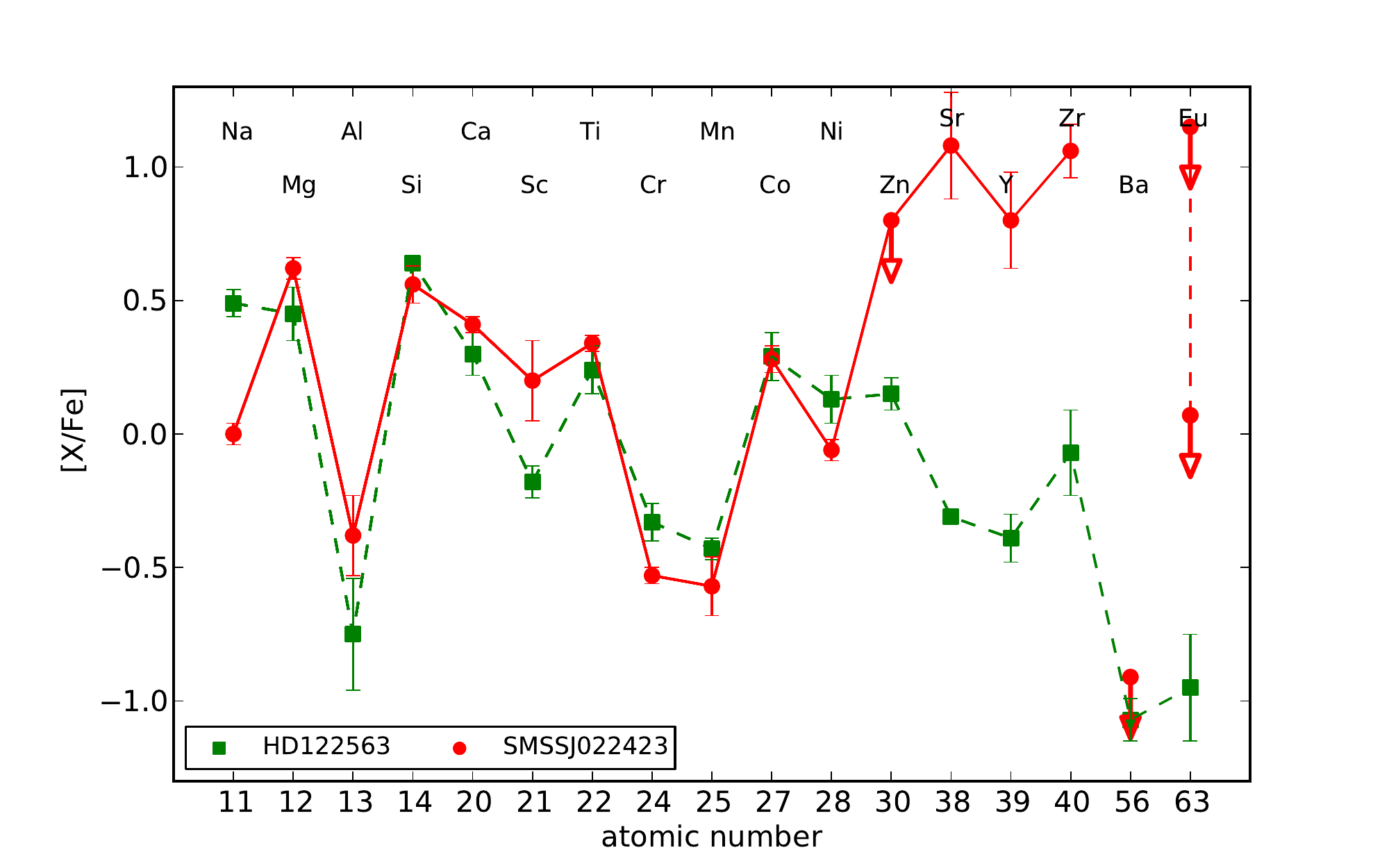} 
     \figcaption{ \label{f_3178}
The LTE element abundance pattern for star 
\5\ relative to that
of HD~122563 (abundances given in Table~\ref{tab:122563}).  Note that
the abundance patterns are quite similar for most elements, save for 
the neutron capture species.  
The large [X/Fe] ratios for Sr, Y, and Zr in \5\ are most striking,
while its Ba abundance
is just an upper limit.  Two upper limits are indicated for [Eu/Fe],
connected by a dashed line: $+$1.15 and $+$0.07, as measured in the
spectrum and as predicted using the relation of
\citet{hansen2014_rproc}, respectively.}
 \end{center}
\end{figure}

Figure~\ref{f_srba} shows that \5\ exhibits one of largest [Sr/Ba]
ratios currently known for a metal-poor star in the Milky Way 
halo.\footnote{We note that stars in the ultra-faint dwarf galaxy Segue-1 exhibit
  extremely low upper limits to their Sr and Ba abundances that point
  to intriguing neutron capture element enrichment episodes that are
  different from the Milky Way halo stars considered here
 \citep{frebel_segue1}.}  
 To our knowledge, only one
other star is known to have [Sr/Ba] $\gtrsim$2: SDSS J1422$+$0031,
with \feh\ = $-$3.03 and [Sr/Ba] = $+$2.2 \citep{aoki2013}.
Together, these two stars are the most extreme examples of the 
growing number of extremely metal-poor stars that show large
($\gtrsim$0.8 dex)
enhancements of the light neutron-capture element Sr relative to
the heavier neutron-capture element  Ba, as have been
found in several studies (e.g.,
\citealt{honda04,aoki05,lai2008,hollek11,aoki2013,cohen2013,placco2013_magII}).
Such stars are
generally taken as evidence for an extra neutron-capture element
production mechanism in addition to the main r-process as the source
of the heaviest elements in the early universe (e.g.,
\citealt{travaglio, honda06, sneden_araa, jacobson13}).  
Mechanisms such as  the Light Element Primary Process 
(LEPP; \citealt{travaglio}), the
weak r-process \citep{ishimaru2005}, 
the weak s-process \citep{heil_weaks} and the truncated r-process \citep{boyd_tr} 
have
been invoked to explain the existence of stars with large enhancements
of Sr, Y, and Zr relative to Ba and Eu. 

We inspected the spectrum of \5 
for the presence of other
neutron-capture species absorption lines, and were able to detect
several Y and Zr lines, but no lines of species belonging to the second
peak (e.g., Ba, La, Ce, Nd) or higher.  Spectrum synthesis of four Y\,II and three
Zr\,II lines resulted in [Y/Fe] = $+$0.80$\pm$0.26 and [Zr/Fe] =
$+$1.06$\pm$0.16 (s.d.).  \5\ is therefore strongly enhanced in first peak
neutron-capture species, with no detectable presence of heavier species.

Figure~\ref{f_3178}
shows the abundance pattern of 
SMSS~J022423.27$-$573705.1 relative to that of
HD~122563, the poster star exhibiting such light neutron-capture element
enhancements with [Sr/Ba] = $+$0.76 \citep{honda06}.  To minimize any systematics in this
comparison, we have carried out our own abundance analysis of
HD~122563, the results of which are presented in
Table~\ref{tab:122563}.  (We refer the reader to \citet{teff_calib} for details
regarding the data, but note that the analysis presented here is
separate from the results in that work.)
These two stars show similar abundance
patterns for most elements save for  Sr, Y and Zr. 
It is clear from this figure that whatever
the source(s) that produced this pattern of heavy elements (i.e., the
LEPP \citep{travaglio}), 
it operated even more strongly in the enrichment that led to the
formation of \5\ than for HD~122563.  

Rapidly rotating, low metallicity massive stars (``spinstars'') have been considered as
a possible source of light neutron-capture elements in the early universe, and models of such have been able to reproduce
the s-process element enhancements of low-metallicity field stars and
globular cluster stars (e.g., \citealt{pignatari},
\citealt{chiappini11}, \citealt{frischknecht}).\footnote{See, however,
  the results of \citet{ness2014} which do not support the spinstar
  origin scenario in the case of globular cluster NGC 6522 \citep{chiappini11}.}  
The
  abundance pattern produced by the 25 M$_{\odot}$,
  $\rm [Fe/H] = -3.8$ model of \citet{frischknecht} agrees relatively
  well with that of SMSS~J022423.27$-$573705.1 
  (Figure~\ref{f_3178}) for the elements in common (Co, Ni, Sr; see
  their Figure 1).  They
  do not give production factors for the elements Cr and Mn, which are
  both low in our star.   Their model also predicts a
  yield of Zn relatively larger than Co and Ni, but we could not
  detect Zn lines in the spectrum of SMSS~J022423.27$-$573705.1.
  Of the three stars in our sample with $\rm [Fe/H] < -3.5$, only one
  star has a detectable Zn line in its spectrum.  An upper limit EW
  measure for
  the Zn~I $\lambda$4810 in the spectrum of
  SMSS~J022423.27$-$573705.1 corresponds to $\rm [Zn/Fe] < +0.8$,
  which, together with the other element abundances, is consistent with
  the pattern from \citet{frischknecht}.  
  
It is not straightforward to
  compare the abundances of elements below the Fe-peak (Mn and lower)
  to the models of \citet{frischknecht}, because these models do not
  include element production in the supernova explosion itself
  (R.\ Hirschi, 2014, private communication).  
As more of
these stars are found, and the abundances of larger numbers of
neutron-capture elements are measured in them, it will be easier to
disentangle the presence of different production mechanisms and to
identify their production sites.

\section{Summary and Conclusions}\label{sec:summary}

We have presented a detailed chemical element abundance analysis of
the first SkyMapper metal-poor star candidates that were observed at high
spectroscopic resolution.  Based on a 1D LTE element
abundance analysis, the stellar parameters and element
abundances for these stars show them to be bona fide metal-poor halo
stars, as indicated by how well they match the abundance patterns of
halo stars in the literature.  

The main finding of this study is the verification of EMP star
candidates selected based on photometry from the SkyMapper Southern
Sky Survey and medium-resolution spectroscopy.  Excluding previously
known extremely metal-poor stars in our sample, we have confirmed 38
new stars to have $\rm [Fe/H] < -3.0$, eight of which have $\rm [Fe/H]
< -3.5$.  More importantly, the EMP candidate selection technique
based on the SkyMapper photometry has been improved over the
course of this program, and indeed the most iron-poor star
known to date (with $\rm [Fe/H] < -7$; \citealt{keller_thestar}), was
confirmed by its high-resolution Magellan-MIKE spectrum 
during the accumulation of the sample presented here.

Concerning the abundances of particular elements or of particular
stars in the study presented here, we have found the following:

\begin{itemize}
\item 
Eight stars previously known in the literature have been recovered by
the SkyMapper survey; six of which were previously known to be
extremely metal-poor.  We find reasonable to excellent agreement with
the results of other studies for four of these objects:
\teff\ within 250 K; \feh\ within 0.4 dex.  One star,
which was not previously identified as metal-poor, turns out to be the
most metal-poor star in our sample, with \feh\ = $-$3.97.

\item
After correcting stellar C abundances for evolutionary effects, 
24 stars are classified as CEMP stars based on the criterion of
\citet{aoki_cemp_2007}.  Considering only stars with
$\rm [Fe/H] \leq -3$, this results in a CEMP fraction of 39\%, in good
agreement with other studies.  Seven stars have $\rm [C/Fe] > 1$ and
are classified as CEMP-no stars.  Of these, five have \feh\ $< -3$.

\item Our most metal-poor star with \feh\ = $-$3.97, 
  has $\rm [Sr/Ba] \gtrsim
  2$, showing an extreme ratio of light to heavy neutron-capture
  element abundances. This indicates that 
  the weak r-process (or other mechanism) can yield more extreme
  light neutron-capture element enhancements than previously thought.

\item One star with $\rm [X/Fe] \leq 0$ for all elements save Si
and Eu, is likely a member of the growing population of
``Fe-enhanced'' metal-poor stars in the literature.

\item Four stars have r-process enhancements $\rm [Eu/Fe] > 1$ and
  are classified as r-II stars, while another 22 
 appear to be at least mildly r-process enhanced
based on their [Eu/Fe] ratios.  The relative fractions of r-I  (22/122
= 18\%) and r-II stars (4/122 = 3\%) are comparable to those found
by \citet{heresII} ($>$14\% and 3\%, respectively).  We caution
however that the metallicity ranges of the two samples are different
(\citealt{heresII} had no stars with \feh\ $< -3.5$), so the similarity
of the r-I/II fractions may be coincidental.
\end{itemize}

These results successfully demonstrate the
capability of the SkyMapper survey to find more stars at the
very metal-poor end of the Milky Way halo MDF, as well as stars exhibiting
interesting abundance signatures.  The increased sample size of these
metal-poor stars will improve our understanding of chemical
enrichment in the early epochs of the universe, as well as reveal
insight into the nature of the Population III stars that were the
first seeds of chemical enrichment.

\acknowledgements{This research has made
  use of the SIMBAD database, operated at CDS, Strasbourg, France and
  of NASA's Astrophysics Data System Bibliographic
  Services.  
This publication also makes use of data products from the Two Micron
All Sky Survey, which is a joint project of the University of 
Massachusetts and the Infrared Processing and Analysis 
Center/California Institute of Technology, funded by the 
National Aeronautics and Space Administration and the 
National Science Foundation.
We thank the referee for helpful suggestions that 
  improved the presentation of this work.
  R.\ Hirschi is thanked for informative discussions regarding
  the models of \citet{frischknecht}.  
  A.F.\ acknowledges support from NSF grant AST-1255160.
  A.C.\ was partially supported by the European Union FP7
  programme through ERC grant 320360. 
  M.S.B, G.D.C. and S.K. acknowledge support from the Australian 
  Research Council through Discovery Projects grant
  DP12010137. M.A. has been supported by an Australian Research
  Council Laureate fellowship (grant FL110100012).  
  K.L.\ acknowledges the European Union FP7-PEOPLE-2012-IEF grant
  No.\ 328098.  B.P.S.\ has been supported by an Australian Research
  Council Laureate fellowship (grant FL0992131). The work of
  J.M.P. and Q.Y. was supported by the MIT UROP program, and
  J.M.W. was supported by the Research Science Institute at MIT.
  Australian access to the Magellan Telescopes was supported through
  the Collaborative Research Infrastructure Strategy of the Australian
Federal Government.}

\textit{Facilities:} \facility{Magellan-Clay (MIKE)}

\begin{deluxetable}{lrrrllrrrrr} 
\tablewidth{0pt} 
\tabletypesize{\scriptsize}
\tablecaption{\label{Tab:obs} Observing Details}
\tablehead{
\colhead{Star} & \colhead{$\alpha$}&\colhead{$\delta$}&
\colhead{$g$}& \colhead{$g-i$} &
\colhead{UT}&
\colhead{$t_{\rm {exp}}$} &
\colhead{Slit size}&
\colhead{$v_{\rm{rad}}$} &
\colhead{S/N pixel$^{-1}$ \@} &
\colhead{S/N pixel$^{-1}$ \@}
\\
\colhead{}&\colhead{(J2000)}&\colhead{(J2000)}&
\colhead{mag}& \colhead{mag} &
\colhead{date\tablenotemark{a}}&
\colhead{sec}&
\colhead{(arcsec)}&
\colhead{km\,s$^{-1}$}&
\colhead{4500\,{\AA}}&
\colhead{6000\,{\AA}}}

\startdata
SMSS~J000113.96$-$363337.9 & 00 01 13.96 & $-$36 33 37.9 & 14.375 &
1.014 &  2013 Jan 08
& 1200 & 0.7 & 243.3 & 49 & 51\\   
SMSS~J001039.86$-$525851.4 & 00 10 39.86 & $-$52 58 51.4 & 14.628 &
1.137 & 2013 Jan 08
& 1500 & 0.7 & 110.0 & 32 & 64\\
SMSS~J001952.15$-$525803.0 & 00 19 52.15 & $-$52 58 03.0 & 15.378 &
1.176 &  2013 Jan 07
& 2720 & 0.7 & 154.8 & 28 & 52\\ 
SMSS~J002148.06$-$471132.1 & 00 21 48.06 & $-$47 11 32.1 & 15.072 &
0.932 & 2013 Jan 07 &
1200 & 0.7 & 204.5 & 20 & 33\\   
SMSS~J003055.81$-$482011.3 & 00 30 55.81 & $-$48 20 11.3 & 13.980 &
1.119 &  2013 Jan 08
& 900   & 0.7 & 109.3 & 38 & 63\\  
SMSS~J003327.36$-$491037.9 & 00 33 27.36 & $-$49 10 37.9 & 15.656 &
0.974 & 2013 Jan 07 &
2400 & 0.7 & 84.7 & 40 & 49\\     
SMSS~J004037.56$-$515025.2 & 00 40 37.56 & $-$51 50 25.2 & 14.617 &
0.939 &  2013 Jan
07 & 2700  & 0.7 & 72.9 & 19 & 33\\
SMSS~J005953.98$-$594329.9 & 00 59 53.98 & $-$59 43 29.9 & 14.925 &
0.381 & 2013 Nov 17 & 10140 & 0.7 & 376.7 & 78 & 105 \\
SMSS~J010332.63$-$534654.3 & 01 03 32.63 & $-$53 46 54.3 & 14.689 &
0.847  &  2013 Jan 08
& 1200 & 1.0 & 96.6 & 51 & 50\\ 
SMSS~J010651.91$-$524410.5 & 01 06 51.91 & $-$52 44 10.5 & 14.134 &
1.025 & 2013 Jan 06 &
1800 & 1.0 & 189.5 & 67 & 83\\    
\enddata
\tablenotetext{a}{UT at start of observation}
\tablenotetext{b}{Suspected binary}
\tablecomments{This table is available in its entirety in machine-readable
format in the online journal.  A portion is shown here for guidance
regarding its form and content.}
\end{deluxetable} 

\begin{deluxetable}{lcccccc} 
\tablecolumns{7} 
\tablewidth{0pt} 
\tabletypesize{\scriptsize}
\tablecaption{\label{Tab:stellpar} Stellar Parameters}
\tablehead{
\colhead{Star}  &
\colhead{$T_{\rm{eff}}$} & 
\colhead{$\log (g)$ }    & 
\colhead{$\mbox{[Fe/H]}$ }  & 
\colhead{$v_{\rm{micr}}$}  &
\colhead{$\log (g)$ } &
\colhead{$\mbox{[Fe/H]}$ }  \\
\colhead{}  &
\colhead{} &
\colhead{LTE} &   
\colhead{LTE }  & 
\colhead{}  &
\colhead{NLTE} &
\colhead{NLTE} \\
\colhead{}&
\colhead{[K]}&
\colhead{[dex]}&
\colhead{[dex]}&
\colhead{[km\,s$^{-1}$]} &
\colhead{[dex]} &
\colhead{[dex]}}
\startdata
SMSS~J000113.96$-$363337.9  &  4810 & 1.60 & $-$2.32 & 1.80 & 1.90 & $-$2.21 \\ 
SMSS~J001039.86$-$525851.4  &  4711 & 1.20 & $-$2.32 & 2.20 & 1.56 & $-$2.18 \\ 
SMSS~J001952.15$-$525803.0 &  4639 & 1.20 & $-$2.56 & 2.40 & 1.56 & $-$2.43 \\ 
SMSS~J002148.06$-$471132.1 &  4765 & 1.40 & $-$3.17 & 2.10 & 1.90 & $-$3.00 \\ 
SMSS~J003055.81$-$482011.3  &  4720 & 1.50 & $-$2.53 & 2.50 & 1.82 & $-$2.41 \\ 
SMSS~J003327.36$-$491037.9  &  4630 & 0.80 & $-$3.36 & 2.35 & 1.37 & $-$3.17  \\ 
SMSS~J004037.56$-$515025.2  &  4468 & 0.55 & $-$3.83 & 2.45 & 1.05 & $-$3.67  \\
SMSS~J005953.98$-$594329.9 & 5413 & 2.95 & $-$3.94 & 1.40  & 3.41 & $-$3.78\\ 
SMSS~J010332.63$-$534654.3  &  4810 & 1.40 & $-$3.03 & 1.80 & 1.90 & $-$2.86 \\ 
SMSS~J010651.91$-$524410.5  &  4486 & 0.65 & $-$3.79 & 2.60 &  1.15 & $-$3.63 \\ 
\enddata
\tablecomments{This table is available in its entirety in machine-readable
format in the online journal.  A portion is shown here for guidance
regarding its form and content.}
\end{deluxetable}

\begin{deluxetable}{lcccccc}
\tablewidth{0pt}
\tabletypesize{\scriptsize}
\tablecaption{\label{tab_ew_stub} Equivalenth Width Measurements}
\tablehead{
\colhead{Star} & \colhead{Wavelength} & \colhead{Species} & \colhead{E.P.} & \colhead{log {\it gf}} & \colhead{EW (m\AA)} & \colhead{log $\epsilon$(X)}
}
\startdata
SMSS~J000113.96$-$363337.9 & 3689.458 & 26.0 & 2.940 & -0.168 & 82.3 & 5.12  \\ 
SMSS~J000113.96$-$363337.9 & 3753.611 & 26.0 & 2.180 & -0.890 & 93.0 & 5.25  \\ 
SMSS~J000113.96$-$363337.9 & 3765.539 & 26.0 & 3.240 & 0.482 & 87.2 & 4.92  \\ 
SMSS~J000113.96$-$363337.9 & 3786.677 & 26.0 & 1.010 & -2.185 & 96.0 & 5.19  \\ 
SMSS~J000113.96$-$363337.9 & 3805.343 & 26.0 & 3.300 & 0.313 & 74.2 & 4.81  \\ 
SMSS~J000113.96$-$363337.9 & 3839.256 & 26.0 & 3.050 & -0.330 & 75.1 & 5.17  \\ 
SMSS~J000113.96$-$363337.9 & 3842.047 & 27.0 & 0.920 & -0.770 & 66.4 & 2.70  \\ 
SMSS~J000113.96$-$363337.9 & 3845.169 & 26.0 & 2.420 & -1.390 & 52.7 & 4.92  \\ 
SMSS~J000113.96$-$363337.9 & 3852.573 & 26.0 & 2.180 & -1.180 & 81.8 & 5.18  \\ 
SMSS~J000113.96$-$363337.9 & 3881.869 & 27.0 & 0.580 & -1.130 & 73.5 & 2.82  \\ 
SMSS~J000113.96$-$363337.9 & 3882.291 & 22.1 & 1.120 & -1.710 & 81.4 & 3.15  \\ 
SMSS~J000113.96$-$363337.9 & 3885.510 & 26.0 & 2.420 & -1.090 & 68.6 & 5.00  \\ 
SMSS~J000113.96$-$363337.9 & 3904.784 & 22.0 & 0.900 & 0.030 & 40.2 & 2.77  \\ 
SMSS~J000113.96$-$363337.9 & 3917.181 & 26.0 & 0.990 & -2.155 & 108.9 & 5.41  \\ 
SMSS~J000113.96$-$363337.9 & 3924.526 & 22.0 & 0.020 & -0.881 & 38.8 & 2.61  \\ 
SMSS~J000113.96$-$363337.9 & 3940.878 & 26.0 & 0.960 & -2.600 & 77.0 &4.96  \\
\enddata 
\tablecomments{This table is available in its entirety in machine-readable
format in the online journal.  A portion is shown here for guidance
regarding its form and content.}
\end{deluxetable}

\begin{deluxetable}{lcrcrr}
\tablewidth{0pt}
\tabletypesize{\scriptsize}
\tablecaption{\label{tab_abund} Element Abundances}
\tablehead{
\colhead{Star} & \colhead{log$\epsilon$(X)} & \colhead{\# lines} & \colhead{s.d.} & \colhead{[X/H]} & \colhead{[X/Fe]}
}
\startdata
\noalign{\vskip +0.5ex}
\multicolumn{6}{c}{C} \cr
\noalign{\vskip +0.8ex} 
\hline
\noalign{\vskip-2ex}\\
SMSS~J000113.96$-$363337.9 & 5.78 & 2 & 0.02 & -2.65 & -0.33\\ 
SMSS~J001039.86$-$525851.4 & 5.88 & 2 & 0.03 & -2.55 & -0.23\\ 
SMSS~J001952.15$-$525803.0 & 5.62 & 2 & 0.02 & -2.81 & -0.25\\ 
SMSS~J002148.06$-$471132.1 & 5.55 & 2 & 0.04 & -2.88 & 0.29\\ 
SMSS~J003055.81$-$482011.3 & 5.95 & 2 & 0.05 & -2.48 & 0.05\\ 
SMSS~J003327.36$-$491037.9 & $<$ 4.67 & 1 & 0.00 & $<$ -3.75 & $<$ -0.40\\ 
SMSS~J004037.56$-$515025.2 & 4.51 & 2 & 0.09 & -3.92 & -0.09\\ 
SMSS~J005953.98$-$594329.9 & 5.70 & 2 & 0.05 & -2.73 & 1.21\\ 
SMSS~J010332.63$-$534654.3 & 5.59 & 2 & 0.09 & -2.84 & 0.19\\ 
SMSS~J010651.91$-$524410.5 & 4.77 & 2 & 0.05 & -3.66 & 0.13\\
\enddata 
\tablecomments{This table is available in its entirety in machine-readable
format in the online journal.  A portion is shown here for guidance
regarding its form and content.}
\end{deluxetable}

\begin{deluxetable}{llcccccc}
\tabletypesize{\scriptsize}
\tablewidth{0pt}
\tablecaption{Abundance Uncertainties due to Atmospheric Parameters\label{tab_unc}}
\tablehead{
\colhead{} &
\colhead{} &
\colhead{} &
\colhead{$\Delta$\teff(K)} &
\colhead{$\Delta$\logg} &
\colhead{$\Delta$\vt} & 
\colhead{$\Delta$[M/H]} & \colhead{}\\
\colhead{Star} & \colhead{[X/Fe]} & \colhead{$\sigma$\tablenotemark{a}} &
\colhead{$+$100 K} &
\colhead{$+$0.3 dex} &
\colhead{$+$0.2 km s$^{-1}$} & \colhead{$-$0.25 dex} & \colhead{Total}}
\startdata
SMSS~J055746.51$-$575057.4 & C\,I & 0.02 & $+$0.09 & $-$0.08 & $+$0.04 & $-$0.06 & $+$0.14 \\
\teff = 5404 K & Na\,I & 0.03 & $+$0.02 & $-$0.08 & $-$0.02 & $-$0.01 & $+$0.09 \\
\logg = 3.05  & Mg\,I & 0.03 & $-$0.01 & $-$0.05 & $+$0.03 & $+$0.00 & $+$0.07\\
\feh = $-$2.50 & Al\,I & 0.20 & $+$0.01 & $+$0.00 & $+$0.00 & $+$0.00 & $+$0.20\\
\vt = 1.45 \kms & Si\,I & 0.20 & $+$0.05 & $-$0.12 & $+$0.02 & $-$0.02 & $+$0.24\\
          & Ca\,I & 0.03 & $-$0.04 & $+$0.00 & $+$0.01 & $-$0.01 & $+$0.05\\
          & Sc\,II & 0.07 & $-$0.05 & $+$0.12 & $-$0.01 & $-$0.01 & $+$0.15\\
          & Ti\,I & 0.03 & $+$0.01 & $+$0.01 & $+$0.02 & $+$0.01 & $+$0.04\\
          & Ti\,II & 0.02 & $-$0.06 & $+$0.11 & $-$0.03 & $-$0.01 & $+$0.13\\
          & Cr\,I & 0.04 & $+$0.01 & $-$0.01 & $-$0.03 & $+$0.00 & $+$0.05\\
          & Mn\,I & 0.06 & $+$0.04 & $-$0.03 & $-$0.09 & $+$0.00 & $+$0.12\\
& Fe\,I\tablenotemark{b} & 0.01 & $+$0.10 & $-$0.02 & $-$0.04 & $-$0.01 &
$+$0.11\\
& Fe\,II\tablenotemark{b} & 0.03 & $+$0.01 & $+$0.12 & $-$0.03 & $-$0.02 &
$+$0.13\\
          & Co\,I & 0.04 & $+$0.02 & $+$0.01 & $-$0.02 & $+$0.00 & $+$0.05\\
          & Ni\,I & 0.04 & $+$0.01 & $-$0.01 & $-$0.05 & $+$0.00 & $+$0.07\\
          & Zn\,I & 0.20 & $-$0.06 & $+$0.08 & $+$0.03 & $+$0.00 & $+$0.23\\
          & Sr\,II & 0.04 & $+$0.00 & $+$0.02 & $-$0.07 & $-$0.01 & $+$0.08\\
          & Ba\,II & 0.10 & $-$0.03 & $+$0.10 & $-$0.06 & $-$0.01 &
$+$0.16\\
          & Eu\,II & 0.20 & $-$0.07 & $+$0.11 & $+$0.05 & $+$0.01 & $+$0.24\\
\hline
SMSS~J004037.56$-$515025.2 & C\,I & 0.13 & $+$0.20 & $-$0.09 & $+$0.00 &
$-$0.02 & $+$0.26\\
\teff = 4468K & Na\,I & 0.11 & $+$0.00 & $+$0.01 & $-$0.04 & $+$0.00 &
$+$0.12\\
\logg = 0.55  & Mg\,I & 0.04 & $-$0.02 & $-$0.05 & $-$0.05 & $+$0.01 & $+$0.08\\
\feh = $-$3.83 & Al\,I & 0.20 & $-$0.01 & $-$0.07 & $-$0.07 & $+$0.00 & $+$0.22\\
\vt = 2.45 \kms & Si\,I & 0.20 & $+$0.00 & $+$0.01 & $+$0.01 & $-$0.01 & $+$0.20\\
          & Ca\,I & 0.02 & $-$0.05 & $+$0.02 & $+$0.02 & $-$0.01 & $+$0.06\\
          & Sc\,II & 0.20 & $-$0.04 & $+$0.11 & $+$0.00 & $-$0.01 & $+$0.23\\
          & Ti\,I & 0.04 & $+$0.01 & $+$0.01 & $+$0.03 & $+$0.00 & $+$0.05\\
          & Ti\,II & 0.02 & $-$0.05 & $+$0.12 & $+$0.00 & $+$0.00 & $+$0.13\\
          & Cr\,I & 0.03 & $+$0.02 & $-$0.03 & $-$0.04 & $+$0.00 & $+$0.06\\ 
          & Mn\,I & 0.16 & $+$0.03 & $-$0.01 & $+$0.00 & $+$0.00 & $+$0.16\\
& Fe\,I\tablenotemark{b} & 0.01 & $+$0.13 & $-$0.05 & $-$0.03 & $-$0.01 &
$+$0.14\\
& Fe\,II\tablenotemark{b} & 0.03 & $+$0.04 & $+$0.09 & $-$0.01 & $+$0.00 &
$+$0.10\\
          & Co\,I & 0.03 & $+$0.02 & $-$0.02 & $-$0.02 & $+$0.00 & $+$0.05\\
          & Ni\,I & 0.04 & $+$0.01 & $-$0.03 & $-$0.05 & $+$0.00 & $+$0.07\\
          & Zn\,I & 0.20 & $-$0.07 & $+$0.09 & $+$0.03 & $-$0.01 & $+$0.23\\
          & Sr\,II & 0.03 & $-$0.02 & $+$0.09 & $-$0.07 & $+$0.02 & $+$0.12\\
          & Ba\,II & 0.13 & $-$0.03 & $+$0.11 & $+$0.00 & $-$0.01 &
$+$0.17\\
          & Eu\,II & 0.20 & $-$0.14 & $+$0.07 & $+$0.03 & $-$0.02 & $+$0.26\\
\enddata
\tablenotetext{a}{The standard error of the mean of individual line element abundances.}
\tablenotetext{b}{These values are [X/H] ratios.}
\end{deluxetable}

\begin{deluxetable}{lccc}
\tabletypesize{\scriptsize}
\tablewidth{0pt}
\tablecaption{Lithium abundances of SkyMapper stars\label{tab_li}}
\tablehead{
\colhead{} & \colhead{EW} & \colhead{A(Li)} & \colhead{A(Li)}\\
\colhead{Star} & \colhead{(m\AA)} & \colhead{(LTE)} & \colhead{(NLTE)}}
\startdata
SMSS~J002148.06$-$471132.1 & 25.1 & 0.92 & 1.02\\
SMSS~J005953.98$-$594329.9 & 70.1 & 2.00 & 1.97\\
SMSS~J010839.58$-$285701.5 & 16.1 & 0.80 & 0.89\\
SMSS~J015941.53$-$781408.7 & 16.2 & 0.76 & 0.85\\
SMSS~J024858.41$-$684306.4 & 12.3 & 0.75 & 0.83\\
SMSS~J031556.09$-$473442.1 & 32.7 & 1.08 & 1.13\\
SMSS~J040148.04$-$743537.3 & 13.0 & 0.73 & 0.80\\
SMSS~J051008.62$-$372019.8 & 15.1 & 1.00 & 1.04\\
SMSS~J062609.83$-$590503.2 & 12.0 & 0.80 & 0.87\\
SMSS~J070257.95$-$600422.4 & 22.3 & 0.91 & 1.02\\
SMSS~J090247.43$-$122755.0 & 10.6 & 0.72 & 0.81\\
SMSS~J105320.99$-$435300.1 & 13.6 & 0.80 & 0.87\\
SMSS~J105438.86$-$435819.9 & 16.3 & 0.92 & 0.99\\
SMSS~J121353.63$-$441911.2 & 24.3 & 0.85 & 0.93\\
SMSS~J125115.37$-$331448.1 & 28.1 & 0.95 & 1.04\\
SMSS~J141547.72$-$414034.0 & 19.1 & 1.00 & 1.06\\
SMSS~J151101.05$-$182103.0 & 30.0 & 1.30 & 1.35\\
SMSS~J155628.74$-$165533.4 & 13.1 & 0.70 & 0.80\\
SMSS~J165219.76$-$253133.7 & 21.5 & 1.00 & 1.08\\
SMSS~J174922.26$-$455103.8 & 24.1 & 0.85 & 0.97\\
SMSS~J190549.33$-$214945.0 & 16.1 & 0.85 & 0.94\\
SMSS~J193617.38$-$790231.4 & 23.0 & 1.16 & 1.21\\
SMSS~J205313.80$-$651830.6 & 17.2 & 0.76 & 0.85\\
SMSS~J215805.81$-$651327.2 & 23.8 & 0.81 & 0.90\\
\enddata
\end{deluxetable}

\begin{deluxetable}{lrrr}
\tabletypesize{\scriptsize}
\tablewidth{0pt}
\tablecaption{Corrected Carbon Abundances\label{tab:cfe_corr}}
\tablehead{
\colhead{Star} & \colhead{$\rm[C/Fe]_{\rm orig}$} & \colhead{corr.} &
\colhead{$\rm[C/Fe]_{\rm corr}$}}
\startdata
SMSS~J000113.96$-$363337.9 & $-$0.33 & 0.42 & 0.09\\ 
SMSS~J001039.86$-$525851.4 &  $-$0.23 & 0.67 & 0.44\\ 
SMSS~J001952.15$-$525803.0 &  $-$0.25 & 0.67 & 0.42\\ 
SMSS~J002148.06$-$471132.1 &  0.29 & 0.49 & 0.78\\ 
SMSS~J003055.81$-$482011.3 &  0.05 & 0.45 & 0.50\\ 
SMSS~J003327.36$-$491037.9 & $< -$0.40 & 0.72 &  $<$ 0.33\\ 
SMSS~J004037.56$-$515025.2 &  $-$0.09 & 0.71 & 0.62 \\ 
SMSS~J005953.98$-$594329.9 &  1.21 & 0.00 & 1.21\\ 
SMSS~J010332.63$-$534654.3 &  0.19 & 0.50 & 0.69\\ 
SMSS~J010651.91$-$524410.5 &  0.13 & 0.72 & 0.85\\
\enddata
\tablecomments{This table is available in its entirety in machine-readable
format in the online journal.  A portion is shown here for guidance
regarding its form and content.}
\end{deluxetable} 

\begin{deluxetable}{lll}
\tabletypesize{\scriptsize}
\tablewidth{0pt}
\tablecaption{Rediscovered EMP Stars from the Literature\label{tab:reobs}}
\tablehead{
\colhead{SMSS ID} &
\colhead{Literature ID(s)} &
\colhead{Ref.}}
\startdata
\1 & RAVE~J003055.8$-$482011 & \citet{rave_dr4}\\
\2 & HE~0057$-$5959 & \citet{norris13_I,yong13_II}\\
\3 & HE~0104$-$5300 & \citet{heresII}\\
\4 & RAVE~J010839.6$-$285701 & \citet{rave_dr4}\\
\5 & RAVE~J022423.3$-$573705 & \citet{rave_dr4}\\
\6 & 2MASS~J10025112$-$0001520 & \citet{MGC03}\\
\7 & CS~229656$-$050 & \citet{cayrel2004}\\
\8 & RAVE~J224844.0$-$543610 & \citet{rave_dr4}\\
\enddata
\end{deluxetable}

\begin{deluxetable}{lcccccccccccccccc}
\tabletypesize{\scriptsize}
\tablewidth{0pt}
\tablecaption{Linear regression results\label{tab_lsq}}
\tablehead{
\colhead{} & 
 \multicolumn{3}{c}{This Study}& \colhead{} &
 \multicolumn{3}{c}{\citet{yong13_II}} & \colhead{} & 
 \multicolumn{3}{c}{\citet{cayrel2004}} & \colhead{} &
 \multicolumn{3}{c}{\citet{cohen2013}}\\
\cline{2-4}\cline{6-8}\cline{10-12}\cline{14-16}
\colhead{[X/Fe]} & 
\colhead{slope} & \colhead{error} &
\colhead{rms} & \colhead{} &
\colhead{slope} & \colhead{error} &
\colhead{rms} & \colhead{} & 
\colhead{slope} & \colhead{error} &
\colhead{rms} & \colhead{} &
\colhead{slope}\tablenotemark{a} & \colhead{error} & 
\colhead{rms}\\
\colhead{} & \colhead{} & \colhead{} & \colhead{(dex)} &
\colhead{}  &
\colhead{} & \colhead{} & \colhead{(dex)} & \colhead{} &
\colhead{} & \colhead{} & \colhead{(dex)} & \colhead{} &
\colhead{} & \colhead{} & \colhead{(dex)}}
\startdata
Na\,I & $+$0.01 & 0.08 & 0.27 & & $+$0.30 & 0.07 & 0.21 & & $+$0.403 &
0.010 & 0.25 & & \nodata & \nodata & \nodata\\
Mg\,I & $-$0.16 & 0.03 & 0.09 & & $-$0.03 & 0.04 & 0.13 & & $+$0.035 &
0.003 & 0.13 & & $+$0.00 & \nodata & 0.17\\
Al\,I & $-$0.18 & 0.07 & 0.26 & & $+$0.11 & 0.06 & 0.17 & & $+$0.047 &
0.005 & 0.18 & & $-$0.06 & \nodata & 0.24\\
Si\,I & $-$0.21 & 0.07 & 0.23 & & $-$0.30 & 0.18 & 0.26 & & $+$0.032 &
0.004 & 0.15 & & $-$0.18 & \nodata & 0.20 \\ 
Ca\,I & $-$0.08 & 0.02 & 0.08 & & $+$0.02 & 0.03 & 0.10 & & $+$0.074 &
0.002 & 0.10 & & $+$0.00 & \nodata & 0.15 \\
Sc\,II & $-$0.09 & 0.05 & 0.17 & & $+$0.08 & 0.05 & 0.13 & & $+$0.034 &
0.002 & 0.11 & & $+$0.04 & \nodata & 0.14 \\
Ti\,I & $-$0.05 & 0.03 & 0.09 & & $-$0.10 & 0.04 & 0.11 & & $-$0.014 &
0.001 & 0.09 & & $+$0.14 & \nodata & 0.14 \\
Ti\,II & $+$0.07 & 0.03 & 0.09 & & $+$0.14 & 0.04 & 0.13 & & $-$0.014 &
0.001 & 0.09 & & $+$0.14 & \nodata & 0.14 \\
Cr\,I & $+$0.23 & 0.03 & 0.10 & & $+$0.15 & 0.03 & 0.10 & & $+$0.117 &
0.000 & 0.05 & & $+$0.24 & \nodata & 0.13 \\
Mn\,I & $+$0.14 & 0.03 & 0.15 & & $+$0.33 & 0.06 & 0.15 & & $+$0.030 &
0.003 & 0.12 & & $+$0.24 & \nodata & 0.16 \\
Co\,I & $-$0.09 & 0.05 & 0.20 & & $-$0.38 & 0.04 & 0.04 & & $-$0.131 &
0.002 & 0.13 & & $-$0.20 & \nodata & 0.16 \\
Ni\,I & $+$0.01 & 0.04 & 0.13 & & $-$0.02 & 0.04 & 0.13 & & $-$0.003
& 0.002 & 0.11 & & $-$0.06 & \nodata & 0.21 \\
Zn\,I & $-$0.49 & 0.09 & 0.15 & & \nodata & \nodata & \nodata & &
$-$0.271 & 0.002 & 0.11 & & $+$0.00 & \nodata & 0.25 \\
\enddata
\tablenotetext{a}{Slope calculated subtracting $<\rm [X/Fe]>$ values at
  $\rm [Fe/H] = -3.0$ from values at $\rm [Fe/H] = -3.5$ in Table 13 of \citet{cohen2013}.}
\end{deluxetable}

\begin{deluxetable}{lcccccccccccccccc}
\tabletypesize{\scriptsize}
\tablewidth{0pt}
\tablecaption{Mean [X/Fe] for different studies\label{tab_means}}
\tablehead{
\colhead{} & 
 \multicolumn{3}{c}{This Study}& \colhead{} &
 \multicolumn{3}{c}{\citet{yong13_II}} & \colhead{} & 
 \multicolumn{3}{c}{\citet{cayrel2004}} & \colhead{} &
 \multicolumn{3}{c}{\citet{cohen2013}\tablenotemark{a}}\\
\cline{2-4}\cline{6-8}\cline{10-12}\cline{14-16}
\colhead{[X/Fe]} & 
\colhead{Mean} & \colhead{N} &
\colhead{$\sigma$/$\sqrt{N}$} & \colhead{} &
\colhead{Mean} & \colhead{N} &
\colhead{$\sigma$/$\sqrt{N}$} & \colhead{} & 
\colhead{Mean} & \colhead{N} &
\colhead{$\sigma$/$\sqrt{N}$} & \colhead{} &
\colhead{Mean} & \colhead{N} &
\colhead{$\sigma$/$\sqrt{N}$}\\
\colhead{} & \colhead{(dex)} & \colhead{} & \colhead{(dex)} &
\colhead{}  &
\colhead{(dex)} & \colhead{} & \colhead{(dex)} & \colhead{} &
\colhead{(dex)} & \colhead{} & \colhead{(dex)} & \colhead{} &
\colhead{(dex)} & \colhead{} & \colhead{(dex)}}
\startdata
Na\,I & $+$0.44 & 83 & 0.03 & & $+$0.24 & 38 & 0.04 & & $+$0.28 & 35 &
0.01 & & $+$0.18 & 49 & 0.07\\
Mg\,I & $+$0.47 & 89 & 0.01 & & $+$0.30 & 60 & 0.02 & & $+$0.29 & 35 & 0.01
& & $+$0.39 & 59 & 0.03 \\
Al\,I & $-$0.71 & 82 & 0.03 & & $-$0.62 & 54 & 0.02 & & $-$0.64 & 35 &
0.01 & & $-$0.78\tablenotemark{b} & 47 & 0.04\\
Si\,I & $+$0.58 & 71 & 0.03 & & $+$0.57 & 14 & 0.07 & & $+$0.48 & 35 &
0.004 & & $+$0.42 & 47 & 0.06\\
Ca\,I & $+$0.38 & 89 & 0.01 & & $+$0.32 & 60 & 0.01 & & $+$0.35 & 35 &
0.003 & & $+$0.22 & 56 & 0.03\\
Sc\,II & $-$0.11 & 84 & 0.02 & & $+$0.15 & 44 & 0.02 & & $+$0.09 & 35 &
0.003 & & $+$0.13 & 46 & 0.03\\
Ti\,I & $+$0.21 & 88 & 0.01 & & $+$0.29 & 55 & 0.02 & & $+$0.28 & 35 &
0.003 & & $+$0.26 & 59 & 0.03\\
Ti\,II & $+$0.27 & 88 & 0.01 & & $+$0.32 & 60 & 0.02 & & $+$0.28 & 35 &
0.003 & & $+$0.26 & 59 & 0.03\\
Cr\,I & $-$0.31 & 88 & 0.01 & & $-$0.34 & 54 & 0.02 & & $-$0.33 & 35 &
0.002 & & $-$0.36 & 59 & 0.02\\
Mn\,I & $-$0.42 & 84 & 0.02 & & $-$0.66 & 37 & 0.03 & & $-$0.48 & 35 &
0.003 & & $-$0.63 & 51 & 0.05\\
Co\,I & $+$0.12 & 89 & 0.02 & & $+$0.13 & 54 & 0.03 & & $+$0.22 & 35 &
0.004 & & $+$0.36 & 41 & 0.05\\
Ni\,I & $+$0.03 & 86 & 0.01 & & $+$0.03 & 56 & 0.02 & & $-$0.002 & 35 &
0.003 & & $-$0.08 & 42 & 0.04\\
Zn\,I & $+$0.32 & 30 & 0.04 & & \nodata & \nodata & \nodata & & $+$0.33 &
35 & 0.005 & & \nodata & \nodata & \nodata\\
\enddata
\tablenotetext{a}{These numbers were taken from Table 16 in
  \citet{cohen2013} for their ``Inner Halo'' sample that have
  distances $4 < D < 15$ kpc, which more likely overlaps with the
  distances spanned by this SkyMapper sample.}
\tablenotetext{b}{Here we have removed the 0.6 dex NLTE 
  correction \citet{cohen2013} applied to their Al abundances in order
  to compare them to the LTE abundances of the other studies.}
\end{deluxetable}

\begin{deluxetable}{lrrrrr}
\tabletypesize{\scriptsize}
\tablewidth{0pt}
\tablecaption{Element Abundances of HD~122563\label{tab:122563}}
\tablehead{
\colhead{Species} & \colhead{log$\epsilon$(X)} & \colhead{\#\ lines} &
\colhead{s.d.} & \colhead{[X/H]} & \colhead{[X/Fe]}}
\startdata
C\,I & 5.53 & 2 & 0.04 & $-$3.20 & $-$0.43\\
Na\,I & 3.96 & 2 & 0.05 & $-$2.28 & $+$0.49\\
Mg\,I & 5.28 & 7 & 0.10 & $-$2.32 & $+$0.45\\
Al\,I & 2.93 & 2 & 0.21 & $-$3.52 & $-$0.60\\
Si\,I & 5.38 & 2 & 0.01 & $-$2.13 & $+$0.64\\
Ca\,I & 3.87 & 22 & 0.08 & $-$2.47 & $+$0.30\\
Sc\,II & 0.20 & 5 & 0.06 & $-$2.91 & $-$0.14\\
Ti\,I & 2.29 & 30 & 0.07 & $-$2.66 & $+$0.11\\
Ti\,II & 2.42 & 46 & 0.09 & $-$2.53 & $+$0.24\\
Cr\,I & 2.54 & 19 & 0.07 & $-$3.10 & $-$0.33\\
Cr\,II & 2.88 & 3 & 0.09 & $-$2.76 & $+$0.01\\
Mn\,I & 2.23 & 7 & 0.04 & $-$3.20 & $-$0.43\\
Fe\,I & 4.73 & 193 & 0.12 & $-$2.77 & \nodata\\
Fe\,II & 4.75 & 27 & 0.11 & $-$2.75 & \nodata\\
Co\,I & 2.51 & 5 & 0.09 & $-$2.48 & $+$0.29\\
Ni\,I & 3.58 & 18 & 0.09 & $-$2.64 & $+$0.13\\
Zn\,I & 1.94 & 2 & 0.06 & $-$2.62 & $+$0.14\\
Sr\,II & $-$0.21 & 2 & 0.01 & $-$3.08 & $-$0.31\\
Y\,II & $-$0.95 & 5 & 0.06 & $-$3.16 & $-$0.39\\
Zr\,II & $-$0.26 & 4 & 0.10 & $-$2.84 & $-$0.07\\
Ba\,II & $-$1.66 & 2 & 0.08 & $-$3.84 & $-$1.07\\
Eu\,II & $-$3.20 & 1 & \nodata & $-$3.72 & $-$0.95\\
\enddata
\end{deluxetable}


\begin{thebibliography}{123}
\expandafter\ifx\csname natexlab\endcsname\relax\def\natexlab#1{#1}\fi

\bibitem[{{Abel} {et~al.}(2002){Abel}, {Bryan}, \& {Norman}}]{abel_sci}
{Abel}, T., {Bryan}, G.~L., \& {Norman}, M.~L. 2002, Science, 295, 93

\bibitem[{{Andrievsky} {et~al.}(2011){Andrievsky}, {Spite}, {Korotin}, {Fran{\c
  c}ois}, {Spite}, {Bonifacio}, {Cayrel}, \& {Hill}}]{andrievsky_sr}
{Andrievsky}, S.~M., {Spite}, F., {Korotin}, S.~A., {Fran{\c c}ois}, P.,
  {Spite}, M., {Bonifacio}, P., {Cayrel}, R., \& {Hill}, V. 2011, \aap, 530,
  A105

\bibitem[{{Andrievsky} {et~al.}(2009){Andrievsky}, {Spite}, {Korotin}, {Spite},
  {Fran{\c c}ois}, {Bonifacio}, {Cayrel}, \& {Hill}}]{andrievsky_ba}
{Andrievsky}, S.~M., {Spite}, M., {Korotin}, S.~A., {Spite}, F., {Fran{\c
  c}ois}, P., {Bonifacio}, P., {Cayrel}, R., \& {Hill}, V. 2009, \aap, 494,
  1083

\bibitem[{{Aoki} {et~al.}(2007){Aoki}, {Beers}, {Christlieb}, {Norris}, {Ryan},
  \& {Tsangarides}}]{aoki_cemp_2007}
{Aoki}, W., {Beers}, T.~C., {Christlieb}, N., {Norris}, J.~E., {Ryan}, S.~G.,
  \& {Tsangarides}, S. 2007, ApJ, 655, 492

\bibitem[{{Aoki} {et~al.}(2013{\natexlab{a}}){Aoki}, {Beers}, {Lee}, {Honda},
  {Ito}, {Takada-Hidai}, {Frebel}, {Suda}, {Fujimoto}, {Carollo}, \&
  {Sivarani}}]{aoki2013}
{Aoki}, W., {Beers}, T.~C., {Lee}, Y.~S., {Honda}, S., {Ito}, H.,
  {Takada-Hidai}, M., {Frebel}, A., {Suda}, T., {Fujimoto}, M.~Y., {Carollo},
  D., \& {Sivarani}, T. 2013{\natexlab{a}}, \aj, 145, 13

\bibitem[{{Aoki} {et~al.}(2008){Aoki}, {Beers}, {Sivarani}, {Marsteller},
  {Lee}, {Honda}, {Norris}, {Ryan}, \& {Carollo}}]{aoki2008}
{Aoki}, W., {Beers}, T.~C., {Sivarani}, T., {Marsteller}, B., {Lee}, Y.~S.,
  {Honda}, S., {Norris}, J.~E., {Ryan}, S.~G., \& {Carollo}, D. 2008, ApJ, 678,
  1351

\bibitem[{{Aoki} {et~al.}(2005){Aoki}, {Honda}, {Beers}, {Kajino}, {Ando},
  {Norris}, {Ryan}, {Izumiura}, {Sadakane}, \& {Takada-Hidai}}]{aoki05}
{Aoki}, W., {Honda}, S., {Beers}, T.~C., {Kajino}, T., {Ando}, H., {Norris},
  J.~E., {Ryan}, S.~G., {Izumiura}, H., {Sadakane}, K., \& {Takada-Hidai}, M.
  2005, ApJ, 632, 611

\bibitem[{{Aoki} {et~al.}(2002){Aoki}, {Norris}, {Ryan}, {Beers}, \&
  {Ando}}]{aoki_mg}
{Aoki}, W., {Norris}, J.~E., {Ryan}, S.~G., {Beers}, T.~C., \& {Ando}, H. 2002,
  ApJ, 576, L141

\bibitem[{{Aoki} {et~al.}(2013{\natexlab{b}}){Aoki}, {Suda}, {Boyd}, {Kajino},
  \& {Famiano}}]{aoki_srba}
{Aoki}, W., {Suda}, T., {Boyd}, R.~N., {Kajino}, T., \& {Famiano}, M.~A.
  2013{\natexlab{b}}, \apjl, 766, L13

\bibitem[{{Asplund}(2005)}]{asplund_araa}
{Asplund}, M. 2005, ARA\&A, 43, 481

\bibitem[{{Asplund} {et~al.}(2009){Asplund}, {Grevesse}, {Sauval}, \&
  {Scott}}]{asplund09}
{Asplund}, M., {Grevesse}, N., {Sauval}, A.~J., \& {Scott}, P. 2009, ARA\&A,
  47, 481

\bibitem[{{Barklem} {et~al.}(2005){Barklem}, {Christlieb}, {Beers}, {Hill},
  {Bessell}, {Holmberg}, {Marsteller}, {Rossi}, {Zickgraf}, \&
  {Reimers}}]{heresII}
{Barklem}, P.~S., {Christlieb}, N., {Beers}, T.~C., {Hill}, V., {Bessell},
  M.~S., {Holmberg}, J., {Marsteller}, B., {Rossi}, S., {Zickgraf}, F.-J., \&
  {Reimers}, D. 2005, A\&A, 439, 129

\bibitem[{{Baumueller} \& {Gehren}(1997)}]{al_nlte}
{Baumueller}, D. \& {Gehren}, T. 1997, A\&A, 325, 1088

\bibitem[{Beers {et~al.}(1992)Beers, Preston, \& Shectman}]{BPSII}
Beers, T.~C., Preston, G.~W., \& Shectman, S.~A. 1992, AJ, 103, 1987

\bibitem[{{Bergemann}(2011)}]{bergemann_ti}
{Bergemann}, M. 2011, \mnras, 413, 2184

\bibitem[{{Bergemann} \& {Gehren}(2008)}]{bergemann_mn}
{Bergemann}, M. \& {Gehren}, T. 2008, A\&A, 492, 823

\bibitem[{{Bergemann} {et~al.}(2012{\natexlab{a}}){Bergemann}, {Hansen},
  {Bautista}, \& {Ruchti}}]{bergemann_sr}
{Bergemann}, M., {Hansen}, C.~J., {Bautista}, M., \& {Ruchti}, G.
  2012{\natexlab{a}}, \aap, 546, A90

\bibitem[{{Bergemann} {et~al.}(2012{\natexlab{b}}){Bergemann}, {Lind},
  {Collet}, {Magic}, \& {Asplund}}]{bergemann12}
{Bergemann}, M., {Lind}, K., {Collet}, R., {Magic}, Z., \& {Asplund}, M.
  2012{\natexlab{b}}, MNRAS, 427, 27

\bibitem[{{Bernstein} {et~al.}(2003){Bernstein}, {Shectman}, {Gunnels},
  {Mochnacki}, \& {Athey}}]{mike}
{Bernstein}, R., {Shectman}, S.~A., {Gunnels}, S.~M., {Mochnacki}, S., \&
  {Athey}, A.~E. 2003, in Society of Photo-Optical Instrumentation Engineers
  (SPIE) Conference Series, ed. M.~{Iye} \& A.~F.~M. {Moorwood}, Vol. 4841,
  1694

\bibitem[{{Bessell} {et~al.}(2011){Bessell}, {Bloxham}, {Schmidt}, {Keller},
  {Tisserand}, \& {Francis}}]{skymapper_filter}
{Bessell}, M., {Bloxham}, G., {Schmidt}, B., {Keller}, S., {Tisserand}, P., \&
  {Francis}, P. 2011, \pasp, 123, 789

\bibitem[{{Bonifacio} {et~al.}(2011){Bonifacio}, {Caffau}, {Fran{\c c}ois},
  {Sbordone}, {Ludwig}, {Spite}, {Molaro}, {Spite}, {Cayrel}, {Hammer}, {Hill},
  {Nonino}, {Randich}, {Stelzer}, \& {Zaggia}}]{bonifacio2011}
{Bonifacio}, P., {Caffau}, E., {Fran{\c c}ois}, P., {Sbordone}, L., {Ludwig},
  H.-G., {Spite}, M., {Molaro}, P., {Spite}, F., {Cayrel}, R., {Hammer}, F.,
  {Hill}, V., {Nonino}, M., {Randich}, S., {Stelzer}, B., \& {Zaggia}, S. 2011,
  Astronomische Nachrichten, 332, 251

\bibitem[{{Bonifacio} {et~al.}(2012){Bonifacio}, {Sbordone}, {Caffau},
  {Ludwig}, {Spite}, {Gonz{\'a}lez Hern{\'a}ndez}, \& {Behara}}]{Bonifacio2012}
{Bonifacio}, P., {Sbordone}, L., {Caffau}, E., {Ludwig}, H.-G., {Spite}, M.,
  {Gonz{\'a}lez Hern{\'a}ndez}, J.~I., \& {Behara}, N.~T. 2012, \aap, 542, A87

\bibitem[{{Boyd} {et~al.}(2012){Boyd}, {Famiano}, {Meyer}, {Motizuki},
  {Kajino}, \& {Roederer}}]{boyd_tr}
{Boyd}, R.~N., {Famiano}, M.~A., {Meyer}, B.~S., {Motizuki}, Y., {Kajino}, T.,
  \& {Roederer}, I.~U. 2012, \apjl, 744, L14

\bibitem[{{Bromm} \& {Larson}(2004)}]{brommARAA}
{Bromm}, V. \& {Larson}, R.~B. 2004, ARA\&A, 42, 79

\bibitem[{{Caffau} {et~al.}(2011){Caffau}, {Bonifacio}, {Fran{\c c}ois},
  {Sbordone}, {Monaco}, {Spite}, {Spite}, {Ludwig}, {Cayrel}, {Zaggia},
  {Hammer}, {Randich}, {Molaro}, \& {Hill}}]{caffau2011}
{Caffau}, E., {Bonifacio}, P., {Fran{\c c}ois}, P., {Sbordone}, L., {Monaco},
  L., {Spite}, M., {Spite}, F., {Ludwig}, H.-G., {Cayrel}, R., {Zaggia}, S.,
  {Hammer}, F., {Randich}, S., {Molaro}, P., \& {Hill}, V. 2011, \nat, 477, 67

\bibitem[{{Caffau} {et~al.}(2013{\natexlab{a}}){Caffau}, {Bonifacio}, {Fran{\c
  c}ois}, {Sbordone}, {Spite}, {Monaco}, {Plez}, {Spite}, {Zaggia}, {Ludwig},
  {Cayrel}, {Molaro}, {Randich}, {Hammer}, \& {Hill}}]{caffau2013}
{Caffau}, E., {Bonifacio}, P., {Fran{\c c}ois}, P., {Sbordone}, L., {Spite},
  M., {Monaco}, L., {Plez}, B., {Spite}, F., {Zaggia}, S., {Ludwig}, H.-G.,
  {Cayrel}, R., {Molaro}, P., {Randich}, S., {Hammer}, F., \& {Hill}, V.
  2013{\natexlab{a}}, \aap, 560, A15

\bibitem[{{Caffau} {et~al.}(2013{\natexlab{b}}){Caffau}, {Bonifacio},
  {Sbordone}, {Francois}, {Monaco}, {Spite}, {Plez}, {Cayrel}, {Christlieb},
  {Clark}, {Glover}, {Klessen}, {Koch}, {Ludwig}, {Spite}, {Steffen}, \&
  {Zaggia}}]{toposI}
{Caffau}, E., {Bonifacio}, P., {Sbordone}, L., {Francois}, P., {Monaco}, L.,
  {Spite}, M., {Plez}, B., {Cayrel}, R., {Christlieb}, N., {Clark}, P.,
  {Glover}, S., {Klessen}, R., {Koch}, A., {Ludwig}, H.-G., {Spite}, F.,
  {Steffen}, M., \& {Zaggia}, S. 2013{\natexlab{b}}, ArXiv e-prints

\bibitem[{{Carney} {et~al.}(1996){Carney}, {Laird}, {Latham}, \&
  {Aguilar}}]{carneyetal:1996}
{Carney}, B.~W., {Laird}, J.~B., {Latham}, D.~W., \& {Aguilar}, L.~A. 1996,
  \aj, 112, 668

\bibitem[{{Casey}(2014)}]{smh}
{Casey}, A.~R. 2014, arXiv:1405.5968

\bibitem[{{Castelli} \& {Kurucz}(2004)}]{castelli_kurucz}
{Castelli}, F. \& {Kurucz}, R.~L. 2004, arXiv:astro-ph/0405087

\bibitem[{{Cayrel} {et~al.}(2004){Cayrel}, {Depagne}, {Spite}, {Hill}, {Spite},
  {Fran{\c c}ois}, {Plez}, {Beers}, {Primas}, {Andersen}, {Barbuy},
  {Bonifacio}, {Molaro}, \& {Nordstr{\"o}m}}]{cayrel2004}
{Cayrel}, R., {Depagne}, E., {Spite}, M., {Hill}, V., {Spite}, F., {Fran{\c
  c}ois}, P., {Plez}, B., {Beers}, T., {Primas}, F., {Andersen}, J., {Barbuy},
  B., {Bonifacio}, P., {Molaro}, P., \& {Nordstr{\"o}m}, B. 2004, A\&A, 416,
  1117

\bibitem[{{Chiappini} {et~al.}(2011){Chiappini}, {Frischknecht}, {Meynet},
  {Hirschi}, {Barbuy}, {Pignatari}, {Decressin}, \& {Maeder}}]{chiappini11}
{Chiappini}, C., {Frischknecht}, U., {Meynet}, G., {Hirschi}, R., {Barbuy}, B.,
  {Pignatari}, M., {Decressin}, T., \& {Maeder}, A. 2011, Nature, 472, 454

\bibitem[{{Christlieb} {et~al.}(2002){Christlieb}, {Bessell}, {Beers},
  {Gustafsson}, {Korn}, {Barklem}, {Karlsson}, {Mizuno-Wiedner}, \&
  {Rossi}}]{HE0107_Nature}
{Christlieb}, N., {Bessell}, M.~S., {Beers}, T.~C., {Gustafsson}, B., {Korn},
  A., {Barklem}, P.~S., {Karlsson}, T., {Mizuno-Wiedner}, M., \& {Rossi}, S.
  2002, Nature, 419, 904

\bibitem[{{Christlieb} {et~al.}(2008){Christlieb}, {Sch{\"o}rck}, {Frebel},
  {Beers}, {Wisotzki}, \& {Reimers}}]{hes4}
{Christlieb}, N., {Sch{\"o}rck}, T., {Frebel}, A., {Beers}, T.~C., {Wisotzki},
  L., \& {Reimers}, D. 2008, A\&A, 484, 721

\bibitem[{{Cohen} {et~al.}(2004){Cohen}, {Christlieb}, {McWilliam}, {Shectman},
  {Thompson}, {Wasserburg}, {Ivans}, {Dehn}, {Karlsson}, \&
  {Melendez}}]{cohen04}
{Cohen}, J.~G., {Christlieb}, N., {McWilliam}, A., {Shectman}, S., {Thompson},
  I., {Wasserburg}, G.~J., {Ivans}, I., {Dehn}, M., {Karlsson}, T., \&
  {Melendez}, J. 2004, ApJ, 612, 1107

\bibitem[{{Cohen} {et~al.}(2013){Cohen}, {Christlieb}, {Thompson}, {McWilliam},
  {Shectman}, {Reimers}, {Wisotzki}, \& {Kirby}}]{cohen2013}
{Cohen}, J.~G., {Christlieb}, N., {Thompson}, I., {McWilliam}, A., {Shectman},
  S., {Reimers}, D., {Wisotzki}, L., \& {Kirby}, E. 2013, \apj, 778, 56

\bibitem[{{Cohen} \& {Huang}(2010)}]{cohen_huang2010}
{Cohen}, J.~G. \& {Huang}, W. 2010, \apj, 719, 931

\bibitem[{{Cui} {et~al.}(2012){Cui}, {Zhao}, {Chu}, {Li}, {Li}, {Zhang}, {Su},
  {Yao}, {Wang}, {Xing}, {Li}, {Zhu}, {Wang}, {Gu}, {Luo}, {Xu}, {Zhang},
  {Liu}, {Zhang}, {Yang}, {Cao}, {Chen}, {Chen}, {Chen}, {Chen}, {Chu}, {Feng},
  {Gong}, {Hou}, {Hu}, {Hu}, {Hu}, {Jia}, {Jiang}, {Jiang}, {Jiang}, {Jin},
  {Li}, {Li}, {Li}, {Liu}, {Liu}, {Lu}, {Mao}, {Men}, {Qi}, {Qi}, {Shi},
  {Tang}, {Tao}, {Wang}, {Wang}, {Wang}, {Wang}, {Wang}, {Wang}, {Wang},
  {Wang}, {Wang}, {Wang}, {Wang}, {Wang}, {Xu}, {Xu}, {Yang}, {Yu}, {Yuan},
  {Yuan}, {Zhai}, {Zhang}, {Zhang}, {Zhang}, {Zhao}, {Zhou}, {Zhou}, {Zhu}, \&
  {Zou}}]{cui_lamost}
{Cui}, X.-Q., {Zhao}, Y.-H., {Chu}, Y.-Q., {Li}, G.-P., {Li}, Q., {Zhang},
  L.-P., {Su}, H.-J., {Yao}, Z.-Q., {Wang}, Y.-N., {Xing}, X.-Z., {Li}, X.-N.,
  {Zhu}, Y.-T., {Wang}, G., {Gu}, B.-Z., {Luo}, A.-L., {Xu}, X.-Q., {Zhang},
  Z.-C., {Liu}, G.-R., {Zhang}, H.-T., {Yang}, D.-H., {Cao}, S.-Y., {Chen},
  H.-Y., {Chen}, J.-J., {Chen}, K.-X., {Chen}, Y., {Chu}, J.-R., {Feng}, L.,
  {Gong}, X.-F., {Hou}, Y.-H., {Hu}, H.-Z., {Hu}, N.-S., {Hu}, Z.-W., {Jia},
  L., {Jiang}, F.-H., {Jiang}, X., {Jiang}, Z.-B., {Jin}, G., {Li}, A.-H.,
  {Li}, Y., {Li}, Y.-P., {Liu}, G.-Q., {Liu}, Z.-G., {Lu}, W.-Z., {Mao}, Y.-D.,
  {Men}, L., {Qi}, Y.-J., {Qi}, Z.-X., {Shi}, H.-M., {Tang}, Z.-H., {Tao},
  Q.-S., {Wang}, D.-Q., {Wang}, D., {Wang}, G.-M., {Wang}, H., {Wang}, J.-N.,
  {Wang}, J., {Wang}, J.-L., {Wang}, J.-P., {Wang}, L., {Wang}, S.-Q., {Wang},
  Y., {Wang}, Y.-F., {Xu}, L.-Z., {Xu}, Y., {Yang}, S.-H., {Yu}, Y., {Yuan},
  H., {Yuan}, X.-Y., {Zhai}, C., {Zhang}, J., {Zhang}, Y.-X., {Zhang}, Y.,
  {Zhao}, M., {Zhou}, F., {Zhou}, G.-H., {Zhu}, J., \& {Zou}, S.-C. 2012,
  Research in Astronomy and Astrophysics, 12, 1197

\bibitem[{{Fran{\c c}ois} {et~al.}(2003){Fran{\c c}ois}, {Depagne}, {Hill},
  {Spite}, {Spite}, {Plez}, {Beers}, {Barbuy}, {Cayrel}, {Andersen},
  {Bonifacio}, {Molaro}, {Nordstr{\"o}m}, \& {Primas}}]{Francois03}
{Fran{\c c}ois}, P., {Depagne}, E., {Hill}, V., {Spite}, M., {Spite}, F.,
  {Plez}, B., {Beers}, T.~C., {Barbuy}, B., {Cayrel}, R., {Andersen}, J.,
  {Bonifacio}, P., {Molaro}, P., {Nordstr{\"o}m}, B., \& {Primas}, F. 2003,
  A\&A, 403, 1105

\bibitem[{{Frebel}(2010)}]{frebel10}
{Frebel}, A. 2010, Astronomische Nachrichten, 331, 474

\bibitem[{{Frebel} {et~al.}(2005){Frebel}, {Aoki}, {Christlieb}, {Ando},
  {Asplund}, {Barklem}, {Beers}, {Eriksson}, {Fechner}, {Fujimoto}, {Honda},
  {Kajino}, {Minezaki}, {Nomoto}, {Norris}, {Ryan}, {Takada-Hidai},
  {Tsangarides}, \& {Yoshii}}]{HE1327_Nature}
{Frebel}, A., {Aoki}, W., {Christlieb}, N., {Ando}, H., {Asplund}, M.,
  {Barklem}, P.~S., {Beers}, T.~C., {Eriksson}, K., {Fechner}, C., {Fujimoto},
  M.~Y., {Honda}, S., {Kajino}, T., {Minezaki}, T., {Nomoto}, K., {Norris},
  J.~E., {Ryan}, S.~G., {Takada-Hidai}, M., {Tsangarides}, S., \& {Yoshii}, Y.
  2005, Nature, 434, 871

\bibitem[{{Frebel} {et~al.}(2013){Frebel}, {Casey}, {Jacobson}, \&
  {Yu}}]{teff_calib}
{Frebel}, A., {Casey}, A.~R., {Jacobson}, H.~R., \& {Yu}, Q. 2013, \apj, 769,
  57

\bibitem[{{Frebel} {et~al.}(2006){Frebel}, {Christlieb}, {Norris}, {Beers},
  {Bessell}, {Rhee}, {Fechner}, {Marsteller}, {Rossi}, {Thom}, {Wisotzki}, \&
  {Reimers}}]{frebel_bmps}
{Frebel}, A., {Christlieb}, N., {Norris}, J.~E., {Beers}, T.~C., {Bessell},
  M.~S., {Rhee}, J., {Fechner}, C., {Marsteller}, B., {Rossi}, S., {Thom}, C.,
  {Wisotzki}, L., \& {Reimers}, D. 2006, ApJ, 652, 1585

\bibitem[Frebel 
\& Norris(2015)]{fnARAA} Frebel, A., \& Norris, J.~E.\ 2015, arXiv:1501.06921

\bibitem[{{Frebel} {et~al.}(2014){Frebel}, {Simon}, \& {Kirby}}]{frebel_segue1}
{Frebel}, A., {Simon}, J.~D., \& {Kirby}, E.~N. 2014, \apj, 786, 74

\bibitem[{{Frischknecht} {et~al.}(2012){Frischknecht}, {Hirschi}, \&
  {Thielemann}}]{frischknecht}
{Frischknecht}, U., {Hirschi}, R., \& {Thielemann}, F.-K. 2012, \aap, 538, L2

\bibitem[{{Gehren} {et~al.}(2004){Gehren}, {Liang}, {Shi}, {Zhang}, \&
  {Zhao}}]{gehren2004_nlte}
{Gehren}, T., {Liang}, Y.~C., {Shi}, J.~R., {Zhang}, H.~W., \& {Zhao}, G. 2004,
  A\&A, 413, 1045

\bibitem[{{Gustafsson} {et~al.}(2008){Gustafsson}, {Edvardsson}, {Eriksson},
  {J{\o}rgensen}, {Nordlund}, \& {Plez}}]{marcs}
{Gustafsson}, B., {Edvardsson}, B., {Eriksson}, K., {J{\o}rgensen}, U.~G.,
  {Nordlund}, {\AA}., \& {Plez}, B. 2008, A\&A, 486, 951

\bibitem[{{Hansen} {et~al.}(2014{\natexlab{a}}){Hansen}, {Montes}, \&
  {Arcones}}]{hansen2014_rproc}
{Hansen}, C.~J., {Montes}, F., \& {Arcones}, A. 2014{\natexlab{a}}, \apj, 797,
  123

\bibitem[{{Hansen} {et~al.}(2014{\natexlab{b}}){Hansen}, {Hansen},
  {Christlieb}, {Yong}, {Bessell}, {Garc{\'{\i}}a P{\'e}rez}, {Beers},
  {Placco}, {Frebel}, {Norris}, \& {Asplund}}]{hansen2014}
{Hansen}, T., {Hansen}, C.~J., {Christlieb}, N., {Yong}, D., {Bessell}, M.~S.,
  {Garc{\'{\i}}a P{\'e}rez}, A.~E., {Beers}, T.~C., {Placco}, V.~M., {Frebel},
  A., {Norris}, J.~E., \& {Asplund}, M. 2014{\natexlab{b}}, \apj, 787, 162

\bibitem[{{Hartwick}(1976)}]{hartwick1976}
{Hartwick}, F.~D.~A. 1976, ApJ, 209, 418

\bibitem[{{Heil} {et~al.}(2009){Heil}, {Juseviciute}, {K{\"a}ppeler},
  {Gallino}, {Pignatari}, \& {Uberseder}}]{heil_weaks}
{Heil}, M., {Juseviciute}, A., {K{\"a}ppeler}, F., {Gallino}, R., {Pignatari},
  M., \& {Uberseder}, E. 2009, 
  Publications of the Astronomical Society of Australia, 26, 243

\bibitem[{{Hirano} {et~al.}(2014){Hirano}, {Hosokawa}, {Yoshida}, {Umeda},
  {Omukai}, {Chiaki}, \& {Yorke}}]{hirano2014}
{Hirano}, S., {Hosokawa}, T., {Yoshida}, N., {Umeda}, H., {Omukai}, K.,
  {Chiaki}, G., \& {Yorke}, H.~W. 2014, \apj, 781, 60

\bibitem[{{Hobbs} {et~al.}(1999){Hobbs}, {Thorburn}, \& {Rebull}}]{hobbs99}
{Hobbs}, L.~M., {Thorburn}, J.~A., \& {Rebull}, L.~M. 1999, \apj, 523, 797

\bibitem[{{Hollek} {et~al.}(2011){Hollek}, {Frebel}, {Roederer}, {Sneden},
  {Shetrone}, {Beers}, {Kang}, \& {Thom}}]{hollek11}
{Hollek}, J.~K., {Frebel}, A., {Roederer}, I.~U., {Sneden}, C., {Shetrone}, M.,
  {Beers}, T.~C., {Kang}, S.-j., \& {Thom}, C. 2011, ApJ, 742, 54

\bibitem[{{Honda} {et~al.}(2006){Honda}, {Aoki}, {Ishimaru}, {Wanajo}, \&
  {Ryan}}]{honda06}
{Honda}, S., {Aoki}, W., {Ishimaru}, Y., {Wanajo}, S., \& {Ryan}, S.~G. 2006,
  ApJ, 643, 1180

\bibitem[{{Honda} {et~al.}(2004){Honda}, {Aoki}, {Kajino}, {Ando}, {Beers},
  {Izumiura}, {Sadakane}, \& {Takada-Hidai}}]{honda04}
{Honda}, S., {Aoki}, W., {Kajino}, T., {Ando}, H., {Beers}, T.~C., {Izumiura},
  H., {Sadakane}, K., \& {Takada-Hidai}, M. 2004, ApJ, 607, 474

\bibitem[{{Howes} {et~al.}(2014){Howes}, {Asplund}, {Casey}, {Keller}, {Yong},
  {Gilmore}, {Lind}, {Worley}, {Bessell}, {Casagrande}, {Marino}, {Nataf},
  {Owen}, {Da Costa}, {Schmidt}, {Tisserand}, {Randich}, {Feltzing},
  {Vallenari}, {Allende Prieto}, {Bensby}, {Flaccomio}, {Korn}, {Pancino},
  {Recio-Blanco}, {Smiljanic}, {Bergemann}, {Costado}, {Damiani}, {Heiter},
  {Hill}, {Hourihane}, {Jofr{\'e}}, {Lardo}, {de Laverny}, {Magrini},
  {Maiorca}, {Masseron}, {Morbidelli}, {Sacco}, {Minniti}, \&
  {Zoccali}}]{howes2014}
{Howes}, L.~M., {Asplund}, M., {Casey}, A.~R., {Keller}, S.~C., {Yong}, D.,
  {Gilmore}, G., {Lind}, K., {Worley}, C., {Bessell}, M.~S., {Casagrande}, L.,
  {Marino}, A.~F., {Nataf}, D.~M., {Owen}, C.~I., {Da Costa}, G.~S., {Schmidt},
  B.~P., {Tisserand}, P., {Randich}, S., {Feltzing}, S., {Vallenari}, A.,
  {Allende Prieto}, C., {Bensby}, T., {Flaccomio}, E., {Korn}, A.~J.,
  {Pancino}, E., {Recio-Blanco}, A., {Smiljanic}, R., {Bergemann}, M.,
  {Costado}, M.~T., {Damiani}, F., {Heiter}, U., {Hill}, V., {Hourihane}, A.,
  {Jofr{\'e}}, P., {Lardo}, C., {de Laverny}, P., {Magrini}, L., {Maiorca}, E.,
  {Masseron}, T., {Morbidelli}, L., {Sacco}, G.~G., {Minniti}, D., \&
  {Zoccali}, M. 2014, \mnras, 445, 4241

\bibitem[{{Ishimaru} {et~al.}(2005){Ishimaru}, {Wanajo}, {Aoki}, {Ryan}, \&
  {Prantzos}}]{ishimaru2005}
{Ishimaru}, Y., {Wanajo}, S., {Aoki}, W., {Ryan}, S.~G., \& {Prantzos}, N.
  2005, Nuclear Physics A, 758, 603

\bibitem[{{Ivans} {et~al.}(2003){Ivans}, {Sneden}, {James}, {Preston},
  {Fulbright}, {H{\"o}flich}, {Carney}, \& {Wheeler}}]{ivans_alphapoor}
{Ivans}, I.~I., {Sneden}, C., {James}, C.~R., {Preston}, G.~W., {Fulbright},
  J.~P., {H{\"o}flich}, P.~A., {Carney}, B.~W., \& {Wheeler}, J.~C. 2003, ApJ,
  592, 906

\bibitem[{{Jacobson} \& {Frebel}(2014)}]{jacobson13}
{Jacobson}, H.~R. \& {Frebel}, A. 2014, Journal of Physics G Nuclear Physics,
  41, 044001

\bibitem[{{Johnson}(2002)}]{johnson2002_23stars}
{Johnson}, J.~A. 2002, ApJS, 139, 219

\bibitem[{{Keller} {et~al.}(2014){Keller}, {Bessell}, {Frebel}, {Casey},
  {Asplund}, {Jacobson}, {Lind}, {Norris}, {Yong}, {Heger}, {Magic}, {da
  Costa}, {Schmidt}, \& {Tisserand}}]{keller_thestar}
{Keller}, S.~C., {Bessell}, M.~S., {Frebel}, A., {Casey}, A.~R., {Asplund}, M.,
  {Jacobson}, H.~R., {Lind}, K., {Norris}, J.~E., {Yong}, D., {Heger}, A.,
  {Magic}, Z., {da Costa}, G.~S., {Schmidt}, B.~P., \& {Tisserand}, P. 2014,
  \nat, 506, 463

\bibitem[{{Keller} {et~al.}(2007){Keller}, {Schmidt}, {Bessell}, {Conroy},
  {Francis}, {Granlund}, {Kowald}, {Oates}, {Martin-Jones}, {Preston},
  {Tisserand}, {Vaccarella}, \& {Waterson}}]{keller}
{Keller}, S.~C., {Schmidt}, B.~P., {Bessell}, M.~S., {Conroy}, P.~G.,
  {Francis}, P., {Granlund}, A., {Kowald}, E., {Oates}, A.~P., {Martin-Jones},
  T., {Preston}, T., {Tisserand}, P., {Vaccarella}, A., \& {Waterson}, M.~F.
  2007, Publications of the Astronomical Society of Australia, 24, 1

\bibitem[{{Kelson}(2003)}]{kelson03}
{Kelson}, D.~D. 2003, \pasp, 115, 688

\bibitem[{{Kim} {et~al.}(2002){Kim}, {Demarque}, {Yi}, \& {Alexander}}]{Y2_iso}
{Kim}, Y.-C., {Demarque}, P., {Yi}, S.~K., \& {Alexander}, D.~R. 2002, ApJS,
  143, 499

\bibitem[{{Kobayashi} {et~al.}(2014){Kobayashi}, {Ishigaki}, {Tominaga}, \&
  {Nomoto}}]{kobayashi2014_alphapoor}
{Kobayashi}, C., {Ishigaki}, M.~N., {Tominaga}, N., \& {Nomoto}, K. 2014,
  \apjl, 785, L5

\bibitem[{{Kobayashi} {et~al.}(2006){Kobayashi}, {Umeda}, {Nomoto}, {Tominaga},
  \& {Ohkubo}}]{kobayashi2006}
{Kobayashi}, C., {Umeda}, H., {Nomoto}, K., {Tominaga}, N., \& {Ohkubo}, T.
  2006, \apj, 653, 1145

\bibitem[{{Kordopatis} {et~al.}(2013){Kordopatis}, {Gilmore}, {Steinmetz},
  {Boeche}, {Seabroke}, {Siebert}, {Zwitter}, {Binney}, {de Laverny},
  {Recio-Blanco}, {Williams}, {Piffl}, {Enke}, {Roeser}, {Bijaoui}, {Wyse},
  {Freeman}, {Munari}, {Carrillo}, {Anguiano}, {Burton}, {Campbell}, {Cass},
  {Fiegert}, {Hartley}, {Parker}, {Reid}, {Ritter}, {Russell}, {Stupar},
  {Watson}, {Bienaym{\'e}}, {Bland-Hawthorn}, {Gerhard}, {Gibson}, {Grebel},
  {Helmi}, {Navarro}, {Conrad}, {Famaey}, {Faure}, {Just}, {Kos}, {Matijevi{\v
  c}}, {McMillan}, {Minchev}, {Scholz}, {Sharma}, {Siviero}, {de Boer}, \& {{\v
  Z}erjal}}]{rave_dr4}
{Kordopatis}, G., {Gilmore}, G., {Steinmetz}, M., {Boeche}, C., {Seabroke},
  G.~M., {Siebert}, A., {Zwitter}, T., {Binney}, J., {de Laverny}, P.,
  {Recio-Blanco}, A., {Williams}, M.~E.~K., {Piffl}, T., {Enke}, H., {Roeser},
  S., {Bijaoui}, A., {Wyse}, R.~F.~G., {Freeman}, K., {Munari}, U., {Carrillo},
  I., {Anguiano}, B., {Burton}, D., {Campbell}, R., {Cass}, C.~J.~P.,
  {Fiegert}, K., {Hartley}, M., {Parker}, Q.~A., {Reid}, W., {Ritter}, A.,
  {Russell}, K.~S., {Stupar}, M., {Watson}, F.~G., {Bienaym{\'e}}, O.,
  {Bland-Hawthorn}, J., {Gerhard}, O., {Gibson}, B.~K., {Grebel}, E.~K.,
  {Helmi}, A., {Navarro}, J.~F., {Conrad}, C., {Famaey}, B., {Faure}, C.,
  {Just}, A., {Kos}, J., {Matijevi{\v c}}, G., {McMillan}, P.~J., {Minchev},
  I., {Scholz}, R., {Sharma}, S., {Siviero}, A., {de Boer}, E.~W., \& {{\v
  Z}erjal}, M. 2013, \aj, 146, 134

\bibitem[{{Kurucz} \& {Bell}(1995)}]{kurucz&bell1995}
{Kurucz}, R.~L. \& {Bell}, B. 1995, {Atomic line list}

\bibitem[{{Lai} {et~al.}(2008){Lai}, {Bolte}, {Johnson}, {Lucatello}, {Heger},
  \& {Woosley}}]{lai2008}
{Lai}, D.~K., {Bolte}, M., {Johnson}, J.~A., {Lucatello}, S., {Heger}, A., \&
  {Woosley}, S.~E. 2008, ApJ, 681, 1524

\bibitem[{{Lawler} \& {Dakin}(1989)}]{lawler_sc}
{Lawler}, J.~E. \& {Dakin}, J.~T. 1989, Journal of the Optical Society of
  America B Optical Physics, 6, 1457

\bibitem[{{Lawler} {et~al.}(2013){Lawler}, {Guzman}, {Wood}, {Sneden}, \&
  {Cowan}}]{lawler_tiI}
{Lawler}, J.~E., {Guzman}, A., {Wood}, M.~P., {Sneden}, C., \& {Cowan}, J.~J.
  2013, \apjs, 205, 11

\bibitem[{{Lee} {et~al.}(2013){Lee}, {Beers}, {Masseron}, {Plez}, {Rockosi},
  {Sobeck}, {Yanny}, {Lucatello}, {Sivarani}, {Placco}, \&
  {Carollo}}]{YSLee_cemp}
{Lee}, Y.~S., {Beers}, T.~C., {Masseron}, T., {Plez}, B., {Rockosi}, C.~M.,
  {Sobeck}, J., {Yanny}, B., {Lucatello}, S., {Sivarani}, T., {Placco}, V.~M.,
  \& {Carollo}, D. 2013, \aj, 146, 132

\bibitem[{{Li} {et~al.}(2010){Li}, {Christlieb}, {Sch{\"o}rck}, {Norris},
  {Bessell}, {Yong}, {Beers}, {Lee}, {Frebel}, \& {Zhao}}]{li_mdf}
{Li}, H.~N., {Christlieb}, N., {Sch{\"o}rck}, T., {Norris}, J.~E., {Bessell},
  M.~S., {Yong}, D., {Beers}, T.~C., {Lee}, Y.~S., {Frebel}, A., \& {Zhao}, G.
  2010, A\&A, 521, A10

\bibitem[{{Li} {et~al.}(2015){Li}, {Zhao}, {Christlieb}, {Wang}, {Wang},
  {Zhang}, {Hou}, \& {Yuan}}]{LAMOST_emp}
{Li}, H.-N., {Zhao}, G., {Christlieb}, N., {Wang}, L., {Wang}, W., {Zhang}, Y.,
  {Hou}, Y., \& {Yuan}, H. 2015, \apj, 798, 110

\bibitem[{{Lind} {et~al.}(2009){Lind}, {Asplund}, \& {Barklem}}]{lind_li}
{Lind}, K., {Asplund}, M., \& {Barklem}, P.~S. 2009, \aap, 503, 541

\bibitem[{{Lind} {et~al.}(2011){Lind}, {Asplund}, {Barklem}, \&
  {Belyaev}}]{lind_na}
{Lind}, K., {Asplund}, M., {Barklem}, P.~S., \& {Belyaev}, A.~K. 2011, \aap,
  528, A103

\bibitem[{{Lind} {et~al.}(2012){Lind}, {Bergemann}, \& {Asplund}}]{lind2012}
{Lind}, K., {Bergemann}, M., \& {Asplund}, M. 2012, MNRAS, 427, 50

\bibitem[{{Liske} {et~al.}(2003){Liske}, {Lemon}, {Driver}, {Cross}, \&
  {Couch}}]{MGC03}
{Liske}, J., {Lemon}, D.~J., {Driver}, S.~P., {Cross}, N.~J.~G., \& {Couch},
  W.~J. 2003, \mnras, 344, 307

\bibitem[{{Mashonkina}(2013)}]{mashonkina_mg}
{Mashonkina}, L. 2013, \aap, 550, A28

\bibitem[{{Mashonkina} {et~al.}(2003){Mashonkina}, {Gehren}, {Travaglio}, \&
  {Borkova}}]{mashonkina2003}
{Mashonkina}, L., {Gehren}, T., {Travaglio}, C., \& {Borkova}, T. 2003, \aap,
  397, 275

\bibitem[{{Mashonkina} {et~al.}(2007){Mashonkina}, {Korn}, \&
  {Przybilla}}]{mashonkina_ca}
{Mashonkina}, L., {Korn}, A.~J., \& {Przybilla}, N. 2007, A\&A, 461, 261

\bibitem[{{McWilliam} {et~al.}(1995){McWilliam}, {Preston}, {Sneden}, \&
  {Searle}}]{McWilliametal}
{McWilliam}, A., {Preston}, G.~W., {Sneden}, C., \& {Searle}, L. 1995, AJ, 109,
  2757

\bibitem[{{Ness} {et~al.}(2014){Ness}, {Asplund}, \& {Casey}}]{ness2014}
{Ness}, M., {Asplund}, M., \& {Casey}, A.~R. 2014, \mnras, 445, 2994

\bibitem[{{Nissen} \& {Schuster}(1997)}]{nissen_schuster}
{Nissen}, P.~E. \& {Schuster}, W.~J. 1997, A\&A, 326, 751

\bibitem[{{Norris} {et~al.}(2013){Norris}, {Bessell}, {Yong}, {Christlieb},
  {Barklem}, {Asplund}, {Murphy}, {Beers}, {Frebel}, \& {Ryan}}]{norris13_I}
{Norris}, J.~E., {Bessell}, M.~S., {Yong}, D., {Christlieb}, N., {Barklem},
  P.~S., {Asplund}, M., {Murphy}, S.~J., {Beers}, T.~C., {Frebel}, A., \&
  {Ryan}, S.~G. 2013, \apj, 762, 25

\bibitem[{{Norris} {et~al.}(2007){Norris}, {Christlieb}, {Korn}, {Eriksson},
  {Bessell}, {Beers}, {Wisotzki}, \& {Reimers}}]{he0557}
{Norris}, J.~E., {Christlieb}, N., {Korn}, A.~J., {Eriksson}, K., {Bessell},
  M.~S., {Beers}, T.~C., {Wisotzki}, L., \& {Reimers}, D. 2007, ApJ, 670, 774

\bibitem[{{Norris} {et~al.}(1996){Norris}, {Ryan}, \& {Beers}}]{norris96data}
{Norris}, J.~E., {Ryan}, S.~G., \& {Beers}, T.~C. 1996, ApJS, 107, 391

\bibitem[{{Norris} {et~al.}(1999){Norris}, {Ryan}, \& {Beers}}]{norrisUBVpho}
---. 1999, ApJS, 123, 639

\bibitem[{{Pietrinferni} {et~al.}(2006){Pietrinferni}, {Cassisi}, {Salaris}, \&
  {Castelli}}]{basti_HB}
{Pietrinferni}, A., {Cassisi}, S., {Salaris}, M., \& {Castelli}, F. 2006, \apj,
  642, 797

\bibitem[{{Pignatari} {et~al.}(2008){Pignatari}, {Gallino}, {Meynet},
  {Hirschi}, {Herwig}, \& {Wiescher}}]{pignatari}
{Pignatari}, M., {Gallino}, R., {Meynet}, G., {Hirschi}, R., {Herwig}, F., \&
  {Wiescher}, M. 2008, ApJL, 687, L95

\bibitem[{{Placco} {et~al.}(2014{\natexlab{a}}){Placco}, {Frebel}, {Beers},
  {Christlieb}, {Lee}, {Kennedy}, {Rossi}, \& {Santucci}}]{placco2013_magII}
{Placco}, V.~M., {Frebel}, A., {Beers}, T.~C., {Christlieb}, N., {Lee}, Y.~S.,
  {Kennedy}, C.~R., {Rossi}, S., \& {Santucci}, R.~M. 2014{\natexlab{a}}, \apj,
  781, 40

\bibitem[{{Placco} {et~al.}(2013){Placco}, {Frebel}, {Beers}, {Karakas},
  {Kennedy}, {Rossi}, {Christlieb}, \& {Stancliffe}}]{placco2013_magI}
{Placco}, V.~M., {Frebel}, A., {Beers}, T.~C., {Karakas}, A.~I., {Kennedy},
  C.~R., {Rossi}, S., {Christlieb}, N., \& {Stancliffe}, R.~J. 2013, \apj, 770,
  104

\bibitem[{{Placco} {et~al.}(2014{\natexlab{b}}){Placco}, {Frebel}, {Beers}, \&
  {Stancliffe}}]{placco14}
{Placco}, V.~M., {Frebel}, A., {Beers}, T.~C., \& {Stancliffe}, R.~J.
  2014{\natexlab{b}}, \apj, 797, 21

\bibitem[{{Placco} {et~al.}(2011){Placco}, {Kennedy}, {Beers}, {Christlieb},
  {Rossi}, {Sivarani}, {Lee}, {Reimers}, \& {Wisotzki}}]{placco_gband}
{Placco}, V.~M., {Kennedy}, C.~R., {Beers}, T.~C., {Christlieb}, N., {Rossi},
  S., {Sivarani}, T., {Lee}, Y.~S., {Reimers}, D., \& {Wisotzki}, L. 2011, \aj,
  142, 188

\bibitem[{{Roederer}(2013)}]{roederer_ncap}
{Roederer}, I.~U. 2013, \aj, 145, 26

\bibitem[{{Roederer} {et~al.}(2010{\natexlab{a}}){Roederer}, {Cowan},
  {Karakas}, {Kratz}, {Lugaro}, {Simmerer}, {Farouqi}, \&
  {Sneden}}]{roederer_ubiqrproc}
{Roederer}, I.~U., {Cowan}, J.~J., {Karakas}, A.~I., {Kratz}, K.-L., {Lugaro},
  M., {Simmerer}, J., {Farouqi}, K., \& {Sneden}, C. 2010{\natexlab{a}}, \apj,
  724, 975

\bibitem[{{Roederer} {et~al.}(2014{\natexlab{a}}){Roederer}, {Cowan},
  {Preston}, {Shectman}, {Sneden}, \& {Thompson}}]{roederer14_9stars}
{Roederer}, I.~U., {Cowan}, J.~J., {Preston}, G.~W., {Shectman}, S.~A.,
  {Sneden}, C., \& {Thompson}, I.~B. 2014{\natexlab{a}}, \mnras, 445, 2970

\bibitem[{{Roederer} {et~al.}(2014{\natexlab{b}}){Roederer}, {Preston},
  {Thompson}, {Shectman}, {Sneden}, {Burley}, \& {Kelson}}]{roederer_313stars}
{Roederer}, I.~U., {Preston}, G.~W., {Thompson}, I.~B., {Shectman}, S.~A.,
  {Sneden}, C., {Burley}, G.~S., \& {Kelson}, D.~D. 2014{\natexlab{b}}, \aj,
  147, 136

\bibitem[{{Roederer} {et~al.}(2010{\natexlab{b}}){Roederer}, {Sneden},
  {Thompson}, {Preston}, \& {Shectman}}]{roederer10}
{Roederer}, I.~U., {Sneden}, C., {Thompson}, I.~B., {Preston}, G.~W., \&
  {Shectman}, S.~A. 2010{\natexlab{b}}, ApJ, 711, 573

\bibitem[{{Ruchti} {et~al.}(2013){Ruchti}, {Bergemann}, {Serenelli},
  {Casagrande}, \& {Lind}}]{ruchti2013}
{Ruchti}, G.~R., {Bergemann}, M., {Serenelli}, A., {Casagrande}, L., \& {Lind},
  K. 2013, \mnras, 429, 126

\bibitem[{Ryan \& Norris(1991)}]{Ryanetal:1991}
Ryan, S.~G. \& Norris, J.~E. 1991, AJ, 101, 1835

\bibitem[{{Ryan} {et~al.}(1996){Ryan}, {Norris}, \& {Beers}}]{Ryan96}
{Ryan}, S.~G., {Norris}, J.~E., \& {Beers}, T.~C. 1996, ApJ, 471, 254

\bibitem[{{Schlaufman} \& {Casey}(2014)}]{sc14}
{Schlaufman}, K.~C. \& {Casey}, A.~R. 2014, ArXiv e-prints

\bibitem[{{Sch{\"o}rck} {et~al.}(2009){Sch{\"o}rck}, {Christlieb}, {Cohen},
  {Beers}, {Shectman}, {Thompson}, {McWilliam}, {Bessell}, {Norris},
  {Mel{\'e}ndez}, {Ram{\'{\i}}rez}, {Haynes}, {Cass}, {Hartley}, {Russell},
  {Watson}, {Zickgraf}, {Behnke}, {Fechner}, {Fuhrmeister}, {Barklem},
  {Edvardsson}, {Frebel}, {Wisotzki}, \& {Reimers}}]{schoerck}
{Sch{\"o}rck}, T., {Christlieb}, N., {Cohen}, J.~G., {Beers}, T.~C.,
  {Shectman}, S., {Thompson}, I., {McWilliam}, A., {Bessell}, M.~S., {Norris},
  J.~E., {Mel{\'e}ndez}, J., {Ram{\'{\i}}rez}, S., {Haynes}, D., {Cass}, P.,
  {Hartley}, M., {Russell}, K., {Watson}, F., {Zickgraf}, F., {Behnke}, B.,
  {Fechner}, C., {Fuhrmeister}, B., {Barklem}, P.~S., {Edvardsson}, B.,
  {Frebel}, A., {Wisotzki}, L., \& {Reimers}, D. 2009, A\&A, 507, 817

\bibitem[{{Skrutskie} {et~al.}(2006){Skrutskie}, {Cutri}, {Stiening},
  {Weinberg}, {Schneider}, {Carpenter}, {Beichman}, {Capps}, {Chester},
  {Elias}, {Huchra}, {Liebert}, {Lonsdale}, {Monet}, {Price}, {Seitzer},
  {Jarrett}, {Kirkpatrick}, {Gizis}, {Howard}, {Evans}, {Fowler}, {Fullmer},
  {Hurt}, {Light}, {Kopan}, {Marsh}, {McCallon}, {Tam}, {Van Dyk}, \&
  {Wheelock}}]{2MASS}
{Skrutskie}, M.~F., {Cutri}, R.~M., {Stiening}, R., {Weinberg}, M.~D.,
  {Schneider}, S., {Carpenter}, J.~M., {Beichman}, C., {Capps}, R., {Chester},
  T., {Elias}, J., {Huchra}, J., {Liebert}, J., {Lonsdale}, C., {Monet}, D.~G.,
  {Price}, S., {Seitzer}, P., {Jarrett}, T., {Kirkpatrick}, J.~D., {Gizis},
  J.~E., {Howard}, E., {Evans}, T., {Fowler}, J., {Fullmer}, L., {Hurt}, R.,
  {Light}, R., {Kopan}, E.~L., {Marsh}, K.~A., {McCallon}, H.~L., {Tam}, R.,
  {Van Dyk}, S., \& {Wheelock}, S. 2006, AJ, 131, 1163

\bibitem[{{Sneden} {et~al.}(2008){Sneden}, {Cowan}, \& {Gallino}}]{sneden_araa}
{Sneden}, C., {Cowan}, J.~J., \& {Gallino}, R. 2008, ARA\&A, 46, 241

\bibitem[{{Sneden}(1973)}]{moog}
{Sneden}, C.~A. 1973, PhD thesis, The University of Texas at Austin

\bibitem[{{Sobeck} {et~al.}(2011){Sobeck}, {Kraft}, {Sneden}, {Preston},
  {Cowan}, {Smith}, {Thompson}, {Shectman}, \& {Burley}}]{sobeck11}
{Sobeck}, J.~S., {Kraft}, R.~P., {Sneden}, C., {Preston}, G.~W., {Cowan},
  J.~J., {Smith}, G.~H., {Thompson}, I.~B., {Shectman}, S.~A., \& {Burley},
  G.~S. 2011, AJ, 141, 175

\bibitem[{{Spite} {et~al.}(2005){Spite}, {Cayrel}, {Plez}, {Hill}, {Spite},
  {Depagne}, {Fran{\c c}ois}, {Bonifacio}, {Barbuy}, {Beers}, {Andersen},
  {Molaro}, {Nordstr{\"o}m}, \& {Primas}}]{spite2005}
{Spite}, M., {Cayrel}, R., {Plez}, B., {Hill}, V., {Spite}, F., {Depagne}, E.,
  {Fran{\c c}ois}, P., {Bonifacio}, P., {Barbuy}, B., {Beers}, T., {Andersen},
  J., {Molaro}, P., {Nordstr{\"o}m}, B., \& {Primas}, F. 2005, A\&A, 430, 655

\bibitem[{{Spite} {et~al.}(2000){Spite}, {Depagne}, {Nordstr{\"o}m}, {Hill},
  {Cayrel}, {Spite}, \& {Beers}}]{spite2000}
{Spite}, M., {Depagne}, E., {Nordstr{\"o}m}, B., {Hill}, V., {Cayrel}, R.,
  {Spite}, F., \& {Beers}, T.~C. 2000, A\&A, 360, 1077

\bibitem[{{Spite} \& {Spite}(2014)}]{spite_ncap2013}
{Spite}, M. \& {Spite}, F. 2014, Astronomische Nachrichten, 335, 65

\bibitem[{{Stacy} \& {Bromm}(2014)}]{stacy14}
{Stacy}, A. \& {Bromm}, V. 2014, \apj, 785, 73

\bibitem[{{Starkenburg} {et~al.}(2010){Starkenburg}, {Hill}, {Tolstoy},
  {Gonz{\'a}lez Hern{\'a}ndez}, {Irwin}, {Helmi}, {Battaglia}, {Jablonka},
  {Tafelmeyer}, {Shetrone}, {Venn}, \& {de Boer}}]{starkenburg10}
{Starkenburg}, E., {Hill}, V., {Tolstoy}, E., {Gonz{\'a}lez Hern{\'a}ndez},
  J.~I., {Irwin}, M., {Helmi}, A., {Battaglia}, G., {Jablonka}, P.,
  {Tafelmeyer}, M., {Shetrone}, M., {Venn}, K., \& {de Boer}, T. 2010, A\&A,
  513, A34

\bibitem[{{Susa} {et~al.}(2014){Susa}, {Hasegawa}, \& {Tominaga}}]{susa2014}
{Susa}, H., {Hasegawa}, K., \& {Tominaga}, N. 2014, \apj, 792, 32

\bibitem[{{Travaglio} {et~al.}(2004){Travaglio}, {Gallino}, {Arnone}, {Cowan},
  {Jordan}, \& {Sneden}}]{travaglio}
{Travaglio}, C., {Gallino}, R., {Arnone}, E., {Cowan}, J., {Jordan}, F., \&
  {Sneden}, C. 2004, ApJ, 601, 864

\bibitem[{{Venn} {et~al.}(2012){Venn}, {Shetrone}, {Irwin}, {Hill}, {Jablonka},
  {Tolstoy}, {Lemasle}, {Divell}, {Starkenburg}, {Letarte}, {Baldner},
  {Battaglia}, {Helmi}, {Kaufer}, \& {Primas}}]{venn2012}
{Venn}, K.~A., {Shetrone}, M.~D., {Irwin}, M.~J., {Hill}, V., {Jablonka}, P.,
  {Tolstoy}, E., {Lemasle}, B., {Divell}, M., {Starkenburg}, E., {Letarte}, B.,
  {Baldner}, C., {Battaglia}, G., {Helmi}, A., {Kaufer}, A., \& {Primas}, F.
  2012, \apj, 751, 102

\bibitem[{{Wood} {et~al.}(2013){Wood}, {Lawler}, {Sneden}, \&
  {Cowan}}]{wood_tiII}
{Wood}, M.~P., {Lawler}, J.~E., {Sneden}, C., \& {Cowan}, J.~J. 2013, \apjs,
  208, 27

\bibitem[{{Yanny} {et~al.}(2009){Yanny}, {Rockosi}, {Newberg}, {Knapp},
  {Adelman-McCarthy}, {Alcorn}, {Allam}, {Allende Prieto}, {An}, {Anderson},
  {Anderson}, {Bailer-Jones}, {Bastian}, {Beers}, {Bell}, {Belokurov},
  {Bizyaev}, {Blythe}, {Bochanski}, {Boroski}, {Brinchmann}, {Brinkmann},
  {Brewington}, {Carey}, {Cudworth}, {Evans}, {Evans}, {Gates}, {G{\"a}nsicke},
  {Gillespie}, {Gilmore}, {Nebot Gomez-Moran}, {Grebel}, {Greenwell}, {Gunn},
  {Jordan}, {Jordan}, {Harding}, {Harris}, {Hendry}, {Holder}, {Ivans},
  {Ivezi{\v c}}, {Jester}, {Johnson}, {Kent}, {Kleinman}, {Kniazev},
  {Krzesinski}, {Kron}, {Kuropatkin}, {Lebedeva}, {Lee}, {French Leger},
  {L{\'e}pine}, {Levine}, {Lin}, {Long}, {Loomis}, {Lupton}, {Malanushenko},
  {Malanushenko}, {Margon}, {Martinez-Delgado}, {McGehee}, {Monet}, {Morrison},
  {Munn}, {Neilsen}, {Nitta}, {Norris}, {Oravetz}, {Owen}, {Padmanabhan},
  {Pan}, {Peterson}, {Pier}, {Platson}, {Re Fiorentin}, {Richards}, {Rix},
  {Schlegel}, {Schneider}, {Schreiber}, {Schwope}, {Sibley}, {Simmons},
  {Snedden}, {Allyn Smith}, {Stark}, {Stauffer}, {Steinmetz}, {Stoughton},
  {SubbaRao}, {Szalay}, {Szkody}, {Thakar}, {Sivarani}, {Tucker}, {Uomoto},
  {Vanden Berk}, {Vidrih}, {Wadadekar}, {Watters}, {Wilhelm}, {Wyse}, {Yarger},
  \& {Zucker}}]{segueI}
{Yanny}, B., {Rockosi}, C., {Newberg}, H.~J., {Knapp}, G.~R.,
  {Adelman-McCarthy}, J.~K., {Alcorn}, B., {Allam}, S., {Allende Prieto}, C.,
  {An}, D., {Anderson}, K.~S.~J., {Anderson}, S., {Bailer-Jones}, C.~A.~L.,
  {Bastian}, S., {Beers}, T.~C., {Bell}, E., {Belokurov}, V., {Bizyaev}, D.,
  {Blythe}, N., {Bochanski}, J.~J., {Boroski}, W.~N., {Brinchmann}, J.,
  {Brinkmann}, J., {Brewington}, H., {Carey}, L., {Cudworth}, K.~M., {Evans},
  M., {Evans}, N.~W., {Gates}, E., {G{\"a}nsicke}, B.~T., {Gillespie}, B.,
  {Gilmore}, G., {Nebot Gomez-Moran}, A., {Grebel}, E.~K., {Greenwell}, J.,
  {Gunn}, J.~E., {Jordan}, C., {Jordan}, W., {Harding}, P., {Harris}, H.,
  {Hendry}, J.~S., {Holder}, D., {Ivans}, I.~I., {Ivezi{\v c}}, {\v Z}.,
  {Jester}, S., {Johnson}, J.~A., {Kent}, S.~M., {Kleinman}, S., {Kniazev}, A.,
  {Krzesinski}, J., {Kron}, R., {Kuropatkin}, N., {Lebedeva}, S., {Lee}, Y.~S.,
  {French Leger}, R., {L{\'e}pine}, S., {Levine}, S., {Lin}, H., {Long}, D.~C.,
  {Loomis}, C., {Lupton}, R., {Malanushenko}, O., {Malanushenko}, V., {Margon},
  B., {Martinez-Delgado}, D., {McGehee}, P., {Monet}, D., {Morrison}, H.~L.,
  {Munn}, J.~A., {Neilsen}, Jr., E.~H., {Nitta}, A., {Norris}, J.~E.,
  {Oravetz}, D., {Owen}, R., {Padmanabhan}, N., {Pan}, K., {Peterson}, R.~S.,
  {Pier}, J.~R., {Platson}, J., {Re Fiorentin}, P., {Richards}, G.~T., {Rix},
  H.-W., {Schlegel}, D.~J., {Schneider}, D.~P., {Schreiber}, M.~R., {Schwope},
  A., {Sibley}, V., {Simmons}, A., {Snedden}, S.~A., {Allyn Smith}, J.,
  {Stark}, L., {Stauffer}, F., {Steinmetz}, M., {Stoughton}, C., {SubbaRao},
  M., {Szalay}, A., {Szkody}, P., {Thakar}, A.~R., {Sivarani}, T., {Tucker},
  D., {Uomoto}, A., {Vanden Berk}, D., {Vidrih}, S., {Wadadekar}, Y.,
  {Watters}, S., {Wilhelm}, R., {Wyse}, R.~F.~G., {Yarger}, J., \& {Zucker}, D.
  2009, \aj, 137, 4377

\bibitem[{{Yong} {et~al.}(2013{\natexlab{a}}){Yong}, {Norris}, {Bessell},
  {Christlieb}, {Asplund}, {Beers}, {Barklem}, {Frebel}, \& {Ryan}}]{yong13_II}
{Yong}, D., {Norris}, J.~E., {Bessell}, M.~S., {Christlieb}, N., {Asplund}, M.,
  {Beers}, T.~C., {Barklem}, P.~S., {Frebel}, A., \& {Ryan}, S.~G.
  2013{\natexlab{a}}, \apj, 762, 26

\bibitem[{{Yong} {et~al.}(2013{\natexlab{b}}){Yong}, {Norris}, {Bessell},
  {Christlieb}, {Asplund}, {Beers}, {Barklem}, {Frebel}, \&
  {Ryan}}]{yong13_III}
---. 2013{\natexlab{b}}, \apj, 762, 27

\bibitem[{{York} {et~al.}(2000){York}, {Adelman}, {Anderson}, {Anderson},
  {Annis}, {Bahcall}, {Bakken}, {Barkhouser}, {Bastian}, {Berman}, {Boroski},
  {Bracker}, {Briegel}, {Briggs}, {Brinkmann}, {Brunner}, {Burles}, {Carey},
  {Carr}, {Castander}, {Chen}, {Colestock}, {Connolly}, {Crocker}, {Csabai},
  {Czarapata}, {Davis}, {Doi}, {Dombeck}, {Eisenstein}, {Ellman}, {Elms},
  {Evans}, {Fan}, {Federwitz}, {Fiscelli}, {Friedman}, {Frieman}, {Fukugita},
  {Gillespie}, {Gunn}, {Gurbani}, {de Haas}, {Haldeman}, {Harris}, {Hayes},
  {Heckman}, {Hennessy}, {Hindsley}, {Holm}, {Holmgren}, {Huang}, {Hull},
  {Husby}, {Ichikawa}, {Ichikawa}, {Ivezi{\'c}}, {Kent}, {Kim}, {Kinney},
  {Klaene}, {Kleinman}, {Kleinman}, {Knapp}, {Korienek}, {Kron}, {Kunszt},
  {Lamb}, {Lee}, {Leger}, {Limmongkol}, {Lindenmeyer}, {Long}, {Loomis},
  {Loveday}, {Lucinio}, {Lupton}, {MacKinnon}, {Mannery}, {Mantsch}, {Margon},
  {McGehee}, {McKay}, {Meiksin}, {Merelli}, {Monet}, {Munn}, {Narayanan},
  {Nash}, {Neilsen}, {Neswold}, {Newberg}, {Nichol}, {Nicinski}, {Nonino},
  {Okada}, {Okamura}, {Ostriker}, {Owen}, {Pauls}, {Peoples}, {Peterson},
  {Petravick}, {Pier}, {Pope}, {Pordes}, {Prosapio}, {Rechenmacher}, {Quinn},
  {Richards}, {Richmond}, {Rivetta}, {Rockosi}, {Ruthmansdorfer}, {Sandford},
  {Schlegel}, {Schneider}, {Sekiguchi}, {Sergey}, {Shimasaku}, {Siegmund},
  {Smee}, {Smith}, {Snedden}, {Stone}, {Stoughton}, {Strauss}, {Stubbs},
  {SubbaRao}, {Szalay}, {Szapudi}, {Szokoly}, {Thakar}, {Tremonti}, {Tucker},
  {Uomoto}, {Vanden Berk}, {Vogeley}, {Waddell}, {Wang}, {Watanabe},
  {Weinberg}, {Yanny}, {Yasuda}, \& {SDSS Collaboration}}]{york_sdss}
{York}, D.~G., {Adelman}, J., {Anderson}, Jr., J.~E., {Anderson}, S.~F.,
  {Annis}, J., {Bahcall}, N.~A., {Bakken}, J.~A., {Barkhouser}, R., {Bastian},
  S., {Berman}, E., {Boroski}, W.~N., {Bracker}, S., {Briegel}, C., {Briggs},
  J.~W., {Brinkmann}, J., {Brunner}, R., {Burles}, S., {Carey}, L., {Carr},
  M.~A., {Castander}, F.~J., {Chen}, B., {Colestock}, P.~L., {Connolly}, A.~J.,
  {Crocker}, J.~H., {Csabai}, I., {Czarapata}, P.~C., {Davis}, J.~E., {Doi},
  M., {Dombeck}, T., {Eisenstein}, D., {Ellman}, N., {Elms}, B.~R., {Evans},
  M.~L., {Fan}, X., {Federwitz}, G.~R., {Fiscelli}, L., {Friedman}, S.,
  {Frieman}, J.~A., {Fukugita}, M., {Gillespie}, B., {Gunn}, J.~E., {Gurbani},
  V.~K., {de Haas}, E., {Haldeman}, M., {Harris}, F.~H., {Hayes}, J.,
  {Heckman}, T.~M., {Hennessy}, G.~S., {Hindsley}, R.~B., {Holm}, S.,
  {Holmgren}, D.~J., {Huang}, C.-h., {Hull}, C., {Husby}, D., {Ichikawa},
  S.-I., {Ichikawa}, T., {Ivezi{\'c}}, {\v Z}., {Kent}, S., {Kim}, R.~S.~J.,
  {Kinney}, E., {Klaene}, M., {Kleinman}, A.~N., {Kleinman}, S., {Knapp},
  G.~R., {Korienek}, J., {Kron}, R.~G., {Kunszt}, P.~Z., {Lamb}, D.~Q., {Lee},
  B., {Leger}, R.~F., {Limmongkol}, S., {Lindenmeyer}, C., {Long}, D.~C.,
  {Loomis}, C., {Loveday}, J., {Lucinio}, R., {Lupton}, R.~H., {MacKinnon}, B.,
  {Mannery}, E.~J., {Mantsch}, P.~M., {Margon}, B., {McGehee}, P., {McKay},
  T.~A., {Meiksin}, A., {Merelli}, A., {Monet}, D.~G., {Munn}, J.~A.,
  {Narayanan}, V.~K., {Nash}, T., {Neilsen}, E., {Neswold}, R., {Newberg},
  H.~J., {Nichol}, R.~C., {Nicinski}, T., {Nonino}, M., {Okada}, N., {Okamura},
  S., {Ostriker}, J.~P., {Owen}, R., {Pauls}, A.~G., {Peoples}, J., {Peterson},
  R.~L., {Petravick}, D., {Pier}, J.~R., {Pope}, A., {Pordes}, R., {Prosapio},
  A., {Rechenmacher}, R., {Quinn}, T.~R., {Richards}, G.~T., {Richmond}, M.~W.,
  {Rivetta}, C.~H., {Rockosi}, C.~M., {Ruthmansdorfer}, K., {Sandford}, D.,
  {Schlegel}, D.~J., {Schneider}, D.~P., {Sekiguchi}, M., {Sergey}, G.,
  {Shimasaku}, K., {Siegmund}, W.~A., {Smee}, S., {Smith}, J.~A., {Snedden},
  S., {Stone}, R., {Stoughton}, C., {Strauss}, M.~A., {Stubbs}, C., {SubbaRao},
  M., {Szalay}, A.~S., {Szapudi}, I., {Szokoly}, G.~P., {Thakar}, A.~R.,
  {Tremonti}, C., {Tucker}, D.~L., {Uomoto}, A., {Vanden Berk}, D., {Vogeley},
  M.~S., {Waddell}, P., {Wang}, S.-i., {Watanabe}, M., {Weinberg}, D.~H.,
  {Yanny}, B., {Yasuda}, N., \& {SDSS Collaboration}. 2000, \aj, 120, 1579

\bibitem[{{Zhao} {et~al.}(2006){Zhao}, {Chen}, {Shi}, {Liang}, {Hou}, {Chen},
  {Zhang}, \& {Li}}]{zhao_lamost}
{Zhao}, G., {Chen}, Y.-Q., {Shi}, J.-R., {Liang}, Y.-C., {Hou}, J.-L., {Chen},
  L., {Zhang}, H.-W., \& {Li}, A.-G. 2006, 
  Chinese Journal of Astronomy and Astrophysics, 6, 265

\end{thebibliography}
\end{document}